\documentclass{iopjournal}
\usepackage[T1]{fontenc}
\usepackage[utf8]{inputenc}
\usepackage{lmodern}
\usepackage{microtype}

\usepackage{amsmath,amssymb,amsfonts,mathtools}
\usepackage{bm}
\usepackage{booktabs}
\usepackage{tabularx}
\usepackage{array}
\usepackage[table]{xcolor}
\usepackage{multirow}
\usepackage{graphicx}
\usepackage{enumitem}

\usepackage{hyperref}
\usepackage[nameinlink,noabbrev]{cleveref}
\usepackage[square,numbers,sort&compress]{natbib}
\usepackage[toc]{appendix}
\usepackage{placeins}

\numberwithin{equation}{section}

\newcommand{\sgn}{\operatorname{sgn}}
\newcommand{\Ree}{\operatorname{Re}}
\begin{document}

\articletype{Paper} 

\title{Non-minimally coupled scalar field dark sector of the universe: in-depth (Einstein frame) case study}

\author{Marcin Postolak$^1$\orcid{0000-0003-4868-6358}}

\affil{$^1$University of Wrocław, Institute of Theoretical Physics, Faculty of Physics and Astronomy, pl. M. Borna 9, Wrocław, 50-204, Poland}

\email{marcin.postolak@uwr.edu.pl, postolak.marcin@gmail.com}


\begin{abstract}
In this study, motivated by recent results from DESI DR2 \cite{DESI:2025zgx,DESI:2025fii} indicating a possible preference for evolving dark energy in some combined data analyses, we analyze spatially flat FLRW interacting scalar-tensor cosmological models with non-minimal coupling (NMC) between the scalar field (SF) and matter/cosmological dust in the Einstein conformal frame. By using modified expansion normalized variables that account for negative values of the scalar field potential, we derive cosmological dynamical system equations and expressions for physical variables. Five specific scalar field models (axions/ALPs, cyclic ekpyrotic, exponential with a constant, quintessence, and SFDM) are examined in depth to determine how they evolve, as they serve as representative candidates used in the literature for the evolving dark energy in the late Universe. Using appropriate mathematical methods (e.g. linear stability, center manifold and Poincar\'e sphere), we present critical points along with their character and physical interpretation with respect to the possible evolution of the Universe. Considering both positive and negative values of the coupling parameter allows us to examine the transfer of energy from the scalar sector to dust and from matter to the scalar field. Considering four representative absolute values of this parameter provides a comprehensive analysis that incorporates all significant types of dynamical system evolution (from mathematical and physical perspectives). The initial conditions for the in-depth numerical analysis were constructed analytically from observationally motivated Planck 2018 \cite{Planck:2018vyg} and DESI DR2 \cite{DESI:2025zgx} CPL parametrizations.
\end{abstract}

\section{Introduction}
\label{sec:intro}

This work is a dynamical system\footnote{More details on the application of dynamical systems methods in cosmology can be found in \cite{Wainwright1997,Coley2003,Bahamonde:2017ize}.} (background level) case study of a wide range of scalar-tensor cosmological models with non-minimal coupling in the Einstein frame for a spatially flat FLRW universe. The aim is to establish a possible cosmological chronology of evolution in each of the specific models (radiation domination, intermediate scaling regimes, matter domination, late-time acceleration, ekpyrotic contraction in the case of negative\footnote{A detailed discussion of negative SF potentials in cosmology can be found in \cite{Felder:2002jk}.} branch of the potential, etc.) in the formalism corresponding to their various critical points.

The DESI DR2 data \cite{DESI:2025zgx,DESI:2025fii} (baryon acoustic oscillations (BAO) measurements) provide geometric observables about the cosmic evolution, which are highly sensitive to the Hubble parameter $H(z)$ and related physical quantities. When these data are compared with external datasets related to the CMB, such as Planck 2018 \cite{Planck:2018vyg}, and supernovae, such as Pantheon+ \cite{Brout:2022vxf}, a statistical preference seems to favor \textit{dynamical/evolving dark energy} (e.g. quintessence \cite{Ratra:1987rm,Wetterich:1987fm,Caldwell:1997ii,Zlatev:1998tr,Steinhardt:1999nw,Brax:1999yv,Chiba:1999wt,Caldwell:2005tm,Linder:2006sv,Amendola_Tsujikawa_2010q,Tsujikawa:2013fta}), usually described by the redshift-dependent equation of state parameter $\omega(z)$, over the standard cosmological $\Lambda$CDM (cosmological constant) paradigm \cite{Perivolaropoulos:2021jda}. Nevertheless, it should be clearly stated that the \textit{issue regarding the nature of dark energy is still far from being definitively resolved} and is the subject of intense debate in the scientific community \cite{DESI:2025wyn,Rezaei:2025vhb,Capozziello:2025qmh}.

An alternative interpretation for the observational results is the \textit{interacting dark energy} (\textit{interacting dark sector}) \cite{Copeland:2006wr,Clifton:2011jh,Wang:2016lxa}. This leads to modifications in the evolution of the cosmological model, affecting both the background level and the formation of the large-scale structure of the Universe (cosmological perturbations). Recent analyses of available data indicate that non-gravitational interactions (beyond minimal metric coupling, e.g., the hypothetical 5th force) between dark energy and dark matter might be compatible with observations and, moreover, have the potential to explain certain cosmological tensions (e.g., Hubble tension) for specific combinations of data sets \cite{He:2010im,DiValentino:2017iww,Barros:2018efl,CosmoVerseNetwork:2025alb}. The publication of data within the DESI DR2 also poses other phenomenological challenges within the formalism of dark interaction, such as the interpretation of the \textit{apparent phantom behavior} of the dark energy equation of state within this formalism \cite{Pan:2025qwy,Guedezounme:2025wav,Li:2026xaz} or the \textit{possible analytical forms of the interaction term} and problems associated with them \cite{vanderWesthuizen:2025vcb,vanderWesthuizen:2025mnw,vanderWesthuizen:2025rip}.

In this article, we explain how interacting scalar-tensor models in the Einstein frame can result in the evolving effective equation of state $\omega_{\mathrm{eff}}(z)$ due to evolution between specific fixed points of the dynamical system. In addition, we determine the nature of the evolution of the scalar sector in individual models based on the CPL parameterization \cite{Chevallier:2000qy,Linder:2002et} and the two basic types of behavior associated with it, i.e., ‘\textit{freezing}’ and ‘\textit{thawing}’ \cite{Linder:2006sv,dePutter:2008wt}.

\section{NMC Einstein frame model}
\label{sec:model}

\subsection{Action and NMC}
\label{subsec:action}

The starting point for our considerations is the following action involving NMC between the SF and matter in the Einstein frame:\footnote{In our convention: $\kappa^{2}\equiv 8\pi G=1$.}
\begin{equation}
\label{eq:action}
S\bigl[g_{\mu\nu},\phi,\chi\bigr]=\int d^{4}x\sqrt{-g}\left[\frac{1}{2\kappa^{2}} R-\frac{1}{2}g^{\mu\nu}\partial_{\mu}\phi\,\partial_{\nu}\phi-\mathcal{V}(\phi)\right]+S_{m}\bigl[f^{2}(\phi)\,g_{\mu\nu},\chi\bigr]\,,
\end{equation}
where:
\begin{equation}
\label{eq:fexp}
f(\phi)=e^{\beta\phi}\,.
\end{equation}
The action \eqref{eq:action} along with the function \eqref{eq:fexp} defining the form of non-minimal coupling is characteristic for the Einstein frame models associated with \textit{coupled quintessence} phenomenology \cite{Amendola:1999er,Wang:2025znm}, \textit{chameleon mechanism} \cite{Khoury:2003rn,Brax:2004qh}, the problem of \textit{separating baryonic matter from dark matter} \cite{Borowiec:2023kmq}, and \textit{dilaton-like coupling} \cite{Wetterich:1987fm,Damour:1994zq,Polchinski_1998dilaton,Gasperini_2007dilaton,Bedroya:2025fwh}. Moreover, this form of action can be obtained via a conformal transformation, which allows for a move from the Jordan frame to the Einstein frame.

In the case we are considering, the matter sector moves along the geodesics that depend on the conformally transformed metric - the \textit{dark sector metric} \cite{Borowiec:2023kmq}:
\begin{equation}
    \tilde{g}_{\mu\nu}\equiv f^{2}(\phi)\,g_{\mu\nu}\,,
\end{equation}
which results in the presence of an additional force (interaction term), the so-called \textit{fifth force} in the Einstein frame. The (non)continuity equation for the matter sector and Klein-Gordon equation for the SF contain such a term that is proportional to the factor:
\begin{equation}
    \beta\equiv\frac{d\ln{f(\phi)}}{d\phi}=\mathrm{const}\,.
\end{equation}

\subsection{Spatially flat FLRW Universe and effective interacting sector}
\label{subsec:background}

For a spatially flat FLRW Universe and a homogeneous scalar field $\phi=\phi(t)$, equations of motion take the following form:
\begin{align}
\label{eq:Friedmann}
3H^2 &= \rho_r+\rho_m+\frac12\dot\phi^2+\mathcal{V}(\phi)\,,
\\
\label{eq:Ray}
\dot H &= -\frac12\left(\frac43\rho_r+\rho_m+\dot\phi^2\right)\le 0\,,
\\
\label{eq:KG}
\ddot\phi+3H\dot\phi+\mathcal{V}'(\phi) & =\beta\,\rho_m\,,
\end{align}
with the continuity (for radiation) and non-continuity equations (for matter):
\begin{equation}
\label{eq:cont_r_m}
\dot\rho_r+4H\rho_r=0 \quad\land\quad \dot\rho_m+3H\rho_m=-\beta\,\dot\phi\,\rho_m\,.
\end{equation}
It has been shown in an earlier work \cite{Borowiec:2023kmq} that the non-continuity equation \eqref{eq:cont_r_m} leads to a general expression for the energy density depending on both the scale factor and the scalar field, which is of the form \cite{Borowiec:2023kmq}:
\begin{equation}
    \rho_{i}(a,\phi)=f^{4}(\phi)\,\rho_{i}\bigl[af(\phi)\bigr]\,,
\end{equation}
and in the case of dust yields:
\begin{equation}
    \rho_{m}(a,\phi)\propto f(\phi)\,a^{-3}=e^{\beta\phi}\,a^{-3}\,.
\end{equation}
\paragraph{Interpretation}
The equation of motion for the scalar field \eqref{eq:KG} (more precisely, its right-hand side) can be interpreted as a kind of energy exchange in the case of the interacting sector model - the scalar field (in our case, playing the role of dark energy in the current epoch) is directly driven by the matter/dust sector (e.g., dark matter).

From a phenomenological point of view, a more natural interpretation is the interaction between the scalar field and dark matter due to very strong constraints on possible baryonic matter interactions \cite{Bertotti:2003rm,Williams:2004qba,Uzan:2010pm,Buen-Abad:2021mvc}. However, the dynamical system we are considering does not distinguish between 'ordinary' matter and dark matter (it only takes into account the existence of cosmological dust).

\section{Autonomous dynamical system}
\label{sec:ds}
\subsection{Variables and constraints}
\label{subsec:variables}

Let us define the time parameter in our dynamical system (number of e-folds):
\begin{equation}
\label{eq:N-variable}
    N\equiv\ln{a}
\end{equation}
and \textit{modified EN\footnote{Expansion normalized variables \cite{Wainwright1997}.} variables}:
\begin{equation}
\label{eq:vars}
x\equiv\frac{\dot\phi}{\sqrt{6}H}\quad\land\quad u\equiv\frac{\mathcal{V}(\phi)}{3H^2} \quad\land\quad \Omega_{r}\equiv\frac{\rho_r}{3H^2} \quad\land\quad \Omega_{m}\equiv\frac{\rho_m}{3H^2}\,.
\end{equation}
The constraint equation is:
\begin{equation}
\label{eq:constraint}
1=x^2+u+\Omega_{r}+\Omega_{m} \quad\implies\quad \Omega_m=1-x^2-u-\Omega_r\,.
\end{equation}
Our choice of the $u$ variable corresponding to the potential energy of the scalar field allows us to take into account the possible negative values of the potential.

Furthermore, in order to close our cosmological dynamical system\footnote{This can be done if the dependence of the parameter $\Gamma_{\phi}$ on the variable $\lambda_{\phi}$ can be expressed in an explicit analytical form. Otherwise, the dynamical system cannot be closed.}, let us also introduce variables describing the slope of the SF potential, namely:
\begin{equation}
\label{eq:lambdaGamma}
\lambda_{\phi}\equiv -\frac{\mathcal{V}'(\phi)}{\mathcal{V}(\phi)}
\quad\land\quad \Gamma_{\phi}\left(\lambda_{\phi}\right)\equiv\frac{\mathcal{V}''(\phi)\,\mathcal{V}(\phi)}{\mathcal{V}'(\phi)^2}\,.
\end{equation}
If the scalar field potentials are exponential:
\begin{equation}
    \mathcal{V}(\phi)\propto e^{-\lambda\phi}\,,
\end{equation}
then:
\begin{equation}
    \lambda_{\phi}=\lambda=\mathrm{const}\,,
\end{equation}
In such a case, the dynamical system reduces to a system with a well-known set of critical points (kinetic energy domination/stiff matter, SF domination, and matter/radiation scaling solutions) with explicit existence and stability conditions (our $\beta\to 0$ limiting case) \cite{Copeland:1997et}.

\subsection{Useful identities and observables}
\label{subsec:obs_map}

From \eqref{eq:Ray} and \eqref{eq:constraint} one obtains:
\begin{equation}
\label{eq:Hdot}
\frac{\dot H}{H^2}=-3x^2-2\,\Omega_{r}-\frac{3}{2}\Omega_m=\frac{1}{2}\Big[3\left(-1-x^2+u\right)-\Omega_r\Big]\,.
\end{equation}
Moreover, using \eqref{eq:KG} and \eqref{eq:constraint} yields:
\begin{equation}
    \frac{\ddot{\phi}}{H \dot{\phi}}=-3+\sqrt{\frac{3}{2}}\frac{\lambda_{\phi}\,u+\beta\,\Omega_{m}}{x}=-3+\sqrt{\frac{3}{2}}\,\frac{\lambda_{\phi}\,u+\beta \left(1-x^2-u-\Omega_{r}\right)}{x}\,.
\end{equation}
The next step is to define the physical parameters that describe our cosmological model, namely:
\begin{equation}\label{eq:Omega-phi-omega-phi}
\Omega_{\phi}\equiv x^2+u \quad\land\quad \omega_\phi=\frac{x^2-u}{x^2+u}\quad\left(\Omega_\phi\neq 0\right)
\end{equation}
and:
\begin{equation}
    \omega_{\mathrm{eff}}\equiv\frac{p_{\mathrm{tot}}}{\rho_{\mathrm{tot}}}=\omega_{r}\,\Omega_{r}+\omega_{m}\,\Omega_{m}+\omega_{\phi}\,\Omega_{\phi}=\frac{1}{3}\Omega_r+x^2-u\,.
\end{equation}
Using the above relationships, we can derive a key formula that describes how the Hubble parameter varies over the time variable:
\begin{equation}
\label{eq:Hdot_weff}
\frac{d\ln{H}}{d N}\equiv\frac{1}{H}\frac{dH}{dN}=\frac{\dot H}{H^2}=-\frac{3}{2}\left(1+\omega_{\mathrm{eff}}\right)\,.
\end{equation}
Therefore, any evolution in phase space described by the effective equation of state parameter, $\omega_{\mathrm{eff}}(N)$, generates a specific evolution of the Hubble parameter, $H(N)$. This yields changes in distance measurements (sound horizon) for the BAO.

\subsection{Final autonomous system}
\label{subsec:DSfinal}

\begin{figure}[htbp]
    \centering
    \includegraphics[width=0.92\linewidth]{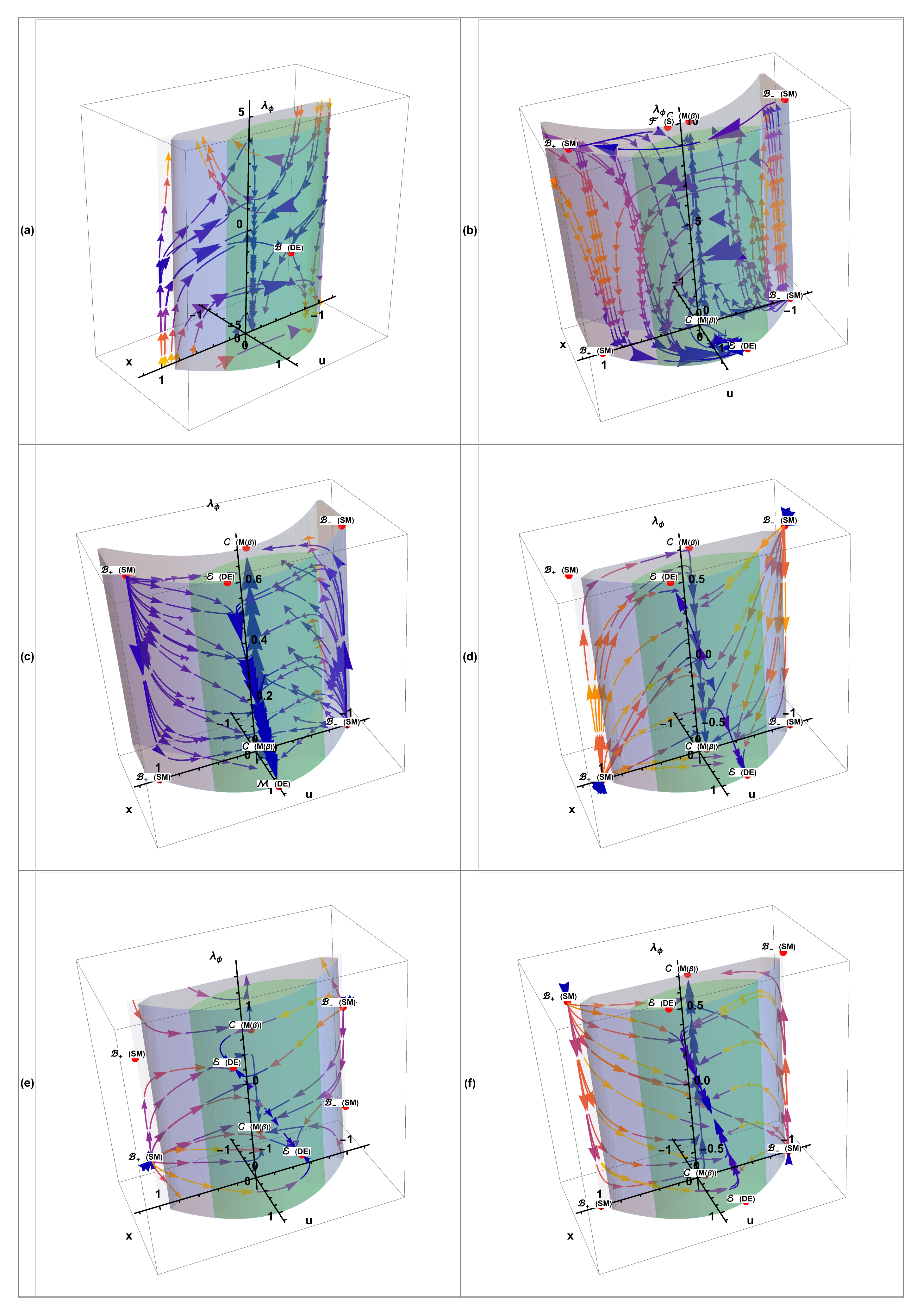}
    \caption{Phase space evolution of the dynamical system \eqref{eq:DSfinal} (without radiation) for specific models considered in this article for $\beta=-10^{-1}$ and DESI DR2+CMB+Pantheon+ free parameters/initial conditions: (a) Axions/ALPs, (b) Cyclic ekpyrotic, (c) Exponential+$\Lambda$, (d) Quintessence for $n=-1$, (e) Quintessence for $n=-2$, (f) SFDM. Green region denotes the accelerated expansion phase.}
    \label{fig:DS-phase-space-models}
\end{figure}

The cosmological models under consideration are defined by a 4D dynamical system\footnote{From a mathematical point of view, the system is actually 5D. However, the exponential form of NMC causes the fifth variable to become constant.} expressed by the following equations:
\begin{subequations}
\label{eq:DSfinal}
\begin{align}
\frac{dx}{dN} &= \frac{1}{2}\left[3x^3-\sqrt{6}\,\beta\,x^2+x\,\left(\Omega_{r}-3u-3\right)+\sqrt{6}\Bigl(\lambda_{\phi} u-\beta \left(u+\Omega_{r}-1\right)\Bigr)\right]\,,
\\
\frac{du}{dN} &= u\Big(3(1+x^2-u)-\sqrt6\,\lambda_\phi x+\Omega_r\Big)\,,
\\
\frac{d \Omega_r}{dN} &= \Omega_r\Big(3(x^2-u)+\Omega_r-1\Big),
\\
\frac{d \lambda_\phi}{dN} &= -\sqrt6\,\Bigl(\Gamma_\phi\left(\lambda_\phi\right)-1\Bigr)\,\lambda_\phi^2\,x\,.
\end{align}
\end{subequations}
Physical trajectories satisfy:
\begin{equation}
    \Omega_{r}\ge 0 \quad\land\quad \Omega_{m}\ge 0\,.
\end{equation}
Fig.~\ref{fig:DS-phase-space-models} illustrates examples showing the evolution of the dynamical system for the models under consideration in this article.
\section{Stability methodology and expanding/contracting phase}
\label{sec:stability_methods}

Let us define:
\begin{equation}
    \vec{X}\equiv\bigl(x,u,\Omega_{r},\lambda_{\phi}\bigr)
\end{equation}
and:
\begin{equation}
    \frac{d\vec{X}}{dN}=\vec{F}\left(\vec{X}\right)\,.
\end{equation}
A fixed (critical) point denoted by $\vec{X}_{*}$ satisfies the condition:
\begin{equation}
    \vec{F}\left(\vec{X}_{*}\right)=0\,.
\end{equation}
Linearization around the critical point gives:
\begin{equation}
\delta\vec{X}'=J\left(\vec{X}_{*}\right)\,\delta\vec{X} \quad\land\quad
J_{ij}=\frac{\partial F_i}{\partial X_j}\Bigg|_{\vec{X}_{*}}\,.
\end{equation}
If $\Ree(\mu_i)\neq 0$ for all the eigenvalues $\mu_i$ of the Jacobian matrix $J$, then the point is called a \textit{hyperbolic fixed point} and its character follows from the signs of $\Ree(\mu_i)$\cite{Perko2001,Wiggins2003}. If one or more of the eigenvalues satisfy the condition $\Ree(\mu)=0$, then the point is called a \textit{non-hyperbolic fixed point}. In such a case one can use the center manifold method (Appendix~\ref{app:center}).

Because our time variable is defined by \eqref{eq:N-variable}, the following conclusions can be drawn for our stability analysis:
\begin{itemize}
\item For expanding branch ($H>0$): \textit{attractor} (\textit{stable} fixed point) requires $\Ree(\mu_i)<0$;
\item For contracting branch ($H<0$): the passage of time corresponds to decreasing values of the $N$ parameter and an \textit{attractor} corresponds to $\Ree(\mu_i)>0$ (e.g., ekpyrotic contraction \cite{Lehners:2008vx}).
\end{itemize}

\section{SF potentials and the closure of dynamical systems}
\label{sec:potentials}

Our analysis is based on five specific self-interaction potentials for the scalar field, $\mathcal{V}(\phi)$, that are well-motivated in theoretical physics, including high-energy physics and cosmological phenomenology. For each model, we derive an explicit form for the variable $\lambda_{\phi}$ and the parameter $\Gamma_{\phi}$ in order to obtain a closed autonomous dynamical system.

\subsection{Axions/ALPs}
\label{subsec:pot_alp}

The \textit{axion} particle arises as a \textit{pseudo-Nambu-Goldstone boson} (pNGB) associated with the \textit{Peccei-Quinn} (PQ) mechanism proposed to solve the so-called \textit{strong CP problem} in quantum chromodynamics (QCD) \cite{Peccei:1977hh,Peccei:1977ur,Weinberg:1977ma,Wilczek:1977pj,DiLuzio:2020wdo}. The generalization of axions are axion-like particles (ALPs) \cite{Arza:2026rsl}, whose existence is postulated in many theoretical extensions of the Standard Model and various versions of UV theories - string theory (e.g., radion and moduli fields). They constitute a well-established cosmological sector \cite{Kim:1986ax,Kuster2008,Kawasaki:2013ae,Choi:2020rgn} connected, in particular, with the phenomenology of \textit{ultralight bosonic dark matter} (UBDM) \cite{Kimball:2023vxk} and \textit{early dark energy} (EDE) \cite{Doran:2006kp,Kamionkowski:2022pkx,Poulin:2023lkg}.

In cosmology, axions and ALPs can be produced non-thermally (e.g., as a result of the vacuum misalignment mechanism \cite{Preskill:1982cy,Dine:1982ah,Abbott:1982af}), resulting in the presence of a coherent classical scalar field behaving effectively like CDM on cosmological scales. This is accompanied by characteristic wave-like effects on smaller scales \cite{Preskill:1982cy,Marsh:2015xka,Poulin:2018dzj}. The specific behavior of axions/ALPs is associated with astrophysical observational limitations \cite{Carenza:2024ehj}. Additionally, the ultralight mass regime of these bosons is aligned with scalar field dark matter (SFDM) and fuzzy cold dark matter (FCDM) models \cite{Hu:2000ke,Ringwald:2012hr,Chadha-Day:2021szb,Adams:2022pbo,Matos:2023usa,OHare:2024nmr}.

For the purposes of our analysis, we consider a (simplified) periodic version of the SF potential with the following form:
\begin{equation}
\label{eq:V_ALP}
\mathcal{V}(\phi)=\Lambda_{c}^{4}\left(1-\cos\phi\right),
\end{equation}
therefore:
\begin{equation}
\lambda_\phi=-\cot{\left(\frac{\phi}{2}\right)} \quad\land\quad \Gamma_{\phi}\left(\lambda_\phi\right)=\frac{1}{2}\left(1-\frac{1}{\lambda_{\phi}^2}\right)
\end{equation}
and:
\begin{equation}
    \frac{d \lambda_{\phi}}{dN}=\sqrt{\frac{3}{2}}\,x\,\left(1+\lambda_\phi^2\right)\,.
\end{equation}

\subsection{Cyclic ekpyrotic}
\label{subsec:pot_cyclic}

The ekpyrotic and cyclic ekpyrotic scenarios aspire to replace the conventional inflationary paradigm \cite{Starobinsky1980,Guth1981,Linde1982,Brandenberger:1999sw,Penrose:1988mg,Martin:2013tda,Martin:2019zia,Martin:2024qnn,Postolak:2024xtm} with a phase of ultra-slow contraction driven by a very steep negative branch of the SF potential. The effective 4D (scalar-tensor) formalism arising from the higher-dimensional brane motivated model \cite{Horava:1995qa,Horava:1996ma,Khoury:2001wf} employs the difference between an exponent with a positive coefficient and an exponent with a negative coefficient \cite{Erickson:2006wc,Lehners:2008vx}, in both cases with explicit dependence on the scalar field.

For this reason, the next model we consider is a potential with two exponential functions of the form:

\begin{equation}
\label{eq:V_cyclic}
\mathcal{V}(\phi)=V_{0}\Bigl(e^{b\phi}-e^{-c\phi}\Bigr)\,,
\end{equation}
from which one obtains:
\begin{equation}
    \lambda_{\phi}=-\frac{b\,e^{(b+c)\phi}+c}{e^{(b+c)\phi}-1}\quad\land\quad \Gamma_{\phi}\left(\lambda_{\phi}\right)=\frac{b\,c}{\lambda_{\phi}^{2}}+\frac{c-b}{\lambda_{\phi}}\,,
\end{equation}
and the autonomous system is described by the expression:
\begin{equation}
\label{eq:lamprime_cyclic}
\frac{d \lambda_\phi}{dN}=-\sqrt6\,x\,\bigl(\lambda_\phi+b\bigr)\bigl(c-\lambda_\phi\bigr)\,.
\end{equation}
The constant branches (in other words, the critical manifolds in the $\lambda_\phi$ sector) are as follows:
\begin{equation}
\label{eq:cyclic-branch}
\lambda_{\phi*}=-b \quad\land\quad \lambda_{\phi*}=c\,.
\end{equation}

\subsection{Exponential with constant (another version of cyclic ekpyrotic)}
\label{subsec:pot_expL}
Another class of models used in cyclic ekpyrotic cosmological models is scalar-tensor theories with a self-interaction SF potential that contains a steep negative exponential function accompanied by an additional plateau (constant) \cite{Steinhardt:2002ih,Lehners:2008vx,Novello:2008ra,Nastase2019}.

This form of potential may also be treated as a class of exponential potentials, which are widely used in modern theoretical cosmology, depending on the values of the free parameters taken into account. In our numerical solutions later in this work, we will use this type of free parameter, which is more consistent with observational data.

The next scalar field potential, shown on the left as exponential and on the right as cyclic ekpyrotic, is described by the following functions:
\begin{equation}
\label{eq:V_expL}
\mathcal{V}(\phi)=2\Lambda\pm V_{0}\,e^{-c\phi} \quad\lor\quad \mathcal{V}(\phi)=V_{0}\bigl(1-e^{-c\phi}\bigr)\,,
\end{equation}
and therefore the dynamical system variables become:
\begin{equation}
    \lambda_{\phi}=\frac{c\,V_{0}}{V_{0}\pm 2\Lambda\,e^{c\phi}}\quad\lor\quad \lambda_{\phi}=\frac{c}{1-e^{c\phi}}\,.
\end{equation}
In both cases, we ultimately obtain the same form for the parameter $\Gamma_{\phi}$:
\begin{equation}
    \Gamma_{\phi}\left(\lambda_{\phi}\right)=\frac{c}{\lambda_{\phi}}
\end{equation}
that implies the evolution differential equation:
\begin{equation}
\label{eq:lamprime_expL}
\frac{d \lambda_\phi}{dN}=-\sqrt6\,x\,\lambda_{\phi}\bigl(c-\lambda_{\phi}\bigr)\,,
\end{equation}
and therefore, the crucial constant branches are:
\begin{equation}
\lambda_{\phi*}=c \quad\land\quad \lambda_{\phi*}=0\,.
\end{equation}

\subsection{Tracking quintessence}
\label{subsec:pot_quint}
Tracking quintessence models attempt to minimize the influence of initial conditions while allowing for the existence of attractors (late-time accelerated expansion of the Universe) and scaling solutions \cite{Copeland:1997et} in the phase space.

From a mathematical point of view, models with scalar field potentials proportional to hyperbolic functions are useful for this purpose because they allow for controlled asymptotic behavior of the potential. This includes exponential-like behavior for large field values and power-law behavior for small SF values. These types of models are closely related to the phenomenology of quintessence and SFDM \cite{Urena-Lopez:2000ewq,Matos:2000ng,Matos:2000ss}.

We use the following relation for this type of cosmological model \cite{Urena-Lopez:2000ewq,Matos:2000ng,Matos:2000ss}:
\begin{equation}
\label{eq:V_quint}
\mathcal{V}(\phi)=V_0\bigl[\sinh{\left(\lambda\phi\right)}\bigr]^n
\quad\land\quad
n\in\mathbb{Z}^{-},
\end{equation}
whose asymptotic behavior corresponds to the above description, namely:
\begin{equation}
\label{eq:asympt-q-V}
    \mathcal{V}(\phi)=V_0\bigl[\sinh{\left(\lambda\phi\right)}\bigr]^n \propto\begin{dcases}
        \left(\lambda\phi\right)^{n}\,,\quad \left|\lambda\phi\right|\ll 1 \,, \\
        e^{\lambda n \phi}\,,\quad\,\,\,\, \left|\lambda\phi\right|\gg 1 \,.
    \end{dcases}
\end{equation}
In this case, the variable and parameter of the dynamical system are represented by the following formulae:
\begin{equation}
    \lambda_{\phi}=-n \lambda\coth{\left(\lambda\phi\right)}\quad\land\quad \Gamma_{\phi}\left(\lambda_{\phi}\right)=1-\frac{1}{n}+n\frac{\lambda^{2}}{\lambda_{\phi}^{2}}\,,
\end{equation}
for which the dynamical equation becomes:
\begin{equation}
\label{eq:lamprime_quint}
\frac{d \lambda_\phi}{dN}=\frac{\sqrt6}{n}\,x\bigl(\lambda_\phi^2-n^2\lambda^2\bigr)\,,
\end{equation}
so that the constant branches are given by the values of:
\begin{equation}
\label{eq:q-branch-l}
\lambda_{\phi*}=\pm n\lambda\,.
\end{equation}

\subsection{SFDM}
\label{subsec:pot_sfdm}

Models corresponding to SFDM take into account SF potential forms that behave quadratically (e.g., canonical massive SF) near their minimum \cite{Kimball:2023vxk}, i.e.:
\begin{equation}
\label{eq:massive-SF}
    \mathcal{V}(\phi)\simeq\frac{1}{2}m_{\phi}^{2}\phi^{2}\,.
\end{equation}
Thus, they generate coherent oscillations of the SF \cite{Turner:1983he}, whose averaged equation of state corresponds to the one known from the cosmological dust/matter \cite{Ferreira:2020fam}:
\begin{equation}
\label{eq:averaged-SF-dust-EoS}
    \left\langle\omega_{\phi}\right\rangle\simeq 0\,.
\end{equation}
On the other hand, for large field values, the potential behaves like an exponential function, giving place to scaling solutions.

An example of such a model is another potential proportional to a hyperbolic function, this time to the hyperbolic cosine \cite{Matos:2000ng,Matos:2000ss}:
\begin{equation}
\label{eq:V_sfdm}
\mathcal{V}(\phi)=V_0\bigl[\cosh{\left(\lambda\phi\right)}-1\bigr]\,.
\end{equation}
After expanding it into a Taylor series around the minimum, the expression takes the form of \eqref{eq:massive-SF} with:
\begin{equation}
    m_{\phi}^{2}\equiv\lambda^{2}\,V_{0}\,.
\end{equation}
Furthermore, the dynamical system involves the following mathematical relationships:
\begin{equation}
\label{eq:l-SFDM}
    \lambda_{\phi}=-\lambda\coth{\left(\frac{\lambda\phi}{2}\right)}\quad\land\quad \Gamma_{\phi}\left(\lambda_{\phi}\right)=\frac{1}{2}\left(1+\frac{\lambda^{2}}{\lambda_{\phi}^{2}}\right)\,,
\end{equation}
so that:
\begin{equation}
\label{eq:lamprime_sfdm}
\frac{d \lambda_\phi}{dN}=\sqrt{\frac{3}{2}}\,x\left(\lambda_\phi^2-\lambda^2\right)\,,
\end{equation}
and the constant $\lambda_{\phi}$ branches are given by:
\begin{equation}
\lambda_{\phi*}=\pm\lambda\,.
\end{equation}

\section{Detailed analysis}
\label{sec:model_by_model}
\subsection{Physical interpretation of critical points and cosmological chronologies}
\label{subsec:phys_chronologies}
In this section, we present a \textit{physical interpretation} of the critical points listed in Tables~\ref{tab:alp_fp}-\ref{tab:sfdm_char} and classify the corresponding possible cosmological chronologies - physically acceptable trajectories in the phase space of our dynamical system \eqref{eq:DSfinal} limited by constraints \eqref{eq:constraint}. In the language of dynamical systems, an \textit{orbit} is the evolution between transitional epochs, which are described by \textit{saddle points} (or \textit{past attractors}). This evolution is followed by the movement of the trajectory toward \textit{future attractors} (or possibly escape to the \textit{boundary} of the designated phase space). These connections are called \textit{heteroclinic orbits} between invariant manifolds corresponding to individual critical points \cite{Wainwright1997,Coley2003,Bahamonde:2017ize,Perko2001,Wiggins2003}.

\paragraph{Expanding vs contracting branch}
Because our time variable is defined by \eqref{eq:N-variable}, the stability conditions depend on the sign of Hubble parameter (see Sec.~\ref{sec:stability_methods}): for $H>0$ (expansion) an attractor requires $\Ree(\mu_i)<0$, while for $H<0$ (contraction) attractor behavior corresponds to $\Ree(\mu_i)>0$. In particular, ekpyrotic solutions are
attractors in the phase of contraction \cite{Lehners:2008vx}.

\paragraph{Universal 'manual' for Models II-V (constant $\lambda_{\phi*}$ branches)}
For Models II, IV, V (and for the constant-$\lambda_{\phi*}=c$ branch of Model III), the fixed points
$\mathcal{A}$-$\mathcal{G}$ represent the following epochs:
\begin{enumerate}[label=(\alph*)]
\item $\mathcal{A}$ (R): radiation domination, $\omega_{\mathrm{eff}}=1/3$ (early-time saddle);
\item $\mathcal{B}_\pm$ (K$_\pm$): kinetic/stiff scalar regime, $\omega_{\mathrm{eff}}=1$ (past source for expansion);
\item $\mathcal{C}$ ($\phi$MDE): coupled matter-like era with nonzero scalar kinetic fraction,
      $\omega_{\mathrm{eff}}=\tfrac{2}{3}\beta^2$ (intermediate saddle for $\beta^2\le 3/2$);
\item $\mathcal{D}$ (R$_\beta$): coupled radiation-like scaling solution ($\omega_{\mathrm{eff}}=1/3$), which may
      become a late-time attractor if $c/\beta>4$ (or $\lambda_{\phi*}/\beta>4$), \textit{\textbf{phenomenologically disfavored}};
\item $\mathcal{E}$ (P): SF dominated solution with
      $\omega_{\mathrm{eff}}=\lambda_{\phi*}^2/3-1$ (accelerated expansion for $\lambda_{\phi*}^2<2$);
\item $\mathcal{F}$ (S): matter--scalar scaling solution with
      $\omega_{\mathrm{eff}}=\beta/(\lambda_{\phi*}-\beta)$ (acceleration possible for $\omega_{\mathrm{eff}}<-1/3$);
\item $\mathcal{G}$ (RS): radiation-scalar scaling solution with $\omega_{\mathrm{eff}}=1/3$ (typically a saddle).
\end{enumerate}

\paragraph{Turnaround boundary and the role of negative potentials}
From the Friedmann equation \eqref{eq:Friedmann}, the trajectories that enter the region $u<0$ (i.e. $\mathcal{V}(\phi)<0$) may reach $\rho_{\mathrm{tot}}=0$ and therefore $H=0$, corresponding to a turnaround from expansion to contraction. The behavior near the $H\to 0$ boundary
and the global structure of the flow is analyzed in Appendix~\ref{app:poincare_full} (Poincar\'e sphere). The present background equations do not produce a genuine bounce, so that, any cyclic completion requires additional physics
beyond the baseline system.

\paragraph{Allowed transitions (expanding branch)}
In the expanding sector, physically acceptable chronologies are constructed based on the following universal model of transition (when the corresponding points exist and are within the range of $\Omega_m\ge 0$):
\begin{equation}
\mathcal{B}_{\pm}\quad\to\quad\mathcal{A}\quad\to\quad\left(\mathcal{G}\lor\mathcal{C}\lor\mathcal{D}\right)\quad\to\quad\left(\mathcal{C}\land/\lor\mathcal{F}\right)\quad\to\quad\left(\mathcal{E}\lor\mathcal{F}\lor\mathcal{M}\right)\,,
\end{equation}
where $\mathcal{M}$ is the de Sitter point of Model III (Table~\ref{tab:expL_fp}). Whether one of these transitions will be realized depends on the attractive regions and the existence/stability conditions listed in the tables.

\subsubsection*{Model I (Axions/ALPs): chronologies and boundary completion}
Model I is described by only two finite $\lambda_\phi$ fixed points (Tables~\ref{tab:alp_fp}-\ref{tab:alp_char}): $\mathcal{A}$ (radiation) and $\mathcal{B}$ (hilltop, de Sitter-like saddle). The physically important late-time regime, oscillations around the minimum with
$\langle\omega_\phi\rangle\simeq 0$, lies on a \emph{singular boundary} of the phase space, because $\lambda_\phi$ diverges as $\phi\to 0$ (see Sec.~\ref{subsec:ALP}). Therefore, all physically admissible expanding chronologies are the following:
\begin{enumerate}[label=\textbf{(I\alph*)}]
\item \textbf{Radiation $\to$ oscillatory (dust-like) axion regime (boundary)}:
\begin{equation}
    \mathcal{A}(R)\quad\to\quad \Bigl(\phi\to 0,\left|\lambda_\phi\right|\to\infty,\left\langle\omega_\phi\right\rangle\simeq 0 \Bigr)\,;
\end{equation}
\item \textbf{Radiation $\to$ (transient) hilltop acceleration $\to$ oscillatory regime (boundary)}:
\begin{equation}
    \mathcal{A}(R)\quad\to\quad\mathcal{B}\left(\omega_{\mathrm{eff}}\simeq -1\right)\quad\to\quad\Bigl(\phi\to 0,\left|\lambda_\phi\right|\to\infty\Bigr)\,;
\end{equation}
\item \textbf{Optional past stiff stage}: trajectories may originate near a kinetic/stiff regime and then approach $\mathcal{A}$ before following (Ia) or (Ib); the late-time completion remains boundary-like because $\mathcal{B}$ is a saddle and no finite $\lambda_\phi$ dust fixed point exists.
\end{enumerate}
Since $\mathcal{V}(\phi)\ge 0$ in Model I, no turnaround via $u<0$ occurs.

\subsubsection*{Model II (Cyclic ekpyrotic): expanding histories and turnaround to contraction completion}
On each constant branch $\lambda_{\phi*}=-b$ or $\lambda_{\phi*}=c$ \eqref{eq:cyclic-branch}, the fixed points $\mathcal{A}$-$\mathcal{G}$ exist subject to Tables~\ref{tab:cyclic_fp}-\ref{tab:cyclic_char}. The set of physically plausible expanding chronologies (within $\Omega_m\ge 0$) is:
\begin{enumerate}[label=\textbf{(II\alph*)}]
\item \textbf{Coupled chronology ending in scalar domination} (if $\mathcal{E}$ is stable):
      \begin{equation}
      \mathcal{A}(R)\quad\to\quad\mathcal{C}(\phi{\mathrm{MDE}})\quad\to\quad\mathcal{E}(P)\qquad\land\qquad \left(\omega_{\mathrm{eff}}<-\frac{1}{3}\iff\lambda_{\phi*}^2<2\right)\,;
      \end{equation}
\item \textbf{Coupled chronology ending in scaling} (if $\mathcal{F}$ is stable):
      \begin{equation}
      \mathcal{A}(R)\quad\to\quad\mathcal{C}(\phi{\rm MDE})\quad\to\quad\mathcal{F}(S)\,;
      \end{equation}
\item \textbf{Chronologies with a radiation-scalar scaling}:
      \begin{equation}
      \mathcal{A}(R)\quad\to\quad\mathcal{G}(RS)\quad\to\quad\mathcal{C}(\phi{\rm MDE})\quad\to\quad(\mathcal{E}\lor\mathcal{F})\,;
      \end{equation}
\item \textbf{Chronologies with a scaling} ($\mathcal{F}$ is a saddle):
      \begin{equation}
      \mathcal{A}\quad\to\quad\mathcal{C}\quad\to\quad\mathcal{F}\quad\to\quad\mathcal{E}\,;
      \end{equation}
\item \textbf{Non-standard coupled radiation late-time behavior} (allowed mathematically):
      \begin{equation}
      \mathcal{A}(R)\quad\to\quad\mathcal{D}(R_\beta)\qquad\iff\qquad \left(\frac{\lambda_{\phi*}}{\beta}>4 \land \beta^2\ge\frac{1}{2}\right)\,;
      \end{equation}
\item \textbf{Optional past stiff stage}:
      \begin{equation}
      \mathcal{B}_{\pm}\left(K_\pm\right)\quad\to\quad\mathcal{A}(R)\quad\to\quad\ldots\quad\to\quad(\mathcal{E}\lor\mathcal{F})\,.
      \end{equation}
\end{enumerate}
Because the SF potential \eqref{eq:V_cyclic} admits $\mathcal{V}<0$, there is an additional, physically distinct completion class:
\begin{enumerate}[label=\textbf{(IIT\alph*)}]
\item \textbf{Turnaround}: any of the expanding chronologies (IIa)-(IIf) may enter $u<0$, followed by approach to the $H\to 0$ boundary (Appendix~\ref{app:poincare_full});
\item \textbf{Contracting completion}: after the turnaround, the contracting branch is typically attracted to an \textit{\textbf{ekpyrotic}} SF-dominated solution for $\lambda_{\phi*}^2\gg 6$, characterized by $u_*<0$ and $\omega_{\mathrm{eff}}\gg 1$ (Sec.~\ref{subsec:cyclic}); alternatively, contraction may approach a kinetic regime if the orbit evolves toward $u\to 0$ with $|x|\to 1$.
\end{enumerate}

\subsubsection*{Model III (Exponential$+\Lambda$): exponential tail chronologies vs. freezing to de Sitter}
Model III possesses the standard exponential branch critical points $\mathcal{A}$-$\mathcal{G}$ at $\lambda_{\phi*}=c$ and an additional de Sitter point $\mathcal{M}$ at $\lambda_\phi=0$ (Tables~\ref{tab:expL_fp}-\ref{tab:expL_char}). Thus, the set of expanding chronologies splits into:
\begin{enumerate}[label=\textbf{(III\alph*)}]
\item \textbf{Freezing to de Sitter}:
      \begin{equation}
      \mathcal{A}(R)\quad\to\quad\mathcal{C}(\phi{\rm MDE})\quad\to\quad\mathcal{M}({\rm dS})\,,
      \end{equation}
      possibly with additional transient scaling saddles, e.g.:
      \begin{equation}
          \mathcal{A}\quad\to\quad\mathcal{G}\quad\to\quad\mathcal{C}\quad\to\quad\mathcal{M}
      \end{equation}
      or:
      \begin{equation}
          \mathcal{A}\quad\to\quad\mathcal{C}\quad\to\quad\mathcal{F}\quad\to\quad\mathcal{M}\,.
      \end{equation}
      The stability of $\mathcal{M}$ is decided by the center manifold, with $\mathcal{V}''(\phi_*)>0$ yielding attraction (Table~\ref{tab:expL_fp}, Appendix~\ref{app:center});
\item \textbf{Exponential tail late time}:
      \begin{equation}
          \mathcal{A}\quad\to\quad\mathcal{C}\quad\to\quad\left(\mathcal{E}\lor\mathcal{F}\right)\,,
      \end{equation}
      with the same possible insertions as in Model II (e.g. via $\mathcal{G}$ or transient $\mathcal{F}$), and with acceleration governed by $c^2<2$ for $\mathcal{E}$;
\item \textbf{Non-standard coupled radiation}:
\begin{equation}
    \mathcal{A}\quad\to\quad{D}\qquad\iff\qquad \frac{c}{\beta}>4\,;
\end{equation}
\item \textbf{Optional past stiff stage}:
\begin{equation}
    \mathcal{B}_\pm\quad\to\quad\mathcal{A}\to\ldots\,.
\end{equation}
\end{enumerate}
If the sign choice in \eqref{eq:V_expL} permits $\mathcal{V}<0$, then the turnaround completion (IITa)-(IITb) applies literally (Appendix~\ref{app:poincare_full}).

\subsubsection*{Model IV (Tracking quintessence): tracker to asymptote chronologies}
For Model IV the slope variable $\lambda_\phi$ evolves according to \eqref{eq:lamprime_quint}, interpolating between a steep region (efficient tracking/scaling) and one of the asymptotic constant branches
$\lambda_{\phi*}=\pm n\lambda$ \eqref{eq:q-branch-l}. On each asymptotic branch the fixed points $\mathcal{A}$-$\mathcal{G}$ exist as in Tables~\ref{tab:quint_fp}-\ref{tab:quint_char}. Hence, the expanding chronologies are the following:
\begin{enumerate}[label=\textbf{(IV\alph*)}]
\item \textbf{Tracker to coupled matter (saddle) to SF domination}:
      \begin{equation}
      \mathcal{A}(R)\quad\to\quad\mathcal{C}(\phi{\rm MDE})\quad\to\quad\mathcal{E}(P) \qquad\land\qquad \left(\omega_{\mathrm{eff}}<-\frac{1}{3}\iff (n\lambda)^2<2\right)\,;
      \end{equation}
\item \textbf{Tracker to scaling late-time}:
      \begin{equation}
      \mathcal{A}(R)\quad\to\quad\mathcal{C}(\phi{\rm MDE})\quad\to\quad\mathcal{F}(S)\,,
      \end{equation}
      possible if $\mathcal{F}$ is stable or when $(n\lambda)^2\gtrsim 2$ prevents accelerated scalar domination;
\item \textbf{Chronologies with radiation scaling}:
      \begin{equation}
      \mathcal{A}\quad\to\quad\mathcal{G}\quad\to\quad\mathcal{C}\quad\to\quad \left(\mathcal{E}\lor\mathcal{F}\right)\,,
      \end{equation}
      and/or with a transient scaling stage:
      \begin{equation}
          \mathcal{C}\quad\to\quad\mathcal{F}\quad\to\quad\mathcal{E}
      \end{equation}
      when $\mathcal{F}$ is a saddle;
\item \textbf{Non-standard coupled radiation}:
\begin{equation}
    \mathcal{A}\quad\to\quad\mathcal{D} \qquad\iff\qquad \frac{\lambda_{\phi*}}{\beta}>4\,;
\end{equation}
\item \textbf{Optional past stiff stage}:
\begin{equation}
    \mathcal{B}_\pm\quad\to\quad\mathcal{A}\quad\to\quad\ldots\,.
\end{equation}
\end{enumerate}
If the orbit enters a negative potential sector ($u<0$), then the turnaround completion described in Model II applies (Appendix~\ref{app:poincare_full}); otherwise the late-time evolution is determined by $\mathcal{E}$ or $\mathcal{F}$ on the asymptotic branch.

\subsubsection*{Model V (SFDM): exponential tail chronologies vs. oscillatory (dust-like) boundary regime}
Model V has two physical regimes (Sec.~\ref{subsec:sfdm}): an exponential tail regime with constant $\lambda_{\phi*}=\pm\lambda$ branches (Tables~\ref{tab:sfdm_fp}-\ref{tab:sfdm_char}), and an oscillatory regime with $\langle\omega_\phi\rangle\simeq 0$ \eqref{eq:averaged-SF-dust-EoS}. As in Model I, the latter corresponds to a
\emph{phase-space boundary} because $\lambda_\phi$ diverges as $\phi\to 0$. Therefore, the expanding chronologies are:
\begin{enumerate}[label=\textbf{(V\alph*)}]
\item \textbf{Exponential tail late-time SF domination} (if $\mathcal{E}$ is stable):
      \begin{equation}
          \mathcal{A}(R)\quad\to\quad\mathcal{C}(\phi{\rm MDE})\quad\to\quad\mathcal{E}(P) \qquad\land\qquad \left(\omega_{\mathrm{eff}}<-\frac{1}{3}\iff \lambda^2<2\right)\,;
      \end{equation}
\item \textbf{Exponential tail late-time scaling} (if $\mathcal{F}$ is stable):
      \begin{equation}
      \mathcal{A}(R)\quad\to\quad\mathcal{C}(\phi{\rm MDE})\quad\to\quad\mathcal{F}(S)\,,
      \end{equation}
      again allowing for interludes via $\mathcal{G}$ and/or transient $\mathcal{F}$;
\item \textbf{SFDM (dust-like) oscillatory completion (boundary)}:
      \begin{equation}
      \mathcal{A}(R)\quad\to\quad\Bigl(\phi\to 0,\left|\lambda_\phi\right|\to\infty,\left\langle\omega_\phi\right\rangle\simeq 0\Bigr)\,,
      \end{equation}
      which represents the coherent oscillations regime of SFDM rather than a finite $\lambda_\phi$ fixed point;
\item \textbf{Non-standard coupled radiation}:
\begin{equation}
    \mathcal{A}\quad\to\quad\mathcal{D} \qquad\iff\qquad \frac{\lambda_{\phi*}}{\beta}>4\,;
\end{equation}
\item \textbf{Optional past stiff stage}:
\begin{equation}
    \mathcal{B}_\pm\quad\to\quad\mathcal{A}\quad\to\quad\ldots\,.
\end{equation}
\end{enumerate}
Since $\mathcal{V}(\phi)\ge 0$, no turnaround via $u<0$ occurs in the standard SFDM case.

\subsection{Phenomenological viability: full vs. partial realizations of the observed cosmic chronology}
\label{subsec:viability_physical_chronology}
The analysis associated with dynamical systems allows us to identify specific epochs of domination (critical points) of specific components of the Universe and physically acceptable chronologies describing its evolution (Sec.~\ref{subsec:phys_chronologies}). Now, we must verify which cosmological models can describe the complete chronology of the Universe and which can only provide a limited description.

\paragraph{Minimal phenomenological requirements}
A minimal 'physical' late-time cosmology should contain:
\begin{enumerate}[label=(\roman*)]
\item An early radiation-dominated epoch (near $\mathcal{A}$ with $\omega_{\mathrm{eff}}=1/3$);
\item An intermediate matter-dominated (or matter-like) epoch with $\omega_{\mathrm{eff}}\simeq 0$ and $\Omega_m\simeq 1$;
\item A late-time accelerated epoch with $\omega_{\mathrm{eff}}<-1/3$ (optimally, we would like to have $\omega_{\mathrm{eff}}\simeq -1$);
\item No late-time radiation attractor ($\mathcal{D}$).
\end{enumerate}
In our formalism, the matter epoch is represented either by the coupled $\phi$MDE saddle point $\mathcal{C}$ or by an effective dust-like oscillatory regime at the phase space boundary (Models I and V). Because the NMC is strongly constrained by the local tests (for BM), the phenomenologically relevant regime is typically the \textit{\textbf{weak/moderate}} coupling $|\beta|\ll 1$ \cite{Bertotti:2003rm,Williams:2004qba,Uzan:2010pm,Buen-Abad:2021mvc}, for which $\mathcal{C}$ is close to standard matter domination:
\begin{equation}
    \omega_{\mathrm{eff}}(\mathcal{C})=\frac{2}{3}\beta^2\ll 1 \quad\land\quad \Omega_\phi(\mathcal{C})=\frac{2}{3}\beta^2\ll 1 \quad\land\quad \Omega_m(\mathcal{C})=1-\frac{2}{3}\beta^2\simeq 1\,,
\end{equation}
see Tables~\ref{tab:cyclic_char}, \ref{tab:expL_char}, \ref{tab:quint_char} and \ref{tab:sfdm_char}. In addition, one must avoid the coupled radiation point $\mathcal{D}$, which is stable (for expansion) if $\lambda_{\phi*}/\beta>4$ and $\beta^2\ge 1/2$ (Tables~\ref{tab:cyclic_fp}, \ref{tab:expL_fp}, \ref{tab:quint_fp}, \ref{tab:sfdm_fp}). Such a region is typically outside the weak coupling regime.

\subsubsection*{Model I (axions/ALPs): typically a \textit{partial} realization}
Model I involves an epoch of radiation domination (point $\mathcal{A}$) and a de Sitter-like hilltop point $\mathcal{B}$ (Tables~\ref{tab:alp_fp}-\ref{tab:alp_char}). However, it should be noted that the latter point is a saddle one, and therefore does not provide an attractor for late-time accelerated expansion.

Furthermore, the important epoch of matter/dust dominance, which is described by oscillations of the scalar field around the potential minimum in this case, is not represented by finite values of the $\lambda_{\phi}$ variable . In other words, it does not correspond to any specific fixed point.

The following conclusions can be drawn from the provided analysis:
\begin{itemize}
\item \textbf{Physical role (partial description):} Model I may be responsible for the realization of the radiation era, the effective matter epoch (oscillatory axion-like (dark) matter), and the transient hilltop-like accelerated expansion, but it \textit{\textbf{does not contain}} (for finite values of $\lambda_{\phi}$) \textit{\textbf{an attractor responsible for the robust dark energy epoch}};
\item \textbf{Physical (LCDM-like) late-time acceleration:} An additional mechanism/component (e.g., cosmological constant/plateau term in SF potential as in Model III) or fine tuning of the initial conditions (initial value of the SF extremely close to the hilltop) is required. However, we will still be facing a temporary period of accelerated expansion of the Universe.\footnote{The fundamental question remains whether the era of late-time accelerated expansion is permanent or temporary.}
\end{itemize}

\subsubsection*{Model II (cyclic ekpyrotic): full realization}
Model II admits standard expanding chronologies involving $\mathcal{A}$ and the matter-like saddle $\mathcal{C}$, with late-time behavior controlled by $\mathcal{E}$ and/or $\mathcal{F}$ on the constant $\lambda_{\phi*}$ branches (Tables~\ref{tab:cyclic_fp}--\ref{tab:cyclic_char}, Sec.~\ref{subsec:phys_chronologies}). However, due to the fact that the potential \eqref{eq:V_cyclic} contains a negative branch, trajectories may enter region of $u<0$ and be driven toward the $H\to 0$ boundary (Appendix~\ref{app:poincare_full}), causing a turnaround (contraction).
Thus:
\begin{itemize}
\item \textbf{Physical role (two distinct interpretations):}
\begin{enumerate}[label=(\alph*),leftmargin=*]
\item \textbf{Standard expanding Universe (full) \emph{only as a restricted sector}:} Requires that the orbit remains in $u\ge 0$ regime throughout the relevant cosmic time and that the coupling is weak enough to provide an acceptable matter-like epoch near $\mathcal{C}$ (typically $|\beta|\ll 1$). In such a sector one can realize $\mathcal{A}\to \mathcal{C}$ chronology, with acceleration obtained either by $\mathcal{E}$ if $\lambda_{\phi*}^2<2$ (Table~\ref{tab:cyclic_char}) or by an accelerating scaling regime $\mathcal{F}$ for:
\begin{equation}
    \omega_{\mathrm{eff}}(\mathcal{F})=\frac{\beta}{\lambda_{\phi*}-\beta}<-\frac{1}{3}\qquad\land\qquad \Omega_{m*}\ge 0\,;
\end{equation}
\item \textbf{Ekpyrotic/cyclic interpretation (partial description):}
For trajectories that explore the negative $u<0$ regime, the model naturally describes an \textit{\textbf{expansion $\to$ turnaround $\to$ contraction}} transition, with an ekpyrotic contracting attractor characterized by $u_*<0$ and $\omega_{\mathrm{eff}}\gg 1$ for $\lambda_{\phi*}^2\gg 6$ (Sec.~\ref{subsec:cyclic}). A genuine bounce/cycle requires a regular extension beyond the considered model (Sec.~\ref{sec:bounce});
\end{enumerate}
\item \textbf{Parameters/initial conditions sensitivity:} stronger coupling and/or initial conditions that sets the orbit into $u<0$ increase the probability of approaching $H\to 0$ (Sec.~\ref{subsec:numerics_cyclic}).
\end{itemize}

\subsubsection*{Model III (exponential$+\Lambda$): a robust \textit{full} realization for weak/moderate coupling}
Model III contains the de Sitter point $\mathcal{M}$ on the plateau branch $\lambda_\phi\to 0$ (Tables~\ref{tab:expL_fp}--\ref{tab:expL_char}), providing a natural late-time accelerating attractor (\textit{\textbf{freezing quintessence}} behavior).

Phenomenological validity is determined by the following aspects:
\begin{itemize}
\item \textbf{Late-time acceleration:} achieved by $\mathcal{M}$ with $\omega_{\mathrm{eff}}=-1$, stable if (Table~\ref{tab:expL_fp}, Appendix~\ref{app:center}):
\begin{equation}
    \mathcal{V}''(\phi_*)>0\,;
\end{equation}
\item \textbf{Matter epoch:} provided by a long transient era near $\mathcal{C}$ ($\phi$MDE saddle), which is close to standard matter domination for $|\beta|\ll 1$ (see above and Table~\ref{tab:expL_char});
\item \textbf{Avoiding nonstandard late-time radiation era:} exclude the parameter region where $\mathcal{D}$ becomes stable (Table~\ref{tab:expL_fp}):
\begin{equation}
    \frac{c}{\beta}>4\quad\land\quad \beta^2\ge \frac{1}{2}\,;
\end{equation}
\item \textbf{Initial conditions:} trajectories must approach the plateau ('freezing') so that $\lambda_\phi\to 0$ and the orbit is captured by $\mathcal{M}$ (Sec.~\ref{subsec:expL}, Sec.~\ref{subsec:numerics_expl}).
\end{itemize}
Under these conditions (notably weak/moderate coupling), Model III offers the \textit{\textbf{clearest dynamical realization}} of the sequence of \textit{radiation} to \textit{matter-like} to \textit{de Sitter acceleration} in the full 4D system.

\subsubsection*{Model IV (tracking quintessence): a \textit{full} realization for $(n\lambda)^2<2$ and $|\beta|\ll 1$}
Model IV was constructed to produce tracking during radiation and matter domination, followed by freezing (accelerated expansion) behavior (Sec.~\ref{subsec:quint}). The crucial parameters are $n\in\mathbb{Z}^-$ and $\lambda$:
\begin{itemize}
\item \textbf{Early-time viability (subdominant SF):} in the steep regime $|\lambda_\phi|\gg 1$ the scalar sector efficiently tracks the background, keeping $\Omega_\phi$ small during radiation and matter epochs (Sec.~\ref{subsec:quint});
\item \textbf{Late-time acceleration:} SF-dominated acceleration at $\mathcal{E}$ requires (Sec.~\ref{subsec:quint}, Table~\ref{tab:quint_char}):
\begin{equation}
      \lambda_{\phi*}^2=\left(n\lambda\right)^2<2\,,
\end{equation}
together with the stability conditions in Table~\ref{tab:quint_fp};
\item \textbf{Matter epoch:} realized by a long transient near $\mathcal{C}$, requiring $|\beta|\ll 1$ for $\omega_{\mathrm{eff}}(\mathcal{C})\simeq 0$ and $\Omega_m(\mathcal{C})\simeq 1$;
\item \textbf{Partial realizations:} if $(n\lambda)^2\gtrsim 2$, the late-time behavior tends toward a non-accelerating scaling regime (Sec.~\ref{subsec:quint}) and the strong coupling may also distort or shorten the matter-like epoch (Sec.~\ref{subsec:numerics_quint}).
\end{itemize}
Therefore, \textit{\textbf{Model IV is a natural candidate for a full realization
in the weak/moderate coupling regime}}, with accelerated expansion controlled primarily by the combination of $(n\lambda)^2$.

\subsubsection*{Model V (SFDM): typically a \textit{partial} realization (DM-like sector)}
Model V contains a quadratic regime near $\phi\simeq 0$ corresponding to coherent oscillations with $\langle\omega_\phi\rangle\simeq 0$ (SFDM-like dust), and exponential tails for $|\lambda\phi|\gg 1$
(Sec.~\ref{subsec:sfdm}). However, the dust-like oscillatory regime is not represented by a finite $\lambda_\phi$ fixed point because $\lambda_\phi\propto -1/\phi$ diverges as $\phi\to 0$; thus the matter-like behavior is realized as a boundary regime, and not as a standard critical point epoch.

In a consequence, one obtains the following conclusions:
\begin{itemize}
\item \textbf{Physical role (partial):} Model V can realize radiation $\mathcal{A}$ and an effective dust-like SFDM epoch (boundary oscillations), making it a plausible \textit{dark matter sector} realization within the present variables;
\item \textbf{Late-time acceleration:} could occur on a constant $\lambda_{\phi*}=\pm\lambda$ branches either via $\mathcal{E}$ if (Table~\ref{tab:sfdm_char}):
\begin{equation}
    \lambda^2<2
\end{equation}
or via an accelerating scaling regime $\mathcal{F}$ for:
\begin{equation}
    \omega_{\mathrm{eff}}=\frac{\beta}{\lambda_{\phi*}-\beta}<-\frac{1}{3} \qquad\land\qquad \Omega_{m*}\ge 0\,.
\end{equation}
In practice, in order to obtain a \textit{full} radiation to matter-like to DE chronology in this framework requires parameter choices and/or initial conditions that allow a long-lived matter-like boundary phase and a subsequent transition to accelerated expansion, or additional DE-like component (e.g., a constant term as in Model III).
\end{itemize}

\paragraph{Model selection}
In the weak/moderate coupling regime favored by phenomenology, the models that most naturally reproduce a \textit{full} observed chronology in the considered formalism are:
\begin{itemize}
\item \textbf{Model III} (exponential$+\Lambda$): radiation $\to$ matter-like ($\mathcal{C}$) $\to$ de Sitter ($\mathcal{M}$), with $\mathcal{M}$ stable for $\mathcal{V}''(\phi_*)>0$;
\item \textbf{Model IV} (tracking quintessence): radiation $\to$ matter-like ($\mathcal{C}$) $\to$ accelerated scalar domination ($\mathcal{E}$) for $(n\lambda)^2<2$.
\end{itemize}
Models I and V are best interpreted as \textit{partial} realizations (effective DM-like sectors with boundary oscillatory epochs),
while Model II is either a restricted expanding model (if $u<0$ is avoided) or a pre-turnaround description of the ekpyrotic/cyclic scenarios (requiring an external bounce completion).

\subsection{Model I: Axions/ALPs}
\label{subsec:ALP}

\paragraph{Fixed points}
Critical points characterized by a finite value of the variable $\lambda_{\phi}$ require:
\begin{equation}
    x_{*}=0\,.
\end{equation}
The critical point corresponding to late-time accelerated expansion (de Sitter-like), in such a case we can talk about hilltop\footnote{Examples include hilltop inflation \cite{Boubekeur:2005zm} and hilltop quintessence \cite{Dutta:2008qn}.} behavior, is a hyperbolic saddle point described by:
\begin{equation}
    u_*=1\quad\land\quad \lambda_{\phi*}=0 \quad\land\quad \phi=\pi
\end{equation}
(see Table~\ref{tab:alp_fp} and Table~\ref{tab:alp_char} for more details). The subsystem $\left(x,\lambda_{\phi}\right)$ contains mixed terms already at the linear level, which leads to one unstable direction. This is consistent with the fact that (see Appendix~\ref{app:center}):
\begin{equation}
    \mathcal{V}''(\pi)<0\,.
\end{equation}
Late-time evolution near the minimum of the scalar field potential is characterized by field oscillations, and therefore cannot be described by finite values of $\lambda_{\phi}$ for the corresponding critical point. In this case, an approach related to averaged behavior may be used \cite{Marsh:2015xka,Turner:1983he}.

\paragraph{Possible cosmological evolution}
The following cosmological chronology may serve as a representative example illustrating the possible evolution of the model under consideration:
\begin{enumerate}[label=(\arabic*)]
    \item Early-time radiation epoch: trajectories in phase space located near the point $\mathcal{A}$ (saddle), for which:
    \begin{equation}
        x^2 \ll 1 \quad\land\quad u \ll 1 \quad\land\quad \Omega_r\simeq 1\,;
    \end{equation}
    \item Freezing and the beginning of scalar field evolution: the SF is overdamped for:
    \begin{equation}
        H\gg m_{\mathrm{eff}}
    \end{equation}
    and begins to evolve when:
    \begin{equation}
        H\sim m_{\mathrm{eff}}\,.
    \end{equation}
    This may lead to a short increase in the $\Omega_{\phi}$ energy density parameter;
    \item SF oscillations near the potential minimum: for small oscillations, $\mathcal{V}(\phi)$ is approximated by relation \eqref{eq:massive-SF}, and coherent oscillations lead to an averaged equation of state characterizing matter/dust \eqref{eq:averaged-SF-dust-EoS}. Nevertheless, no fixed point with a finite value of $\lambda_{\phi}$ corresponds to such behavior, because it diverges at:
    \begin{equation}
        \phi=0 \pmod{2\pi}\,;
    \end{equation}
    \item Transitional phase of accelerated expansion near the hilltop: if the initial misalignment is close to the unstable maximum:
    \begin{equation}
    \label{eq:hilltop-a}
        \phi\simeq\pi\qquad\iff\qquad \bigl(\lambda_{\phi}\simeq 0\quad\land\quad x\simeq 0\bigr)\,,
    \end{equation}
    then the SF sector behaves effectively analogously to the cosmological constant (de Sitter-like fixed point $\mathcal{B}$ in Table~\ref{tab:alp_fp} and Table~\ref{tab:alp_char}):
    \begin{equation}
        \omega_{\mathrm{eff}}\simeq -1\,.
    \end{equation}
    Due to the fact that \eqref{eq:hilltop-a} corresponds to the maximum of $\mathcal{V}(\phi)$, this critical point is, in a sense, automatically a saddle point (late-time accelerated expansion is a temporary phase with a finite lifetime). We should note that hilltop quintessence models usually require fine-tuning of the initial conditions and an ultralight effective mass of the SF (even smaller than in SFDM models) to prevent oscillations too early.
\end{enumerate}
\paragraph{Why are there only two finite critical points?}
For the self-interaction potential under consideration \eqref{eq:V_ALP}, the explicit dependence of the parameter $\Gamma_{\phi}$ on the scalar field is expressed as follows:
\begin{equation}
\Gamma(\phi)=\frac{1}{1+\sec{\phi}}\,,
\label{eq:alp_lambda_gamma}
\end{equation}
therefore, for a finite $\phi$ values it never satisfies the condition:
\begin{equation}
    \Gamma_{\phi}=1\,.
\end{equation}
Consequently, within the finite values of the variable $\lambda_{\phi}$ in the considered phase space of the cosmological dynamical system, the condition of stationarity of the system:
\begin{equation}
    \frac{d \lambda_{\phi}}{dN}=0
\end{equation}
enforces the condition:
\begin{equation}
    x=0\quad\lor\quad \lambda_{\phi}=0
\end{equation}
and the remaining conditions of stationarity imply that the variable $u$ corresponding to the potential energy of the field must vanish:
\begin{equation}
    u\propto\mathcal{V}(\phi)\to 0\,.
\end{equation}
Namely, the late-time epoch of cosmological evolution, which is significant from a physical point of view, corresponds to the oscillation of the field around the minimum, for which:
\begin{equation}
    \mathcal{V}(\phi)\propto\phi^{2}\to 0\,,
\end{equation}
but at the same time in such a case:
\begin{equation}
    \lambda_{\phi}\propto -\cot{\phi}\to -\infty \quad\land\quad\phi>0\,.
\end{equation}
This implies that the minimum lies on the singular boundary of the phase space and the background level analysis takes into account only the aforementioned fixed points of the system. Thus, matter domination epoch (field oscillations) does not occur as an explicit critical point \cite{Copeland:2006wr,Hui:2016ltb}.

\begin{table*}[htbp]
\caption{\label{tab:alp_fp}Fixed points for the NMC axion/ALP model.}
\centering
\small
\setlength{\tabcolsep}{6pt}
\renewcommand{\arraystretch}{1.18}
\begin{tabularx}{\linewidth}{>{\centering\arraybackslash}p{0.09\linewidth}
                             >{\centering\arraybackslash}X
                             >{\centering\arraybackslash}p{0.20\linewidth}
                             >{\centering\arraybackslash}p{0.15\linewidth}
                             >{\centering\arraybackslash}p{0.10\linewidth}}
\toprule
Point & Eigenvalues $\{\mu_i\}$ & Existence & Stability (expansion) & Epoch \\
\midrule
$\mathcal{A}$ &
$\{-1,\ 1,\ 4,\ 0\}$ &
Always &
Saddle (non-hyperbolic; $\lambda_\phi$ is a center direction at linear order) &
R \\
$\mathcal{B}$ &
$\left\{-4,\ -3,\ \frac{-3-\sqrt{15}}{2},\ \frac{-3+\sqrt{15}}{2}\right\}$ &
$\lambda_\phi=0,\ u=1$ ($\phi=\pi$) &
Saddle (one unstable direction; hilltop) &
DE-like \\
\bottomrule
\end{tabularx}
\end{table*}

\begin{table*}[htbp]
\caption{\label{tab:alp_char}Detailed characterization of critical points for the NMC axion/ALP model.}
\centering
\small
\setlength{\tabcolsep}{5.6pt}
\renewcommand{\arraystretch}{1.18}
\begin{tabular}{cccccccccc}
\toprule
Point & $x$ & $u$ & $\lambda_\phi$ & $\Omega_r$ & $\omega_{\mathrm{eff}}$ & Accel. & $\Omega_\phi$ & $\omega_\phi$ & $\Omega_m$ \\
\midrule
$\mathcal{A}$ & $0$ & $0$ & arbitrary & $1$ & $1/3$ & No & $0$ & -- & $0$ \\
$\mathcal{B}$ & $0$ & $1$ & $0$ & $0$ & $-1$ & Yes & $1$ & $-1$ & $0$ \\
\bottomrule
\end{tabular}
\end{table*}

\FloatBarrier

\subsection{Model II: Cyclic ekpyrotic}
\label{subsec:cyclic}

\paragraph{Physical regimes}
The dynamical system evolution equation \eqref{eq:lamprime_cyclic} yields two exponential branches expressed by \eqref{eq:cyclic-branch}. They may correspond to distinct physical regimes, such as the ekpyrotic contraction of the Universe, which occurs when:
\begin{equation}
    \lambda_{\phi*}^{2}\gg 6\,,
\end{equation}
so that the scalar dominated fixed point corresponds to:
\begin{equation}
    u_{*}\propto\mathcal{V}(\phi)<0\,.
\end{equation}

\paragraph{Possible model evolution}
\begin{enumerate}[label=(\arabic*)]
    \item Early-time radiation domination - saddle fixed point $\mathcal{A}$;
    \item Intermediate epoch of scaling evolution ($\phi$MDE) described by point $\mathcal{C}$ and/or matter-SF scaling solution (S) corresponding to critical point $\mathcal{F}$ depending on parameter values;
    \item Late-time evolution: SF potential energy domination (P) corresponding to point $\mathcal{E}$, leading to accelerated expansion for:
    \begin{equation}
        \lambda_{\phi*}^{2}<2
    \end{equation}
    or scaling regime (S) represented by point $\mathcal{F}$ and associated with the effective equation of state:
    \begin{equation}
        \omega_{\mathrm{eff}}=\frac{\beta}{\lambda_{\phi*}-\beta}\,;
    \end{equation}
    \item Turnaround: in the case of trajectories reaching the region of negative SF potential values ($u<0$), the total energy density can reach zero - the Universe passes from the stage of expansion to the stage of contraction:
    \begin{equation}
        \rho_{\mathrm{tot}}\equiv\sum_{i=r,m,\phi}\rho_{i}=0\quad\implies\quad H=0\,.
    \end{equation}
\end{enumerate}
A detailed analysis of all critical points for the model is provided in Table~\ref{tab:cyclic_fp} and Table~\ref{tab:cyclic_char}.

\paragraph{Ekpyrotic contraction}
For:
\begin{equation}
    \lambda_{\phi*}^2\gg 6\,,
\end{equation}
the scalar dominated fixed point is characterized by:
\begin{equation}
    u_*<0 \quad\land\quad \omega_{\mathrm{eff}}\gg 1
\end{equation}
and constitutes an ekpyrotic attractor in the contracting phase (change of sign). The general behavior of the model in the vicinity of $H\to 0$ and $u<0$ one can find in Appendix~\ref{app:poincare_full}.

\paragraph{Note on $\mu_{\lambda}$}
From the relation \eqref{eq:lamprime_cyclic} one can obtain:
\begin{subequations}
\begin{equation}
    \mu_{\lambda}\equiv\frac{\partial\lambda_{\phi}'}{\partial\lambda_\phi}\Bigg|_{\lambda_\phi=\lambda_{\phi*}}=-\sqrt{6}\,x_{*}\left(c+b\right)\quad\land\quad\lambda_{\phi*}=-b\,,
\end{equation}
\begin{equation}
    \mu_{\lambda}=\sqrt{6}\,x_{*}\left(c+b\right)\quad\land\quad\lambda_{\phi*}=c\,,
\end{equation}
\end{subequations}
and:
\begin{equation}
    \mu_{\lambda}=0 \quad\land\quad x_{*}=0\,.
\end{equation}

\paragraph{Convention for $\mu_{\lambda}$}
In the tables of critical points, $\mu_{\lambda}$ denotes the eigenvalue associated with the $\lambda_\phi$ direction in the full 4D system \eqref{eq:DSfinal}. Explicit expressions for the models III-V are collected in Appendix~\ref{app:eigs}.

\begin{table*}[htbp]
\caption{\label{tab:cyclic_fp}Fixed points for the NMC cyclic ekpyrotic model on a constant $\lambda_{\phi*}$ branch ($\lambda_{\phi*}=-b$ or $\lambda_{\phi*}=c$).}
\centering
\small
\setlength{\tabcolsep}{6pt}
\renewcommand{\arraystretch}{1.16}
\begin{tabularx}{\linewidth}{>{\centering\arraybackslash}p{0.08\linewidth}
                             >{\centering\arraybackslash}X
                             >{\centering\arraybackslash}p{0.24\linewidth}
                             >{\centering\arraybackslash}p{0.14\linewidth}
                             >{\centering\arraybackslash}p{0.10\linewidth}}
\toprule
Point & Eigenvalues $\{\mu_i\}$ & Existence & Stability (expansion) & Epoch \\
\midrule
$\mathcal{A}$ (R) &
$\{-1,\,1,\,4,\,0\}$ &
Always &
Saddle (non-hyperbolic) &
R \\
$\mathcal{B}_\pm$ (K$_\pm$) &
$\{2,\ 3\mp\sqrt6\beta,\ 6\mp\sqrt6\lambda_{\phi*},\ \mu_{\lambda}\}$ &
$\lambda_\phi=\lambda_{\phi*}$ &
Unstable &
SM \\
$\mathcal{C}$ ($\phi$MDE) &
$\{\beta^2-\frac32,\ 2\beta^2-1,\ 2\beta^2-2\beta\lambda_{\phi*}+3,\ \mu_{\lambda}\}$ &
$\beta^2\le\frac32$ &
Saddle &
M$(\beta)$ \\
$\mathcal{D}$ (R$_\beta$) &
$\{4-\lambda_{\phi*}/\beta,\ -\frac12\pm \frac12\sqrt{2/\beta^2-3},\ \mu_{\lambda}\}$ &
$\beta^2\ge\frac12$ &
Stable if $\lambda_{\phi*}/\beta>4$ &
R \\
$\mathcal{E}$ (P) &
$\{\lambda_{\phi*}^2-4,\ \frac12(\lambda_{\phi*}^2-6),\ \lambda_{\phi*}^2-\beta\lambda_{\phi*}-3,\ \mu_{\lambda}\}$ &
$\lambda_\phi=\lambda_{\phi*}$ &
Stable if all $<0$ &
DE/ekpy \\
$\mathcal{F}$ (S) &
$\{\mu_1,\ \mu_{\pm},\ \mu_{\lambda}\}$ &
$\Omega_{m*}\ge 0$ &
Parameter-dependent &
scaling \\
$\mathcal{G}$ (RS) &
$\{\frac{\lambda_{\phi*}-4\beta}{\lambda_{\phi*}},\ -\frac12\pm \frac{1}{2\lambda_{\phi*}}\sqrt{64-15\lambda_{\phi*}^2},\ \mu_{\lambda}\}$ &
$\lambda_{\phi*}^2\ge 4$ &
Typically saddle &
R \\
\bottomrule
\end{tabularx}
\end{table*}

\begin{table*}[htbp]
\caption{\label{tab:cyclic_char}Detailed characterization of critical points for the NMC cyclic ekpyrotic model on a constant $\lambda_{\phi*}$ branch.}
\centering
\small
\setlength{\tabcolsep}{5.2pt}
\renewcommand{\arraystretch}{1.14}
\begin{tabular}{cccccccccc}
\toprule
Point & $x$ & $u$ & $\lambda_\phi$ & $\Omega_r$ & $\omega_{\mathrm{eff}}$ & Accel. & $\Omega_\phi$ & $\omega_\phi$ & $\Omega_m$ \\
\midrule
$\mathcal{A}$ & $0$ & $0$ & arbitrary & $1$ & $1/3$ & No & $0$ & -- & $0$ \\
$\mathcal{B}_+$ & $+1$ & $0$ & $\lambda_{\phi*}$ & $0$ & $1$ & No & $1$ & $1$ & $0$ \\
$\mathcal{B}_-$ & $-1$ & $0$ & $\lambda_{\phi*}$ & $0$ & $1$ & No & $1$ & $1$ & $0$ \\
$\mathcal{C}$ & $\sqrt{\frac{2}{3}}\beta$ & $0$ & $\lambda_{\phi*}$ & $0$ & $\frac{2}{3}\beta^2$ & No & $\frac{2}{3}\beta^2$ & $1$ & $1-\frac{2}{3}\beta^2$ \\
$\mathcal{D}$ & $\frac{1}{\sqrt6\beta}$ & $0$ & $\lambda_{\phi*}$ & $1-\frac{1}{2\beta^2}$ & $1/3$ & No & $\frac{1}{6\beta^2}$ & $1$ & $\frac{1}{3\beta^2}$ \\
$\mathcal{E}$ & $\frac{\lambda_{\phi*}}{\sqrt6}$ & $1-\frac{\lambda_{\phi*}^2}{6}$ & $\lambda_{\phi*}$ & $0$ & $\frac{\lambda_{\phi*}^2}{3}-1$ & $\lambda_{\phi*}^2<2$ & $1$ & $\frac{\lambda_{\phi*}^2}{3}-1$ & $0$ \\
$\mathcal{F}$ & $\frac{\sqrt6}{2(\lambda_{\phi*}-\beta)}$ &
$\frac{2\beta^2-2\beta\lambda_{\phi*}+3}{2(\lambda_{\phi*}-\beta)^2}$ &
$\lambda_{\phi*}$ & $0$ &
$\frac{\beta}{\lambda_{\phi*}-\beta}$ &
$\omega_{\mathrm{eff}}<-\frac13$ &
$\frac{\beta^2-\beta\lambda_{\phi*}+3}{(\lambda_{\phi*}-\beta)^2}$ &
$\frac{x^2-u}{x^2+u}$ &
$\frac{\lambda_{\phi*}^2-\beta\lambda_{\phi*}-3}{(\lambda_{\phi*}-\beta)^2}$ \\
$\mathcal{G}$ & $\frac{2\sqrt6}{3\lambda_{\phi*}}$ & $\frac{4}{3\lambda_{\phi*}^2}$ & $\lambda_{\phi*}$ &
$1-\frac{4}{\lambda_{\phi*}^2}$ & $1/3$ & No & $\frac{4}{\lambda_{\phi*}^2}$ & $1/3$ & $0$ \\
\bottomrule
\end{tabular}
\end{table*}

\FloatBarrier

\subsection{Model III: Exponential with a constant}
\label{subsec:expL}

\paragraph{Freezing trajectories and de Sitter epoch}
The branch corresponding to:
\begin{equation}
    \lambda_{\phi}\to 0
\end{equation}
constitutes an asymptotically constant part of the potential \eqref{eq:V_expL}, namely:
\begin{equation}
    \mathcal{V}(\phi)\simeq 2\Lambda\,,
\end{equation}
From a physical point of view, it is responsible for late-time accelerated expansion in which the phase space trajectories can approach:
\begin{equation}
    \omega_{\mathrm{eff}}\to -1
\end{equation}
after a transient scaling epoch. In terms of models that describe quintessence scenarios, this is a typical ‘freezing’ behavior \cite{Chevallier:2000qy,Linder:2002et,dePutter:2008wt,Copeland:2006wr,Linder:2006sv,Linder:2024rdj,Linder:2025zxb,LinderStambul2025}. The full classification of critical points with explicit expressions can be found in Table~\ref{tab:expL_fp} and Table~\ref{tab:expL_char}.

\begin{table*}[htbp]
\caption{\label{tab:expL_fp}Fixed points for the NMC exponential$+\Lambda$ model.}
\centering
\small
\setlength{\tabcolsep}{6pt}
\renewcommand{\arraystretch}{1.16}
\begin{tabularx}{\linewidth}{>{\centering\arraybackslash}p{0.08\linewidth}
                             >{\centering\arraybackslash}X
                             >{\centering\arraybackslash}p{0.24\linewidth}
                             >{\centering\arraybackslash}p{0.14\linewidth}
                             >{\centering\arraybackslash}p{0.10\linewidth}}
\toprule
Point & Eigenvalues $\{\mu_i\}$ & Existence & Stability (expansion) & Epoch \\
\midrule
$\mathcal{A}$ (R) &
$\{-1,\,1,\,4,\,0\}$ &
Always &
Saddle (non-hyperbolic) &
R \\
$\mathcal{B}_\pm$ (K$_\pm$) &
$\{2,\ 3\mp\sqrt6\beta,\ 6\mp\sqrt6 c,\ \mu_{\lambda}\}$ &
$\lambda_\phi=c$ &
Unstable &
SM \\
$\mathcal{C}$ ($\phi$MDE) &
$\{\beta^2-\frac32,\ 2\beta^2-1,\ 2\beta^2-2\beta c+3,\ \mu_{\lambda}\}$ &
$\beta^2\le\frac32$ &
Saddle &
M$(\beta)$ \\
$\mathcal{D}$ (R$_\beta$) &
$\{4-c/\beta,\ -\frac12\pm \frac12\sqrt{2/\beta^2-3},\ \mu_{\lambda}\}$ &
$\beta^2\ge\frac12$ &
Stable if $c/\beta>4$ &
R \\
$\mathcal{E}$ (P$_c$) &
$\{c^2-4,\ \frac12(c^2-6),\ c^2-\beta c-3,\ \mu_{\lambda}\}$ &
$\lambda_\phi=c$ &
Stable if all $<0$ &
DE / stiff \\
$\mathcal{F}$ (S$_c$) &
$\{\mu_1,\ \mu_{\pm},\ \mu_{\lambda}\}$ &
$\Omega_{m*}\ge0$ &
Parameter-dependent &
scaling \\
$\mathcal{G}$ (RS$_c$) &
$\{\frac{c-4\beta}{c},\ -\frac12\pm \frac{1}{2c}\sqrt{64-15c^2},\ \mu_{\lambda}\}$ &
$c^2\ge 4$ &
Typically saddle &
R \\
$\mathcal{M}$ (dS) &
$\{-3,\ -3,\ -4,\ 0\}$ (linear); center manifold decides $\lambda_\phi$ &
$\lambda_\phi=0,\ u=1$ &
Stable if $\mathcal{V}''(\phi_*)>0$ &
DE \\
\bottomrule
\end{tabularx}
\end{table*}

\begin{table*}[htbp]
\caption{\label{tab:expL_char}Detailed characterization of critical points for the NMC exponential$+\Lambda$ model.}
\centering
\small
\setlength{\tabcolsep}{5.2pt}
\renewcommand{\arraystretch}{1.14}
\begin{tabular}{cccccccccc}
\toprule
Point & $x$ & $u$ & $\lambda_\phi$ & $\Omega_r$ & $\omega_{\mathrm{eff}}$ & Accel. & $\Omega_\phi$ & $\omega_\phi$ & $\Omega_m$ \\
\midrule
$\mathcal{A}$ & $0$ & $0$ & arbitrary & $1$ & $1/3$ & No & $0$ & -- & $0$ \\
$\mathcal{B}_+$ & $+1$ & $0$ & $c$ & $0$ & $1$ & No & $1$ & $1$ & $0$ \\
$\mathcal{B}_-$ & $-1$ & $0$ & $c$ & $0$ & $1$ & No & $1$ & $1$ & $0$ \\
$\mathcal{C}$ & $\sqrt{\frac{2}{3}}\beta$ & $0$ & $c$ & $0$ & $\frac{2}{3}\beta^2$ & No & $\frac{2}{3}\beta^2$ & $1$ & $1-\frac{2}{3}\beta^2$ \\
$\mathcal{D}$ & $\frac{1}{\sqrt6\beta}$ & $0$ & $c$ & $1-\frac{1}{2\beta^2}$ & $1/3$ & No & $\frac{1}{6\beta^2}$ & $1$ & $\frac{1}{3\beta^2}$ \\
$\mathcal{E}$ & $\frac{c}{\sqrt6}$ & $1-\frac{c^2}{6}$ & $c$ & $0$ & $\frac{c^2}{3}-1$ & $c^2<2$ & $1$ & $\frac{c^2}{3}-1$ & $0$ \\
$\mathcal{F}$ & $\frac{\sqrt6}{2(c-\beta)}$ &
$\frac{2\beta^2-2\beta c+3}{2(c-\beta)^2}$ &
$c$ & $0$ &
$\frac{\beta}{c-\beta}$ &
$\omega_{\mathrm{eff}}<-\frac13$ &
$\frac{\beta^2-\beta c+3}{(c-\beta)^2}$ &
$\frac{x^2-u}{x^2+u}$ &
$\frac{c^2-\beta c-3}{(c-\beta)^2}$ \\
$\mathcal{G}$ & $\frac{2\sqrt6}{3c}$ & $\frac{4}{3c^2}$ & $c$ &
$1-\frac{4}{c^2}$ & $1/3$ & No & $\frac{4}{c^2}$ & $1/3$ & $0$ \\
$\mathcal{M}$ & $0$ & $1$ & $0$ & $0$ & $-1$ & Yes & $1$ & $-1$ & $0$ \\
\bottomrule
\end{tabular}
\end{table*}

\FloatBarrier

\subsection{Model IV: Tracking quintessence}
\label{subsec:quint}

\paragraph{The role of evolving $\lambda_\phi$ variable}
The asymptotic branches of the $\lambda_{\phi}$ variable for the SF potential \eqref{eq:V_quint} are given by \eqref{eq:q-branch-l}. However, this type of cosmological models allows for a dynamical, rapid change in the value of this variable - tracking/freezing behavior \cite{Steinhardt:1999nw,Copeland:2006wr,Linder:2006sv}.

For this particular model, it should be noted that the parameter describing the slope of the SF potential takes the following explicit form:
\begin{equation}
    \Gamma_{\phi}(\phi)=1-\frac{1}{n\cosh^2(\lambda\phi)}\,,
\end{equation}
so that for $n\in\mathbb{Z}^{-}$ yields the standard condition for the existence of tracker/attractor solutions \cite{Steinhardt:1999nw,Copeland:2006wr} insensitive to the initial conditions of the system:
\begin{equation}
    \Gamma_{\phi}>1 \quad\land\quad \lim_{|\phi|\to\infty}\Gamma_{\phi}(\phi)=1\,.
\end{equation}
The asymptotic behavior of the potential expressed by the relation \eqref{eq:asympt-q-V} results in the decrease of $\lambda_{\phi}$ from:
\begin{equation}
    \left|\lambda_{\phi}\right|\gg 1
\end{equation}
(steep region with efficient tracking/scaling during radiation and matter domination) into the constant value given by \eqref{eq:q-branch-l} (exponential regime). During this transition, the decreasing slope of the potential allows $\omega_{\phi}$ to deviate from the background equation of state and approach the frozen value \cite{Linder:2006sv}:
\begin{equation}
    \omega_\phi\simeq -1\,.
\end{equation}
Therefore, the late-time acceleration era in the case of scalar field potential energy dominance requires an asymptotically shallow slope:
\begin{equation}
    \lambda_{\phi*}^2 = (n\lambda)^{2}<2\,.
\end{equation}
In other cases, the evolution of the model tends toward a non-accelerating scaling regime \cite{Copeland:2006wr}. The remaining critical points are described in Table~\ref{tab:quint_fp} and Table~\ref{tab:quint_char}.

\begin{table*}[htbp]
\caption{\label{tab:quint_fp}Fixed points for the NMC tracking quintessence model on a constant $\lambda_{\phi*}$ branch ($\lambda_{\phi*}=\pm n\lambda$).}
\centering
\small
\setlength{\tabcolsep}{6pt}
\renewcommand{\arraystretch}{1.16}
\begin{tabularx}{\linewidth}{>{\centering\arraybackslash}p{0.08\linewidth}
                             >{\centering\arraybackslash}X
                             >{\centering\arraybackslash}p{0.24\linewidth}
                             >{\centering\arraybackslash}p{0.14\linewidth}
                             >{\centering\arraybackslash}p{0.10\linewidth}}
\toprule
Point & Eigenvalues $\{\mu_i\}$ & Existence & Stability (expansion) & Epoch \\
\midrule
$\mathcal{A}$ (R) &
$\{-1,\,1,\,4,\,0\}$ &
Always &
Saddle (non-hyperbolic) &
R \\
$\mathcal{B}_\pm$ (K$_\pm$) &
$\{2,\ 3\mp\sqrt6\beta,\ 6\mp\sqrt6\lambda_{\phi*},\ \mu_{\lambda}\}$ &
$\lambda_\phi=\lambda_{\phi*}$ &
Unstable &
SM \\
$\mathcal{C}$ ($\phi$MDE) &
$\{\beta^2-\frac32,\ 2\beta^2-1,\ 2\beta^2-2\beta\lambda_{\phi*}+3,\ \mu_{\lambda}\}$ &
$\beta^2\le\frac32$ &
Saddle &
M$(\beta)$ \\
$\mathcal{D}$ (R$_\beta$) &
$\{4-\lambda_{\phi*}/\beta,\ -\frac12\pm \frac12\sqrt{2/\beta^2-3},\ \mu_{\lambda}\}$ &
$\beta^2\ge\frac12$ &
Stable if $\lambda_{\phi*}/\beta>4$ &
R \\
$\mathcal{E}$ (P) &
$\{\lambda_{\phi*}^2-4,\ \frac12(\lambda_{\phi*}^2-6),\ \lambda_{\phi*}^2-\beta\lambda_{\phi*}-3,\ \mu_{\lambda}\}$ &
$\lambda_\phi=\lambda_{\phi*}$ &
Stable if all $<0$ &
DE / stiff \\
$\mathcal{F}$ (S) &
$\{\mu_1,\ \mu_{\pm},\ \mu_{\lambda}\}$ &
$\Omega_{m*}\ge 0$ &
Parameter-dependent &
scaling \\
$\mathcal{G}$ (RS) &
$\{\frac{\lambda_{\phi*}-4\beta}{\lambda_{\phi*}},\ -\frac12\pm \frac{1}{2\lambda_{\phi*}}\sqrt{64-15\lambda_{\phi*}^2},\ \mu_{\lambda}\}$ &
$\lambda_{\phi*}^2\ge 4$ &
Typically saddle &
R \\
\bottomrule
\end{tabularx}
\end{table*}

\begin{table*}[htbp]
\caption{\label{tab:quint_char}Detailed characterization of the fixed points for the NMC tracking quintessence model on a constant $\lambda_{\phi*}$ branch.}
\centering
\small
\setlength{\tabcolsep}{5.2pt}
\renewcommand{\arraystretch}{1.14}
\begin{tabular}{cccccccccc}
\toprule
Point & $x$ & $u$ & $\lambda_\phi$ & $\Omega_r$ & $\omega_{\mathrm{eff}}$ & Accel. & $\Omega_\phi$ & $\omega_\phi$ & $\Omega_m$ \\
\midrule
$\mathcal{A}$ & $0$ & $0$ & arbitrary & $1$ & $1/3$ & No & $0$ & -- & $0$ \\
$\mathcal{B}_+$ & $+1$ & $0$ & $\lambda_{\phi*}$ & $0$ & $1$ & No & $1$ & $1$ & $0$ \\
$\mathcal{B}_-$ & $-1$ & $0$ & $\lambda_{\phi*}$ & $0$ & $1$ & No & $1$ & $1$ & $0$ \\
$\mathcal{C}$ & $\sqrt{\frac{2}{3}}\beta$ & $0$ & $\lambda_{\phi*}$ & $0$ & $\frac{2}{3}\beta^2$ & No & $\frac{2}{3}\beta^2$ & $1$ & $1-\frac{2}{3}\beta^2$ \\
$\mathcal{D}$ & $\frac{1}{\sqrt6\beta}$ & $0$ & $\lambda_{\phi*}$ & $1-\frac{1}{2\beta^2}$ & $1/3$ & No & $\frac{1}{6\beta^2}$ & $1$ & $\frac{1}{3\beta^2}$ \\
$\mathcal{E}$ & $\frac{\lambda_{\phi*}}{\sqrt6}$ & $1-\frac{\lambda_{\phi*}^2}{6}$ & $\lambda_{\phi*}$ & $0$ & $\frac{\lambda_{\phi*}^2}{3}-1$ & $\lambda_{\phi*}^2<2$ & $1$ & $\frac{\lambda_{\phi*}^2}{3}-1$ & $0$ \\
$\mathcal{F}$ & $\frac{\sqrt6}{2(\lambda_{\phi*}-\beta)}$ &
$\frac{2\beta^2-2\beta\lambda_{\phi*}+3}{2(\lambda_{\phi*}-\beta)^2}$ &
$\lambda_{\phi*}$ & $0$ &
$\frac{\beta}{\lambda_{\phi*}-\beta}$ &
$\omega_{\mathrm{eff}}<-\frac13$ &
$\frac{\beta^2-\beta\lambda_{\phi*}+3}{(\lambda_{\phi*}-\beta)^2}$ &
$\frac{x^2-u}{x^2+u}$ &
$\frac{\lambda_{\phi*}^2-\beta\lambda_{\phi*}-3}{(\lambda_{\phi*}-\beta)^2}$ \\
$\mathcal{G}$ & $\frac{2\sqrt6}{3\lambda_{\phi*}}$ & $\frac{4}{3\lambda_{\phi*}^2}$ & $\lambda_{\phi*}$ &
$1-\frac{4}{\lambda_{\phi*}^2}$ & $1/3$ & No & $\frac{4}{\lambda_{\phi*}^2}$ & $1/3$ & $0$ \\
\bottomrule
\end{tabular}
\end{table*}

\FloatBarrier

\subsection{Model V: SFDM}
\label{subsec:sfdm}

\paragraph{Two dynamical regimes}
This model has a quadratic regime near:
\begin{equation}
    \phi\simeq 0
\end{equation}
(oscillations as in the case of axions/ALPs) and the exponential regime:
\begin{equation}
    \mathcal{V}(\phi)\propto e^{|\lambda\phi|}
\end{equation}
for:
\begin{equation}
    \left|\lambda\phi\right|\gg 1
\end{equation}
(e.g., asymptotically constant $\lambda_{\phi*}$ branches) with the usual scaling dynamics.

It should be emphasized that, as in the case of axions/ALPs, there is no critical point in the dynamical system corresponding to the era of matter/cosmological dust domination. It follows from the Taylor expansion of $\lambda_{\phi}$ \eqref{eq:l-SFDM}:
\begin{equation}
    \lambda_{\phi}(\phi)\propto -\frac{1}{\phi}
\end{equation}
diverges as:
\begin{equation}
    \phi\to 0\,.
\end{equation}
The complete characteristics of all critical points are provided in Table~\ref{tab:sfdm_fp} and Table~\ref{tab:sfdm_char}.

\begin{table*}[htbp]
\caption{\label{tab:sfdm_fp}Fixed points for the NMC SFDM model on a constant $\lambda_{\phi*}$ branch ($\lambda_{\phi*}=\pm\lambda$).}
\centering
\small
\setlength{\tabcolsep}{6pt}
\renewcommand{\arraystretch}{1.16}
\begin{tabularx}{\linewidth}{>{\centering\arraybackslash}p{0.08\linewidth}
                             >{\centering\arraybackslash}X
                             >{\centering\arraybackslash}p{0.24\linewidth}
                             >{\centering\arraybackslash}p{0.14\linewidth}
                             >{\centering\arraybackslash}p{0.10\linewidth}}
\toprule
Point & Eigenvalues $\{\mu_i\}$ & Existence & Stability (expansion) & Epoch \\
\midrule
$\mathcal{A}$ (R) &
$\{-1,\,1,\,4,\,0\}$ &
Always &
Saddle (non-hyperbolic) &
R \\
$\mathcal{B}_\pm$ (K$_\pm$) &
$\{2,\ 3\mp\sqrt6\beta,\ 6\mp\sqrt6\lambda_{\phi*},\ \mu_{\lambda}\}$ &
$\lambda_\phi=\lambda_{\phi*}$ &
Unstable &
SM \\
$\mathcal{C}$ ($\phi$MDE) &
$\{\beta^2-\frac32,\ 2\beta^2-1,\ 2\beta^2-2\beta\lambda_{\phi*}+3,\ \mu_{\lambda}\}$ &
$\beta^2\le\frac32$ &
Saddle &
M$(\beta)$ \\
$\mathcal{D}$ (R$_\beta$) &
$\{4-\lambda_{\phi*}/\beta,\ -\frac12\pm \frac12\sqrt{2/\beta^2-3},\ \mu_{\lambda}\}$ &
$\beta^2\ge\frac12$ &
Stable if $\lambda_{\phi*}/\beta>4$ &
R \\
$\mathcal{E}$ (P) &
$\{\lambda_{\phi*}^2-4,\ \frac12(\lambda_{\phi*}^2-6),\ \lambda_{\phi*}^2-\beta\lambda_{\phi*}-3,\ \mu_{\lambda}\}$ &
$\lambda_\phi=\lambda_{\phi*}$ &
Stable if all $<0$ &
DE / stiff \\
$\mathcal{F}$ (S) &
$\{\mu_1,\ \mu_{\pm},\ \mu_{\lambda}\}$ &
$\Omega_{m*}\ge 0$ &
Parameter-dependent &
scaling \\
$\mathcal{G}$ (RS) &
$\{\frac{\lambda_{\phi*}-4\beta}{\lambda_{\phi*}},\ -\frac12\pm \frac{1}{2\lambda_{\phi*}}\sqrt{64-15\lambda_{\phi*}^2},\ \mu_{\lambda}\}$ &
$\lambda_{\phi*}^2\ge 4$ &
Typically saddle &
R \\
\bottomrule
\end{tabularx}
\end{table*}

\begin{table*}[htbp]
\caption{\label{tab:sfdm_char}Detailed characterization of the fixed points for the NMC SFDM model on a constant $\lambda_{\phi*}$ branch.}
\centering
\small
\setlength{\tabcolsep}{5.2pt}
\renewcommand{\arraystretch}{1.14}
\begin{tabular}{cccccccccc}
\toprule
Point & $x$ & $u$ & $\lambda_\phi$ & $\Omega_r$ & $\omega_{\mathrm{eff}}$ & Accel. & $\Omega_\phi$ & $\omega_\phi$ & $\Omega_m$ \\
\midrule
$\mathcal{A}$ & $0$ & $0$ & arbitrary & $1$ & $1/3$ & No & $0$ & -- & $0$ \\
$\mathcal{B}_+$ & $+1$ & $0$ & $\lambda_{\phi*}$ & $0$ & $1$ & No & $1$ & $1$ & $0$ \\
$\mathcal{B}_-$ & $-1$ & $0$ & $\lambda_{\phi*}$ & $0$ & $1$ & No & $1$ & $1$ & $0$ \\
$\mathcal{C}$ & $\sqrt{\frac{2}{3}}\beta$ & $0$ & $\lambda_{\phi*}$ & $0$ & $\frac{2}{3}\beta^2$ & No & $\frac{2}{3}\beta^2$ & $1$ & $1-\frac{2}{3}\beta^2$ \\
$\mathcal{D}$ & $\frac{1}{\sqrt6\beta}$ & $0$ & $\lambda_{\phi*}$ & $1-\frac{1}{2\beta^2}$ & $1/3$ & No & $\frac{1}{6\beta^2}$ & $1$ & $\frac{1}{3\beta^2}$ \\
$\mathcal{E}$ & $\frac{\lambda_{\phi*}}{\sqrt6}$ & $1-\frac{\lambda_{\phi*}^2}{6}$ & $\lambda_{\phi*}$ & $0$ & $\frac{\lambda_{\phi*}^2}{3}-1$ & $\lambda_{\phi*}^2<2$ & $1$ & $\frac{\lambda_{\phi*}^2}{3}-1$ & $0$ \\
$\mathcal{F}$ & $\frac{\sqrt6}{2(\lambda_{\phi*}-\beta)}$ &
$\frac{2\beta^2-2\beta\lambda_{\phi*}+3}{2(\lambda_{\phi*}-\beta)^2}$ &
$\lambda_{\phi*}$ & $0$ &
$\frac{\beta}{\lambda_{\phi*}-\beta}$ &
$\omega_{\mathrm{eff}}<-\frac13$ &
$\frac{\beta^2-\beta\lambda_{\phi*}+3}{(\lambda_{\phi*}-\beta)^2}$ &
$\frac{x^2-u}{x^2+u}$ &
$\frac{\lambda_{\phi*}^2-\beta\lambda_{\phi*}-3}{(\lambda_{\phi*}-\beta)^2}$ \\
$\mathcal{G}$ & $\frac{2\sqrt6}{3\lambda_{\phi*}}$ & $\frac{4}{3\lambda_{\phi*}^2}$ & $\lambda_{\phi*}$ &
$1-\frac{4}{\lambda_{\phi*}^2}$ & $1/3$ & No & $\frac{4}{\lambda_{\phi*}^2}$ & $1/3$ & $0$ \\
\bottomrule
\end{tabular}
\end{table*}

\FloatBarrier

\section{Bounce/cyclic evolution and its influence on the dynamical system}
\label{sec:bounce}
The necessary condition for a non-singular cosmological bounce is:
\begin{equation}
    H=0 \quad\land\quad \dot{H}>0\,.
\end{equation}
However, as can be easily noticed, the 2nd Friedmann equation \eqref{eq:Ray} yields\footnote{For spatially flat FLRW cosmological models with a canonical SF and a ‘standard’ matter sector (cosmological dust and radiation/relativistic particles)}:
\begin{equation}
    \dot{H}\le 0\,.
\end{equation}
Therefore, the transition from the contraction phase to expansion (cosmological bounce) is \textit{impossible} in the case of our equations of motion \eqref{eq:Friedmann}-\eqref{eq:KG}.

The logical conclusion is that any models attempting to model a hypothetical bounce phase or more sophisticated models trying to model cyclic cosmological bounces require additional modifications, including \cite{Battefeld:2014uga,Brandenberger:2016vhg,Postolak:2024xtm}\footnote{There is another approach to periodicity in the evolution of the Universe - CCC (Conformal Cyclic Cosmology) \cite{Penrose1979,Penrose:2006zz,Penrose2010}. However, it does not assume successive stages of expansion and contraction (there is no bounce in cosmological sense).}:
\begin{enumerate}[label=(\arabic*)]
    \item Stable NEC violation (ghost condensate, Galileon, Horndeski) - New Ekpyrotic Cosmology \cite{Buchbinder:2007ad,Lehners:2008vx,Nastase2019};
    \item Positive spatial curvature \cite{Tolman1934};
    \item Modified and quantum gravity/cosmology formalisms \cite{Ashtekar:2006rx,Poplawski:2011jz,Cai:2012va,Lehners_2026book};
    \item Higher-dimensional physics/cosmology (e.g., \textit{string theory} and \textit{superstring theory} \cite{Polchinski_1998a,Polchinski_1998b,Green_Schwarz_Witten_2012a,Green_Schwarz_Witten_2012b}) \cite{Khoury:2001wf,Steinhardt:2001st,Gasperini_2007book,Nastase:2019mhe,Cicoli:2023opf,Lehners_2026chapter}.
\end{enumerate}
The latest results obtained within the DESI DR2 \cite{DESI:2025zgx,DESI:2025fii} have renewed the discussion on the possible \textit{phantom behavior} of dark energy (in our case, the scalar field). Nevertheless, in the phenomenological perspective, it is worth to emphasize and distinguish two important aspects:
\begin{enumerate}[label=(\arabic*)]
    \item Genuine NEC violation in the gravitational (classical field theory) sense;
    \item Effective NEC violation resulting from the initial assumptions regarding the theoretical framework or possible interactions of dark energy (dark sector) \cite{Caldwell:2025inn,Guedezounme:2025wav}
\end{enumerate}
From the perspective of a dynamical system (as considered in our formalism), passing through the zero value of the Hubble parameter ruins our e-fold time variable, $N=\ln{a}$. To avoid this issue, we can introduce a positive normalization factor along with a modified (compact) Hubble variable, which can be defined as follows:
\begin{equation}
    \mathcal{H}\equiv\frac{H}{D}\in \left(-1,1\right)\quad\land\quad D>0\,
\end{equation}
therefore, the $H=0$ hypersurface becomes finite.
The analysis at infinity in Appendix~\ref{app:poincare_full} shows that for:
\begin{equation}
    u<0 \quad\land\quad \Omega_r\ge 0
\end{equation}
the reduced system is generically drawn toward the $H\to 0$ boundary. The cosmological bounce completion makes this boundary dynamical rather than singular.

\section{Numerical background evolution}
\label{sec:numerical_evolution}

\subsection{Description of diagrams}
\label{subsec:read_plots}

For each cosmological model, we integrate the autonomous dynamical system \eqref{eq:DSfinal} (with the appropriate $\lambda_\phi$ differential equation from Sec.~\ref{sec:potentials}) and reconstruct the physical parameters:
\begin{equation}
\label{eq:plot_defs}
\Omega_\phi \equiv x^2+u \quad\land\quad \Omega_m \equiv 1-x^2-u-\Omega_r \quad\land\quad \omega_{\mathrm{eff}}\equiv\frac{1}{3}\Omega_r +x^2-u \quad\land\quad \omega_\phi \equiv \frac{x^2-u}{x^2+u}\quad(\Omega_\phi\neq 0)\,.
\end{equation}
The horizontal dashed blue line indicates the boundary value of the equation of state parameter that determines accelerated expansion:
\begin{equation}
    \omega=-\frac{1}{3}\,.
\end{equation}

\paragraph{Parameter values describing NMC}
In our analysis we consider the following values of the $\beta$ parameter:
\begin{equation}
\label{eq:beta_scan_set}
\beta\in\left\{\pm\sqrt{\frac32},\pm\frac{1}{\sqrt{2}},\pm10^{-1},\pm10^{-2}\right\}\,.
\end{equation}
These values can be characterized as follows:
\begin{itemize}
    \item Weak coupling:
    \begin{equation}
        \left|\beta\right|=10^{-2}\,;
    \end{equation}
    \item Moderate coupling:
    \begin{equation}
        \left|\beta\right|=10^{-1}\,;
    \end{equation}
    \item Strong coupling (relevant from a mathematical point of view for our dynamical system - existence/stability conditions in Sec.~\ref{sec:model_by_model} and Appendix~\ref{app:eigs}):
    \begin{equation}
        \left|\beta\right|=\frac{1}{\sqrt{2}}\quad\lor\quad \left|\beta\right|=\sqrt{\frac{3}{2}}\,.
    \end{equation}
\end{itemize}
\paragraph{Scope of the $\beta$ scan (phenomenology vs. dynamical systems completeness).}
The local Solar System tests impose strong constraints on the NMC of the SF to BM via \textit{parameterized post-Newtonian (PPN) formalism} \cite{Will:2014kxa,Will_2018} constraints \cite{Bertotti:2003rm,Williams:2004qba}:
\begin{equation}
    \begin{dcases}
        \gamma^{\mathrm{PPN}}-1=\left(2.1\pm 2.3\right)\times 10^{-5} \\
        \beta^{\mathrm{PPN}}-1=\left(1.2\pm 1.1\right)\times 10^{-4}
    \end{dcases}
\end{equation}
that in the case under consideration implies the following \cite{Borowiec:2023kmq,Coc:2006rt} (for more details see Sec. 3.6 in \cite{Borowiec:2023kmq}):
\begin{equation}
    \gamma^{\mathrm{PPN}}-1\equiv -\frac{2\alpha_{\mathrm{PPN},0}^{2}}{1+\alpha_{\mathrm{PPN},0}^{2}}=-\frac{4\beta^{2}}{1+2\beta^{2}}=-2+\frac{2}{1+2\beta^{2}} \quad\implies\quad \left|\beta\right|\lesssim 2.4\times 10^{-3}\,.
\end{equation}
Therefore, if the SF couples directly to BM, only rather small values of $\beta$ parameter are allowed locally, unless an additional screening or suppression mechanism is present.

The situation is different if the coupling is interpreted as a coupling between the scalar field and dark matter only, while baryons remain minimally coupled. In that case the Solar-System PPN bound does not directly constrain the dark-sector coupling. Instead, the relevant constraints are cosmological. For an unscreened long-range scalar interaction confined to the dark sector, the scalar-mediated force between DM particles is commonly parametrized as \cite{Li:2009sy}:
\begin{equation}
    \frac{F_{\phi}}{F_{N}}=2\beta^{2}\,,
\end{equation}
up to conventions for the normalization of the scalar coupling. Linear cosmology analyses of long-range dark fifth forces acting only on DM, using CMB and BAO data, constrain the dark 5th force strength to be below the percent level relative to gravity \cite{Archidiacono:2022iuu}. In the above normalization, this gives the rough estimate of:
\begin{equation}
    2\beta^2\lesssim 10^{-2} \quad\implies\quad \left|\beta\right|\lesssim 7\times10^{-2}\,.
\end{equation}
In a consequence, the values of $|\beta|=10^{-2}$ and $|\beta|=10^{-1}$ should be interpreted as weak or moderate illustrative couplings for exploring the phase-space structure. The latter value is already close to or above the typical unscreened dark 5th force bounds. The larger values, $|\beta|=1/\sqrt{2}$ and $|\beta|=\sqrt{3/2}$, are included only to demonstrate the complete structure of the dynamical systems, including changes in the existence and stability of critical points, as well as the emergence of coupled scaling regimes. Therefore, they should be viewed as mathematically instructive rather than observationally viable benchmark values.
\subsection{Physical constraints, negative potentials, and the meaning of $\Omega_\phi<0$}
\label{subsec:physicality_negative_Omegaphi}

Physical trajectories in our Hubble-normalized variables satisfy the following condition:
\begin{equation}
\label{eq:phys_constraints}
\Omega_{r}\ge 0 \quad\land\quad \Omega_{m}\ge 0 \quad\implies\quad 1-x^2-u-\Omega_r\ge 0
\end{equation}
as already denoted in \eqref{eq:DSfinal}. However, because we use the sign-safe variable $u$, the SF density parameter contribution:
\begin{equation}
\label{eq:rho_phi_sign}
\Omega_\phi = x^2+u \ \propto\ \rho_\phi \equiv \frac12\dot\phi^2+\mathcal{V}(\phi)
\end{equation}
can take negative values if:
\begin{equation}
    \left(\mathcal{V}(\phi)<0\quad\land\quad \dot{\phi}\simeq 0\right) \quad\lor\quad |u|>x^{2}\,.
\end{equation}
This is a regime associated with the cosmology of negative scalar field potentials, in which a transition from expansion to contraction (turning point) and phases of ekpyrotic contraction may occur \cite{Felder:2002jk,Heard:2002dr,Lehners:2008vx}.

\paragraph{Interpretation of $\Omega_\phi<0$}
Temporary stages in which:
\begin{equation}
    \Omega_\phi<0
\end{equation}
is not an algebraic inconsistency of the system - this means that the scalar sector makes a negative contribution to the energy balance in Friedmann equation \eqref{eq:Friedmann}, compensated by positive values for the energy density of radiation and matter.

This leads to the following two important conclusions:
\begin{enumerate}[label=(\arabic*)]
\item \textbf{$\boldsymbol{\omega_\phi}$ can peak or diverge}: Since $\omega_\phi$ is defined only if $\Omega_\phi\neq 0$, any crossing through $\Omega_\phi=0$ generates a divergence in $\omega_\phi$ (vertical peak). Nevertheless, this is not a physical singularity (it is an
artifact of expressing the SF pressure-to-density ratio when the denominator changes its sign). On the other hand, $\omega_{\mathrm{eff}}$ remains finite because it is reconstructed directly from $(x,u,\Omega_r)$ via \eqref{eq:plot_defs};
\item \textbf{Approaching to the $\boldsymbol{H\to 0}$ boundary becomes generic in the $u<0$ sector}: Negative SF potentials can cause the total energy density to approach zero, but they also allow for:
\begin{equation}
    H^2\to 0
\end{equation}
at finite scale factor values (turnaround/recollapse). In Hubble-normalized variables, such a behavior corresponds to divergences due to the fact that:
\begin{equation}
    \Omega_{i}\propto H^{-2}\,.
\end{equation}
We examine this behavior by the Poincar\'e compactification in Appendix~\ref{app:poincare_full} - in the physical
$u<0$ sector with $\Omega_r\ge 0$, the reduced flow is generically drawn toward the $H\to 0$ boundary, so that the $N=\ln a$ formulation becomes singular and a bounce/cyclic completion must be formulated in a Hubble parameter regular extension \cite{Felder:2002jk,Battefeld:2014uga,Brandenberger:2016vhg}.
\end{enumerate}

\subsection{NMC qualitative features}
\label{subsec:beta_qualitative}

The NMC parameter $\beta$ appears in the dust/matter continuity equation \eqref{eq:cont_r_m} and in the Klein-Gordon equation \eqref{eq:KG}. Therefore, increasing $|\beta|$ stimulates:
\begin{enumerate}[label=(\arabic*)]
    \item Deviations from the usual cosmological chronology: $R\to M\to DE$;
    \item Importance of coupled scaling regimes described in the stability analysis of the dynamical systems.
\end{enumerate}
A specific example of such behavior is the saddle point $\phi$MDE (Sec.~\ref{sec:model_by_model} and Appendix~\ref{app:eigs}), which yields the following for constant branches of the variable $\lambda_{\phi}$:
\begin{equation}
\label{eq:phiMDE_signature}
\Omega_{\phi}\Big|_{\mathrm{\phi MDE}}=\omega_{\mathrm{eff}}\Big|_{\mathrm{\phi MDE}}=\frac{2}{3}\beta^2\,,
\end{equation}

so that the $\beta$ values \eqref{eq:beta_scan_set} correspond to the following hierarchy:
\begin{subequations}
    \begin{equation}
        |\beta|=10^{-2}\quad\implies\quad\Omega_\phi\simeq 6.7\times 10^{-5}\,,
    \end{equation}
    \begin{equation}
        |\beta|=10^{-1}\quad\implies\quad\Omega_\phi\simeq 6.7\times 10^{-3}\,,
    \end{equation}
    \begin{equation}
        |\beta|=\frac{1}{\sqrt{2}}\quad\implies\quad\Omega_\phi=\frac{1}{3}\,,
    \end{equation}
    \begin{equation}
        |\beta|=\sqrt{\frac{3}{2}}\quad\implies\quad\Omega_\phi=1\,.
    \end{equation}
\end{subequations}
Therefore, the strong NMC of the SF with matter (for the saddle point) causes a significant part of the energy of the Universe to be contained in the scalar sector. This, in turn, may lead to the modification (and, in extreme cases, even the disappearance) of the traditional epoch of matter domination necessary to create the large-scale structure of the Universe. 

For intermediate and small values of $\beta$, the evolution of the models corresponds to trajectories similar to the LCDM model near the current epoch of evolution.

\paragraph{Physical meaning of positive and negative $\beta$ values}
The non-minimal coupling between the scalar field and matter implicates an explicit energy flow/transfer between these components. Using \eqref{eq:cont_r_m} and covariant conservation law for the total energy-momentum tensor:
\begin{equation}
    \nabla_{\mu}T^{\mu\nu}_{(\mathrm{tot})}=0\quad\land\quad T^{\mu\nu}_{(\mathrm{tot})}\equiv T^{\mu\nu}_{(\mathrm{m})}+T^{\mu\nu}_{(\mathrm{r})}+T^{\mu\nu}_{(\phi)}\,,
\end{equation}
together with the SF equation of motion \eqref{eq:KG} yields:
\begin{subequations}
\label{eq:Q_def}
\begin{align}
\dot\rho_m + 3H\rho_m &= -Q\,, \label{eq:Q_m}\\
\dot\rho_\phi + 3H\left(1+\omega_\phi\right)\rho_\phi &= +Q\,, \label{eq:Q_phi}\\
Q &\equiv \beta\,\dot\phi\,\rho_m = \sqrt{6}\,\beta\,x\,H\,\rho_m\,, \label{eq:Q_source}
\end{align}
\end{subequations}
where $Q$ denotes the energy transfer between the sectors.

For the expansion phase ($H>0$), taking into account positive energy density values for matter, the direction of energy transfer is explicitly determined by $\sgn(Q)=\sgn(\beta\,x)$:
\begin{equation}\label{eq:E-flow}
    Q=
\begin{cases}
>0, & \beta\dot{\phi}>0\,,\\
<0, & \beta\dot{\phi}<0\,.
\end{cases}
\end{equation}
The first case in \eqref{eq:E-flow} represents the flow of energy from the matter sector to the scalar sector (matter dilutes faster than $a^{-3}$), while in the second case, energy moves from the scalar sector to matter (matter dilutes slowly than $a^{-3}$). Therefore, the sign of the parameter $\beta$ does not only correspond to the adopted convention. For the selected direction of energy flow, it determines whether non-minimal coupling injects energy into the scalar sector, resulting in an increase of $\Omega_{\phi}$ and a departure from standard LCDM evolution, or extracts energy away from it.

Based on the above considerations, we take into account both cases of energy transfer (different signs for the parameter $\beta$) in this study. Strong coupling is mainly taken into account due to the aforementioned mathematical structure of the dynamical system. A detailed description of the possible mathematical forms of the energy transfer function and the restrictions resulting from observations can be found in \cite{Wang:2016lxa,He:2010im,DiValentino:2017iww,vanderWesthuizen:2025rip}).

\subsection{Analytical initial conditions motivated by Planck 2018 and DESI DR2}
\label{subsec:init_from_w0_wa}
In order to obtain physical (consistent with observational data) numerical trajectories of the dynamical system considered in our study, it is necessary to express the initial (current: $a=1$ or equivalently $N=0$) values of the dynamical variables\footnote{We use the $\Omega_{r,0}\simeq 9.2\times 10^{-5}$ value for radiation from Planck 2018 data \cite{Planck:2018vyg}.}:
\begin{equation}
    \left(x_{0},u_{0},\lambda_{\phi,0}\right)
\end{equation}
in terms of the equation of state.

\paragraph{From $\left(\Omega_{\phi,0},w_{\phi,0}\right)$ to $\left(x_0,u_0\right)$}
Applying explicit definitions for $\Omega_{\phi}$ and $\omega_{\phi}$ \eqref{eq:Omega-phi-omega-phi} allows us to obtain analytical forms for the initial conditions $x_{0}$ and $u_{0}$:
\begin{subequations}\label{eq:init_x0_u0_from_wphi0}
    \begin{equation}
        x_{0}^{2}=\frac{1}{2}\left(1+\omega_{\phi,0}\right)\,\Omega_{\phi,0} \quad\implies\quad x_0=\pm\sqrt{\frac{1}{2}\left(1+\omega_{\phi,0}\right)\,\Omega_{\phi,0}}\,,
    \end{equation}
    \begin{equation}
        u_0=\frac{1}{2}\left(1-\omega_{\phi,0}\right)\,\Omega_{\phi,0}\,,
    \end{equation}
\end{subequations}
where the sign in front of $x_0$ fixes the initial rolling direction of the SF.

\paragraph{CPL parametrization for the equation of state}
In the late Universe:
\begin{equation}
    \Omega_{r,0}\ll 1
\end{equation}
and therefore:
\begin{equation}
\label{eq:init_weff_approx}
\omega_{\mathrm{eff},0}\simeq\Omega_{\phi,0}\,\omega_{\phi,0}\,,
\end{equation}
Using the CPL\footnote{Chevallier-Polarski-Linder.} \cite{Chevallier:2000qy,Linder:2002et} parametrization for the effective equation of state:
\begin{equation}
    \omega(a)=\omega_{0}+\omega_{a}(1-a)\,,
\end{equation}
which can, but does not have to, be interpreted as a Taylor expansion (up to the 1st order) of an unknown function, $\omega(a)$, around $a_{0}=1$ with:
\begin{equation}
    \omega_{0}\equiv\omega(1)=\mathrm{const}\quad\land\quad \omega_{a}\equiv -\frac{d \omega(a)}{da}\Bigg|_{a=1}=\mathrm{const}\,,
\end{equation}
one can easily obtain the condition for $\omega_{\phi,0}$, namely:
\begin{equation}
\label{eq:init_wphi0_from_weff0}
\omega_{\phi,0}\simeq\frac{\omega_{0}}{\Omega_{\phi,0}}
\quad\land\quad \Omega_{r,0}\simeq 0\,.
\end{equation}
Therefore, the relations \eqref{eq:init_x0_u0_from_wphi0}-\eqref{eq:init_wphi0_from_weff0} fully describe $x_{0}$ and $u_{0}$.

\paragraph{Matter content}
Once $(x_0,u_0)$ are fixed, the constraint equation \eqref{eq:constraint} yields:
\begin{equation}
\label{eq:init_Omm0}
\Omega_{m,0}=1-x_0^2-u_0-\Omega_{r,0}\simeq 1-\Omega_{\phi,0}\,.
\end{equation}

\paragraph{Slow-roll condition for $\lambda_{\phi,0}$}
By assuming a slow-roll regime for the scalar field in the current epoch of evolution:
\begin{equation}
    \ddot{\phi}\simeq 0
\end{equation}
the Klein-Gordon equation \eqref{eq:KG} produces the following field equation:
\begin{equation}
    3H\dot\phi\simeq\beta\,\rho_m-\mathcal{V}'(\phi)\,,
\end{equation}
so that the $\lambda_{\phi,0}$ value explicitly depends on the NMC parameter $\beta$:
\begin{equation}
\label{eq:init_lambda0_slowroll}
x_0 \simeq \frac{1}{\sqrt6}\bigl(\lambda_{\phi,0}u_0+\beta\,\Omega_{m,0}\bigr) \quad\implies\quad \lambda_{\phi,0}^{(\mathrm{SR})}\simeq \frac{\sqrt{6}\,x_0-\beta\,\Omega_{m,0}}{u_0}
\end{equation}
for $u_0\neq 0$. In this case, we \textit{\textbf{do not take into account}} the $\omega_{a}$ parameter from the data.

\paragraph{Non-slow-roll condition for $\lambda_{\phi,0}$}
In order to completely apply the results obtained from observations (e.g., DESI DR2 \cite{DESI:2025zgx,DESI:2025fii}), we must take into account the parameter $\omega_{a}$ corresponding to the current slope of the equation of state for the scalar sector:
\begin{equation}
\label{eq:init_wphi_prime0}
\omega'_{\phi}\equiv\frac{d \omega_{\phi}}{dN}=a\frac{d \omega_{\phi}}{da} \quad\implies\quad \omega'_{\phi,0}=-\omega_{a}\,.
\end{equation}
This leads to an analytical form describing the initial value for the variable $\lambda_{\phi}$ (nowadays slope of the SF potential):
\begin{equation}
\label{eq:init_lambda0_from_wa}
\lambda_{\phi,0}^{(\mathrm{NSR})}=\frac{\sqrt{6}\,x_{0}-\beta\,\Omega_{m,0}}{\Omega_{\phi,0}}-\frac{\omega_{a}\,\Omega_{\phi,0}}{2\sqrt{6}\,x_{0}\,u_{0}}\,,
\end{equation}
valid for:
\begin{equation}
    \Omega_{\phi,0}\equiv x_0^2+u_0\neq 0 \quad\land\quad x_{0}\,u_{0}\neq 0\,.
\end{equation}
It is easy to see that the relation \eqref{eq:init_lambda0_from_wa} corresponds to the reduced expression \eqref{eq:init_lambda0_slowroll} when the term containing $\omega_{a}$ is negligible.

\paragraph{Effective slope and quasi-de Sitter limit}
Due to the fact that the Klein-Gordon equation is sourced by $\beta\rho_m-\mathcal{V}'(\phi)$, it is useful to define an effective slope parameter:
\begin{equation}
\label{eq:init_lambda_eff_def}
\lambda_{\mathrm{eff}}\equiv\lambda_{\phi}+\beta\,\frac{\Omega_m}{u}\quad\implies\quad \lambda_\phi\,u+\beta\,\Omega_m=u\,\lambda_{\mathrm{eff}}\,.
\end{equation}
In the quasi-de Sitter regime (domination of the SF potential):
\begin{equation}
    |x|\ll 1 \quad\land\quad u\simeq \Omega_\phi
\end{equation}
one obtains:
\begin{equation}
\label{eq:init_lambda_eff_wphi}
\omega_\phi \simeq -1+\frac{\lambda_{\mathrm{eff}}^2}{3}
\quad\implies\quad \lambda_{\mathrm{eff},0}\simeq \sqrt{3\bigl(1+\omega_{\phi,0}\bigr)}
\end{equation}
for:
\begin{equation}
    u_0\simeq \Omega_{\phi,0}\simeq 1\,.
\end{equation}
The equation for the effective slope \eqref{eq:init_lambda_eff_wphi} provides a useful consistency check for the chosen initial conditions near a de Sitter-like phase.
\paragraph{Initial numerical conditions under consideration}
We consider three specific types of initial conditions in our numerical analysis related to the created diagrams:
\begin{enumerate}[label=(\arabic*)]
    \item \textit{\textbf{Slow-roll conditions for Planck 2018}} data (we only take into account the value of $\omega_{0}$, see Appendix~\ref{app:num-plots} for numerical plots):
    \begin{equation}
        \omega_{0}=-0.957\,;
    \end{equation}
    \item \textit{\textbf{Non-slow-roll CPL conditions for Planck 2018}} data (we take into account the value of $\omega_{0}$ and $\omega_{a}$, see Appendix~\ref{app:num-plots} for numerical plots):
    \begin{equation}
        \omega_{0}=-0.957 \quad\land\quad \omega_{a}=-0.29\,;
    \end{equation}
    \item \textit{\textbf{Non-slow-roll CPL conditions for DESI DR2+CMB+Pantheon+}} data (most geometric (background level) dataset) data (we take into account the value of $\omega_{0}$ and $\omega_{a}$):
    \begin{equation}
        \omega_{0}=-0.838 \quad\land\quad \omega_{a}=-0.62\,.
    \end{equation}
\end{enumerate}
Additionally, if possible, we use the effective slope of the SF potential \eqref{eq:init_lambda_eff_wphi} to determine possible values for the free parameters. Comparing the numerical solutions obtained for the physical parameters may provide some clues about the '\textit{inconsistencies}' in the results obtained in \textit{\textbf{Planck 2018}} and \textit{\textbf{DESI DR2}}, which are related to \textbf{\textit{cosmological tensions}} \cite{CosmoVerseNetwork:2025alb}.

\subsection{Model I: Axions/ALPs}
\label{subsec:numerics_axion}
\begin{figure}[htbp]
\centering
\includegraphics[width=1\linewidth]{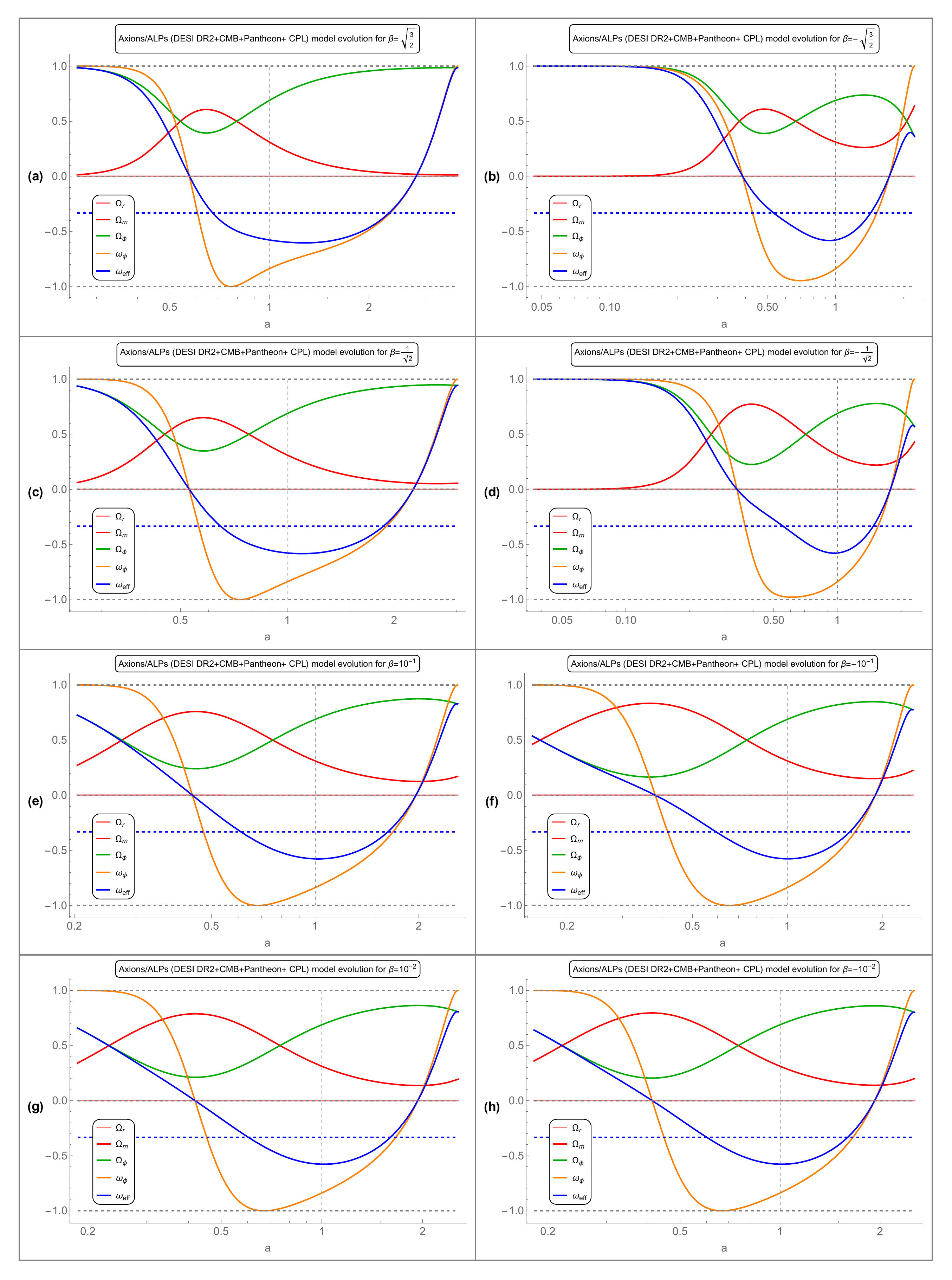}
\caption{Evolution of the NMC axions/ALPs model in the non-slow-roll CPL approximation for DESI+CMB+Pantheon+ initial data.}
\label{fig:evol_desi_cplnsl_pantheon_axions}
\end{figure}

The resulting evolution diagrams (Fig.~\ref{fig:evol_desi_cplnsl_pantheon_axions}, Fig.~\ref{fig:evol_planck2018_cplsl_axions}, Fig.~\ref{fig:evol_planck2018_cplnsl_axions}) demonstrate the following qualitative behavior:
\begin{itemize}
    \item For weak and moderate NMC parameters ($|\beta|=10^{-2},10^{-1}$) the transition from matter domination to SF domination (accelerated expansion) is preserved, with $\omega_{\mathrm{eff}}$ crossing $-1/3$ once $\Omega_\phi$ becomes dominant;
    \item For strong NMC regimes ($|\beta|=\sqrt{\frac{3}{2}},\frac{1}{\sqrt{2}}$) there is a reshaping of the intermediate epoch (consistent with the large $\phi$MDE signatures \eqref{eq:phiMDE_signature});
    \item Any sharp peaks in $\omega_\phi$ should be interpreted as ill-defined due to the $\Omega_\phi=0$ value (Sec.~\ref{subsec:physicality_negative_Omegaphi}).
\end{itemize}

\subsection{Model II: Cyclic ekpyrotic}
\label{subsec:numerics_cyclic}

The cyclic ekpyrotic SF potential \eqref{eq:V_cyclic} contains a negative branch and therefore probes the regime:
\begin{equation}
    u<0\,,
\end{equation}
so that $\Omega_\phi$ can approach or even cross zero value \cite{Felder:2002jk,Heard:2002dr,Lehners:2008vx}. It can be seen in the obtained diagrams by:
\begin{itemize}
    \item Possible epochs characterized by very low values of $\Omega_\phi$;
    \item Trajectories for which $\omega_\phi$ peaks.
\end{itemize}
Moreover, if NMC is sufficiently strong and/or trajectories explore the negative SF potential sector, the flow can be driven toward the:
\begin{equation}
    H\to 0
\end{equation}
boundary analyzed in Appendix~\ref{app:poincare_full}.

\begin{figure}[htbp]
\centering
\includegraphics[width=1\linewidth]{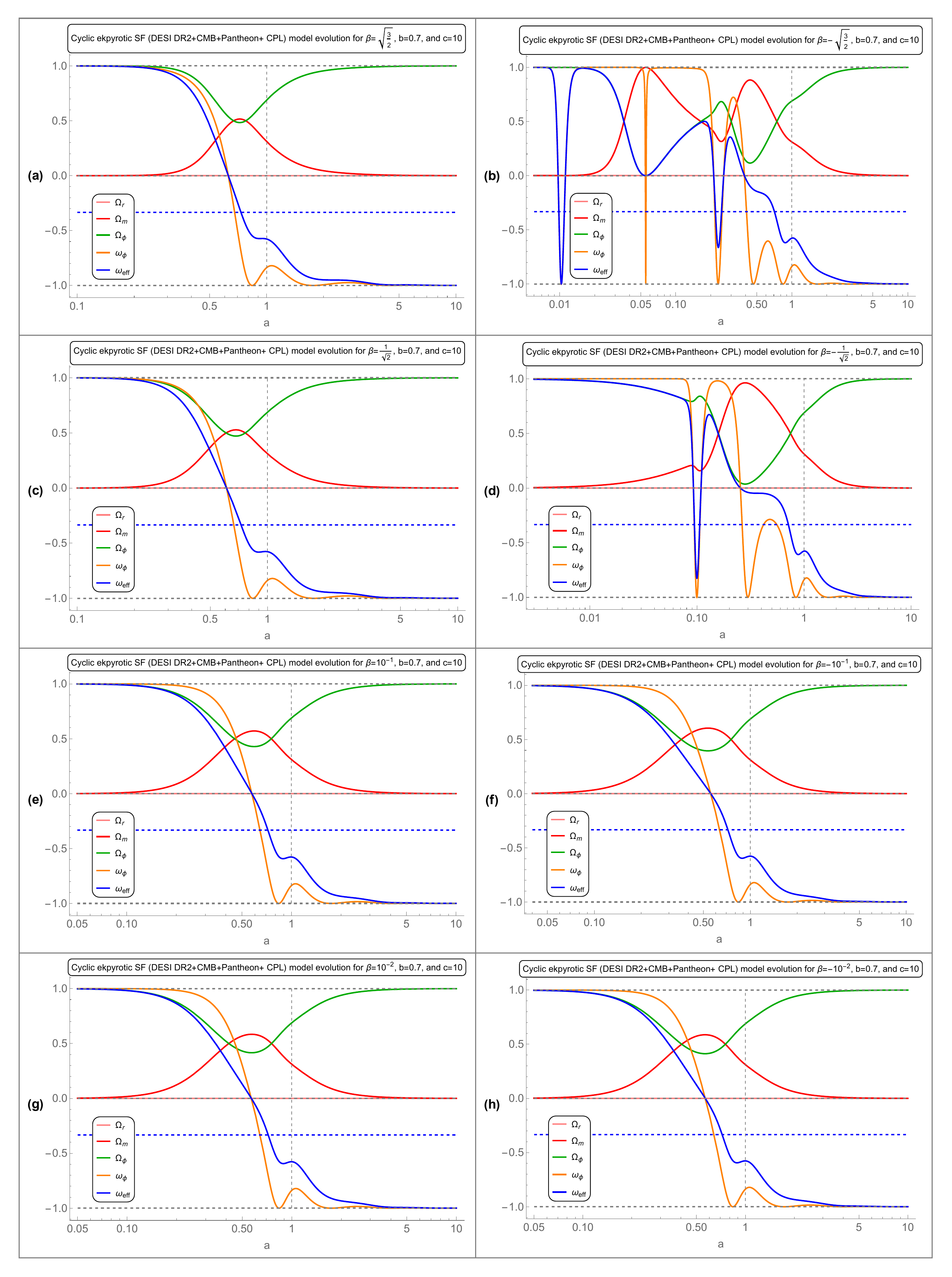}
\caption{Evolution of the NMC cyclic ekpyrotic model in the non-slow-roll CPL approximation for DESI+CMB+Pantheon+ initial data.}
\label{fig:evol_desi_cplnsl_pantheon_cyclic}
\end{figure}

The resulting evolution diagrams (Fig.~\ref{fig:evol_desi_cplnsl_pantheon_cyclic}, Fig.~\ref{fig:evol_planck2018_cplsl_cyclic}, Fig.~\ref{fig:evol_planck2018_cplnsl_cyclic}) demonstrate the following qualitative behavior:
\begin{itemize}
    \item For weak/moderate NMC ($|\beta|=10^{-2},10^{-1}$), one typically observes a traditional cosmic chronology with an intermediate coupled scaling imprint consistent with the saddle points $\mathcal{C}$ and $\mathcal{F}$ in Sec.~\ref{subsec:cyclic};
    \item For strong NMC ($|\beta|=\frac{1}{\sqrt{2}},\sqrt{\frac{3}{2}}$), trajectories may probe regions with negative $u$ variable values and/or $\Omega_\phi\simeq 0$ (resulting in sharp fluctuations of $\omega_\phi$). They should be interpreted as signatures of branches with negative potential, rather than as phantom microphysics;
    \item For negative $\Omega_\phi$ values dynamical system variables tend to grow because they scale as $H^{-2}$ - approaching $H\to 0$. In this regime our formalism must be replaced by a regular extension for possible bounce/cyclic behavior of the model (see Sec.~\ref{sec:bounce} and Appendix~\ref{app:poincare_full}) \cite{Battefeld:2014uga,Brandenberger:2016vhg}.
\end{itemize}

\subsection{Model III: Exponential with constant}
\label{subsec:numerics_expl}
\begin{figure}[htbp]
\centering
\includegraphics[width=1\linewidth]{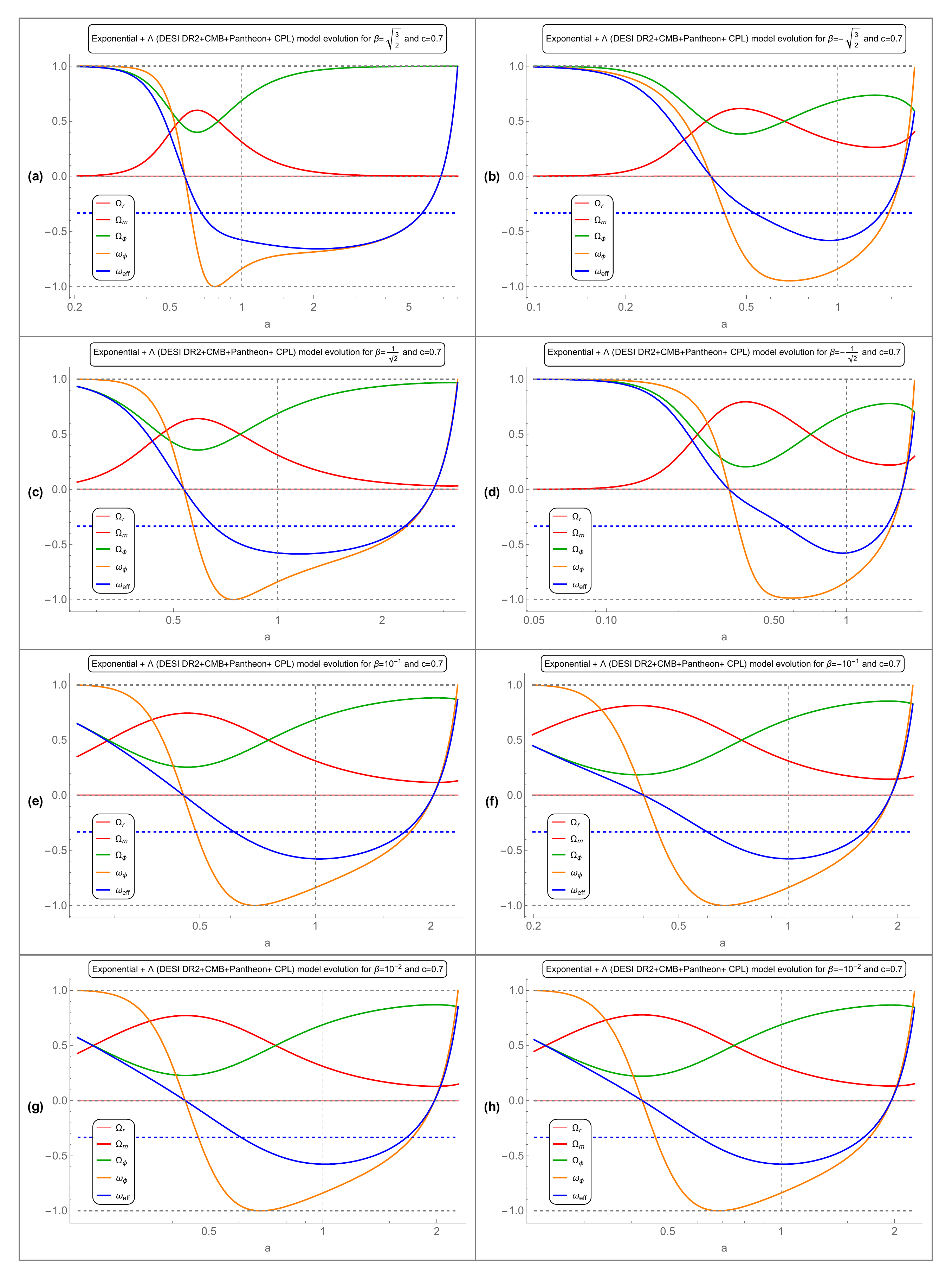}
\caption{Evolution of the NMC exponential with constant model in the non-slow-roll CPL approximation for DESI+CMB+Pantheon+ initial data.}
\label{fig:evol_desi_cplnsl_pantheon_expl}
\end{figure}

In this model, the resulting evolution diagrams (Fig.~\ref{fig:evol_desi_cplnsl_pantheon_expl}, Fig.~\ref{fig:evol_planck2018_cplsl_expl}, Fig.~\ref{fig:evol_planck2018_cplnsl_expl}) demonstrate the following qualitative behavior:
\begin{itemize}
    \item For weak and moderate NMC one obtains the late-time accelerated expansion (point $\mathcal{M}$ in Table~\ref{tab:expL_fp});
    \item Strong NMC (depending on initial data) may prolong the coupled scaling regime or even drive the cosmological system away from:
    \begin{equation}
        \lambda_\phi\to 0
    \end{equation}
    plateau for some period of time. It can yield a delayed or distorted approach into the equation of state:
    \begin{equation}
        \omega_{\mathrm{eff}}\simeq -1\,;
    \end{equation}
    \item As was mentioned before, any $\omega_\phi$ peaks should be checked with the $\Omega_\phi$ values (Sec.~\ref{subsec:physicality_negative_Omegaphi}).
\end{itemize}

\subsection{Model IV: Tracking quintessence}
\label{subsec:numerics_quint}

\begin{figure}[htbp]
\centering
\includegraphics[width=1\linewidth]{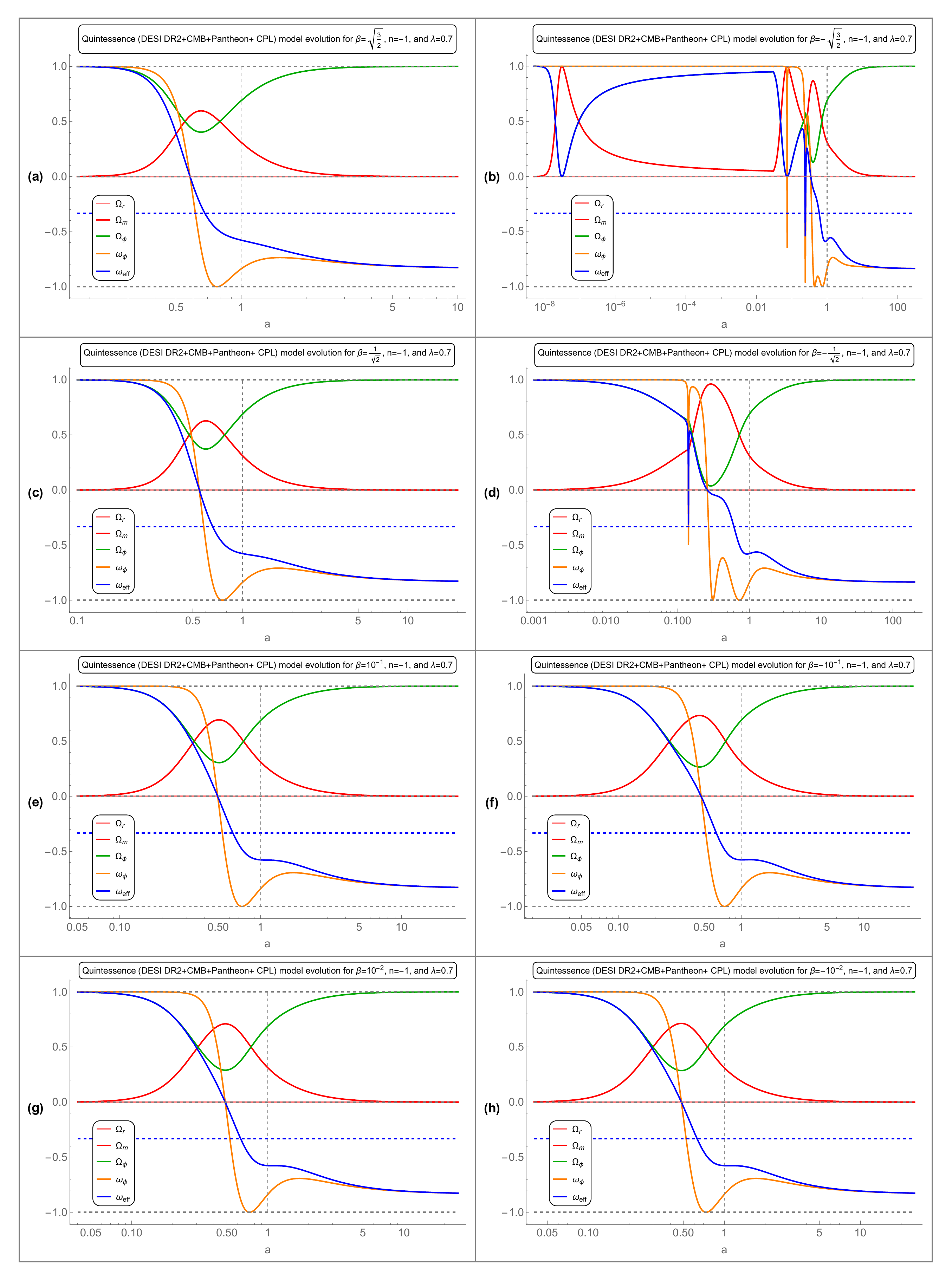}
\caption{Evolution of the NMC quintessence (n=-1) model in the non-slow-roll CPL approximation for DESI+CMB+Pantheon+ initial data.}
\label{fig:evol_desi_cplnsl_pantheon_quint_n1}
\end{figure}

\begin{figure}[htbp]
\centering
\includegraphics[width=1\linewidth]{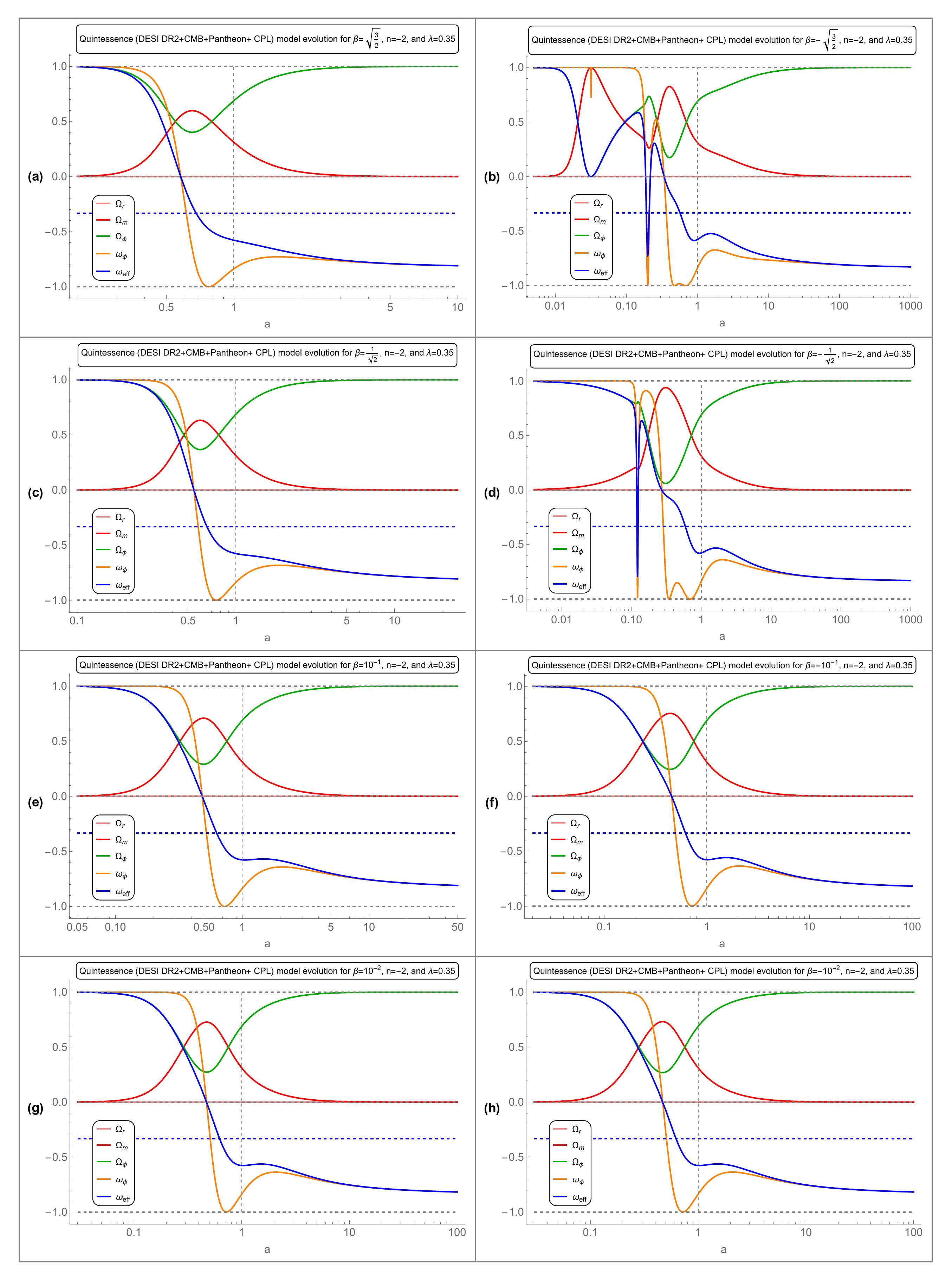}
\caption{Evolution of the NMC quintessence (n=-2) model in the non-slow-roll CPL approximation for DESI+CMB+Pantheon+ initial data.}
\label{fig:evol_desi_cplnsl_pantheon_quint_n2}
\end{figure}

The resulting evolution diagrams (Fig.~\ref{fig:evol_desi_cplnsl_pantheon_quint_n1}, Fig.~\ref{fig:evol_desi_cplnsl_pantheon_quint_n2}, Fig.~\ref{fig:evol_planck2018_cplsl_quint_n1} - Fig.~\ref{fig:evol_planck2018_cplnsl_quint_n2}) demonstrate the following qualitative behavior:
\begin{itemize}
    \item For weak and moderate NMC the model corresponds to evolution as in the case of tracking/freezing behavior: $\Omega_\phi$ grows at late times and $\omega_{\mathrm{eff}}$ crosses below $-1/3$ once the scalar sector dominates;
    \item For strong NMC one could observe extended coupled scaling regimes and/or sharp $\omega_\phi$ evolution;
    \item For n=-1, such an evolution causes $\Omega_{\phi}$ to pass through (or near) zero as a result of a change in the sign of the variable $u$.
\end{itemize}

\subsection{Model V: SFDM}
\label{subsec:numerics_sfdm}

\begin{figure}[htbp]
\centering
\includegraphics[width=1\linewidth]{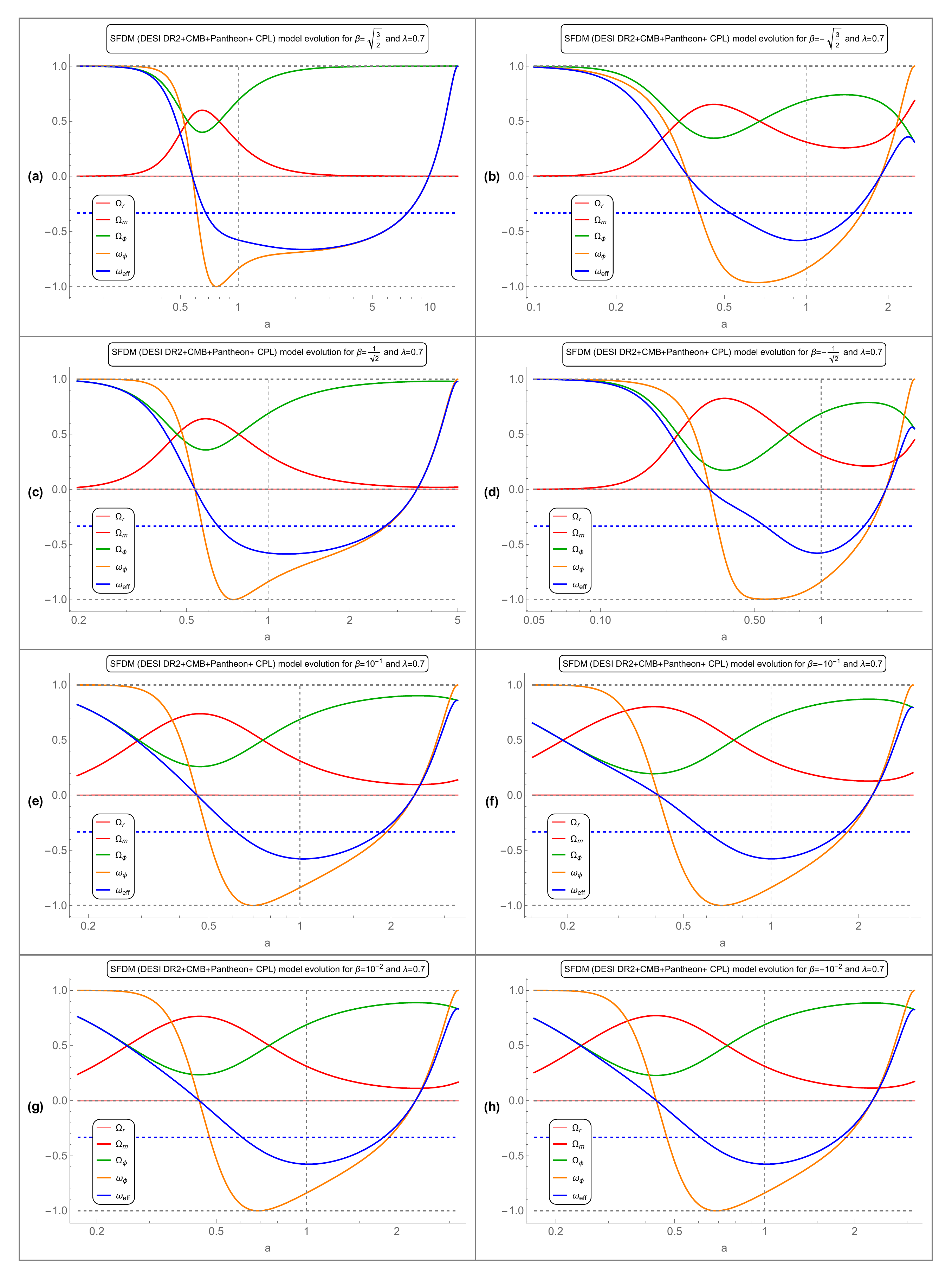}
\caption{Evolution of the NMC SFDM model in the non-slow-roll CPL approximation for DESI+CMB+Pantheon+ initial data.}
\label{fig:evol_desi_cplnsl_pantheon_sfdm}
\end{figure}

The resulting evolution diagrams (Fig.~\ref{fig:evol_desi_cplnsl_pantheon_sfdm}, Fig.~\ref{fig:evol_planck2018_cplsl_sfdm}, Fig.~\ref{fig:evol_planck2018_cplnsl_sfdm}) demonstrate the following qualitative behavior:
\begin{itemize}
    \item For weak and moderate NMC, one obtains the late-time acceleration if trajectories approach the epoch of SF potential domination;
    \item Strong couplings can delay, advance, or transiently interrupt the accelerated expansion phase, demonstrating the increased importance of coupled scaling solutions.
\end{itemize}

\subsection{Summary}
\label{subsec:numerical_summary}

Analysis of the diagrams obtained during the analysis indicates the following aspects:
\begin{enumerate}[label=(\arabic*)]
    \item \textbf{Weak/moderate NMC preserves usual cosmic chronology}. For $|\beta|=10^{-2}$ and often also $|\beta|=10^{-1}$, the background evolution remains close to the conventional LCDM evolution, with moderate interaction driven distortions of $\omega_{\mathrm{eff}}(a)$;
    \item \textbf{Strong coupling generically imprints (or replaces) matter domination by coupled scaling regimes}. For $|\beta|=\frac{1}{\sqrt{2}}$ and especially $|\beta|=\sqrt{\frac{3}{2}}$, the SF fraction expected near coupled saddle points could be large \eqref{eq:phiMDE_signature}, and the evolution may deviate strongly from a standard matter epoch, consistent with the fixed points structure;
    \item \textbf{Negative SF potentials allow $\Omega_\phi\le 0$ and could drive the flow toward $H\to 0$}. If trajectories of the models explore the negative $u$ branch, then one expects:
    \begin{equation}
        \Omega_\phi\simeq 0 \quad\lor\quad \Omega_\phi<0\,,
    \end{equation}
    therefore, $\omega_\phi$ peaks, or in extreme cases drift toward the $H\to 0$ boundary (Appendix~\ref{app:poincare_full}) \cite{Felder:2002jk,Heard:2002dr}. For an expanding phase matching observations up to $a\simeq 1$, such behavior must be absent (or at least safely outside the observational window).
\end{enumerate}
\section{DESI DR2 interpretation in the interacting scalar sector}
\label{sec:desi}
\paragraph{Interpretational caveat.}
Sec.~\ref{sec:desi} provides a \emph{phenomenological} connection between the dynamical systems evolution and BAO-sensitive background observables. We emphasize that we do not perform a statistical parameter estimation against DESI/Planck/Pantheon+ likelihoods here; rather, CPL $(w_0,w_a)$ values are used to construct physically motivated \emph{illustrative} initial conditions for representative trajectories. A determination of the allowed $\beta$ region (and potential parameters) requires a dedicated global fit, ideally including perturbations/LSS, which we leave for future work.
\subsection{BAO constraints at the background level}
BAO measurements are related to relationships that include the Hubble parameter. Through equation \eqref{eq:Hdot_weff}, any evolution of the effective equation of state $\omega_{\mathrm{eff}}(N)$ affects the evolution of the Hubble parameter $H(z)$ and therefore changes the distances within BAO. Due to the fact that BAOs are geometric in nature, the dynamical system working at the background level directly determines what behaviors of models for late epochs allow for a departure from the standard expansion characteristic of the $\Lambda$CDM model.

\subsection{Why interacting scalar sector generates evolving $\omega_{\mathrm{eff}}(z)$}
In our formalism:
\begin{equation}
\omega_{\mathrm{eff}}(N)=\frac{1}{3}\Omega_{r}(N)+x^{2}(N)-u(N)\,,
\end{equation}
and the other dynamical system variables evolve according to \eqref{eq:DSfinal}. 
If a given trajectory does not lie exactly on an invariant submanifold for the critical point for all values of $N$, then the effective equation of state changes over time. Orbits move between early-time saddle fixed points and late-time stable points.

\subsection{Thawing vs. freezing behavior}
The thawing trajectory is the one for which the SF is initially frozen with
\begin{equation}
    \omega_\phi\simeq -1
\end{equation}
and then evolves away from that value at late times, on the other hand, freezing trajectories evolve toward the de Sitter-like epoch \cite{Caldwell:2005tm,Barboza:2008rh,Copeland:2006wr,Linder:2006sv}:
\begin{equation}
    \omega_\phi\to -1\,.
\end{equation}
In the language of our dynamical system, this means that for domination of the SF potential one obtains:
\begin{equation}
    u\simeq 1 \quad\land\quad |x|\ll 1\,
\end{equation}
therefore:
\begin{equation}
    1+\omega_\phi\simeq 2x^2\,.
\end{equation}
Therefore, thawing behavior corresponds to an increase in $|x|$ from an initially small value to larger values.

\section{Conclusions}
\label{sec:conclusions}
In this paper, we present an analysis of a dynamical system incorporating a non-minimal coupling of exponential form between the scalar field and matter/dust within the Einstein frame. We have taken into account potentials with exclusively positive values as well as those that can take negative values.

Our main results are:
\begin{enumerate}[label=(\arabic*)]
\item Developing a dynamical variable corresponding to the potential energy of a scalar field, taking into account both positive and negative values, and obtaining an explicit evolution of key physical parameters for individual models;
\item We have performed a precise analysis of the stability of five specific physically motivated SF potential forms (on expanding and contracting branches). We clarified which regimes are physically relevant within the finite values of the variable $\lambda_{\phi}$ and what specific mathematical methods can be used to characterize them;
\item By using a Poincar\'e compactification of the reduced constant $\lambda_{\phi}$ system we have obtained a complete classification of critical sets at infinity. In the physical $u<0$ sector the reduced Hubble-normalized flow is generically drawn toward the equator at infinity, corresponding to $r\to\infty$ and thus $H\to0$. This provides a precise explanation of the way in which negative potentials for the scalar field can naturally lead to a transition from the expansion phase to the collapse phase (turnaround) and why the use of ‘e-fold time’ becomes singular when $H\to0$.
\item We performed a detailed numerical analysis of the equations describing our dynamical system using observationally motivated illustrative initial conditions derived from Planck 2018 and DESI DR2 CPL parameterizations for four selected values of the parameter describing the strength of the NMC. We have described how changing the value and sign of this parameter can affect individual cosmological models.
\end{enumerate}

\section*{Declaration of competing interest}
The authors declare that they have no known competing financial interests or personal relationships that could have appeared to
influence the work reported in this paper.

\section*{Data availability}
No data was used for the research described in the article. The Wolfram Mathematica notebook used to create the diagrams for the article can be made available upon special request.

\section*{Acknowledgments}
The author would like to thank \textit{Roksana Szwarc}, \textit{Eleonora Di Valentino}, \textit{William Giar\`e}, \textit{Carsten van de Bruck} and \textit{Dong Ha Lee} for their accurate observations and comments in the initial phase of preparing the article. In addition, the author dedicates the article to \textit{Roksana Szwarc}, his parents: \textit{Bo\.zena} and \textit{Krzysztof}, his sister \textit{Dominika}, in loving memory of \textit{Grandma Stefcia}, and last but not least: \textit{Antonio}, \textit{Bob}, \textit{Bob II}, \textit{Groovy}, \textit{Milena} and \textit{Stuart}.

This article is based upon work from COST Action CA21136 – “Addressing observational tensions in cosmology with systematics and fundamental physics (\href{https://cosmoversetensions.eu/}{CosmoVerse})”, supported by COST (European Cooperation in Science and Technology).

This article is based upon work from COST Action \href{https://cosmicwispers.eu/}{COSMIC WISPers} CA21106, supported by COST (European Cooperation in Science and Technology).

\appendix

\section{Jacobian eigenvalues used in the tables (explicit forms)}
\label{app:eigs}

For constant $\lambda_{\phi*}$ branches, let us treat $\lambda_\phi=\lambda_{\phi*}$ as a parameter in the reduced $(x,u,\Omega_r)$ system.

\paragraph{Explicit formula of $\mu_{\lambda}$ for models III-V}
For tables that list the fourth eigenvalue $\mu_{\lambda}$ associated with the $\lambda_\phi$ direction in the full $(x,u,\Omega_r,\lambda_\phi)$ dynamical system, we define:
\begin{equation}
    \mu_{\lambda}\equiv
\left.\frac{\partial \lambda_\phi'}{\partial \lambda_\phi}\right|_{*}\,.
\end{equation}
Using \eqref{eq:DSfinal} and the explicit closures adopted in Sec.~\ref{sec:potentials}, one obtains:
\begin{itemize}
\item \textbf{Model III (exponential$+\Lambda$):}
\begin{equation}
\mu_{\lambda}=\left.-\sqrt6\,x_*(c-2\lambda_{\phi*})\right.\,.
\end{equation}
In particular, on the exponential branch $\lambda_{\phi*}=c$ one has:
\begin{equation}
    \mu_{\lambda}=+\sqrt6\,c\,x_*\,,
\end{equation}
while on the plateau branch $\lambda_{\phi*}=0$ one obtains:
\begin{equation}
    \mu_{\lambda}=-\sqrt6\,c\,x_*\,;
\end{equation}
\item \textbf{Model IV (tracking quintessence):}
\begin{equation}
\mu_{\lambda}=\left.\frac{2\sqrt6}{n}\,\lambda_{\phi*}\,x_*\right.\,.
\end{equation}
Thus, on the constant branches $\lambda_{\phi*}=\pm n\lambda$ one obtains:
\begin{equation}
    \mu_{\lambda}=\pm 2\sqrt6\,\lambda\,x_*\,;
\end{equation}
\item \textbf{Model V (SFDM):}
\begin{equation}
\mu_{\lambda}=\left.2\sqrt{\frac32}\,\lambda_{\phi*}\,x_*\right.\,.
\end{equation}
Therefore, on the constant branches $\lambda_{\phi*}=\pm\lambda$ it yields:
\begin{equation}
    \mu_{\lambda}=\pm \sqrt6\,\lambda\,x_*\,.
\end{equation}
\end{itemize}
Note also that for all fixed points with $x_*=0$ (e.g. radiation domination and de Sitter-like points) one obtains:
\begin{equation}
    \mu_{\lambda}=0
\end{equation}
at linear order.

\paragraph{Reduced eigenvalues.}
The eigenvalues for the reduced 3D dynamical system (for fixed $\lambda_{\phi*}$) are as follows:
\begin{itemize}
    \item Radiation (R):
    \begin{equation}
        \{-1,1,4\}\,;
    \end{equation}
    \item Kinetic (K$_\pm$):
    \begin{equation}
        \left\{2,\ 3\mp\sqrt6\beta,\ 6\mp\sqrt6\lambda_{\phi*}\right\}\,;
    \end{equation}
    \item $\phi$MDE:
    \begin{equation}
        \left\{\beta^2-\frac32,\ 2\beta^2-1,\ 2\beta^2-2\beta\lambda_{\phi*}+3\right\}\,;
    \end{equation}
    \item Coupled radiation scaling:
    \begin{equation}
        \left\{\,4-\frac{\lambda_{\phi*}}{\beta},\ -\frac12\pm \frac12\sqrt{\frac{2}{\beta^2}-3}\,\right\}\,;
    \end{equation}
    \item Scalar dominated (P):
    \begin{equation}
        \left\{\,\lambda_{\phi*}^2-4,\ \frac12(\lambda_{\phi*}^2-6),\ \lambda_{\phi*}^2-\beta\lambda_{\phi*}-3\,\right\}\,;
    \end{equation}
    \item Matter-scalar scaling (S):
    \begin{equation}
        x_*=\frac{\sqrt6}{2(\lambda_{\phi*}-\beta)} \quad\land\quad u_*=\frac{2\beta^2-2\beta\lambda_{\phi*}+3}{2(\lambda_{\phi*}-\beta)^2} \quad\land\quad \Omega_{r*}=0
    \end{equation}
    with:
    \begin{equation}
        \mu_1=\frac{\lambda_{\phi*}-4\beta}{\beta-\lambda_{\phi*}}
    \end{equation}
    and:
    \begin{equation}
        \mu_{\pm}=
-\frac{6\beta-3\lambda_{\phi*}\pm \sqrt{3}\sqrt{\Delta(\beta,\lambda_{\phi*})}}{4(\beta-\lambda_{\phi*})}\,,
    \end{equation}
    where:
    \begin{equation}
        \Delta(\beta,\lambda_{\phi*})=16\beta^3\lambda_{\phi*}-32\beta^2\lambda_{\phi*}^2+60\beta^2+16\beta\lambda_{\phi*}^3-36\beta\lambda_{\phi*}-21\lambda_{\phi*}^2+72\,;
    \end{equation}
    \item Radiation--scalar scaling (RS):
    \begin{equation}
        x_*=\frac{2\sqrt6}{3\lambda_{\phi*}} \quad\land\quad u_*=\frac{4}{3\lambda_{\phi*}^2} \quad\land\quad \Omega_{r*}=1-\frac{4}{\lambda_{\phi*}^2}
    \end{equation}
    and:
    \begin{equation}
        \left\{\frac{\lambda_{\phi*}-4\beta}{\lambda_{\phi*}},\ -\frac12\pm\frac{1}{2\lambda_{\phi*}}\sqrt{64-15\lambda_{\phi*}^2}\right\}\,.
    \end{equation}
\end{itemize}

\section{Center manifold stability for non-hyperbolic points}
\label{app:center}

At fixed points with $x_*=0$, the $\lambda_\phi$ direction is a center at linear order because $\lambda_\phi'\propto x$ in \eqref{eq:DSfinal}.Therefore, stability must be specified by the reduced scalar dynamics near the relevant region of $\mathcal{V}(\phi)$.

\paragraph{Axions/ALPs hilltop}
At the hilltop fixed point:
\begin{equation}
    (x_*,u_*,\Omega_{r*},\lambda_{\phi*})=(0,1,0,0)\,,
\end{equation}
the full linearization of \eqref{eq:DSfinal} does not require a center manifold treatment due to the fact that the $(x,\lambda_\phi)$
subsystem mixes at linear order and yields the following eigenvalues:
\begin{equation}
    \mu_\pm=\frac{-3\pm\sqrt{15}}{2}\,,
\end{equation}
so that the point is a hyperbolic saddle with one unstable direction (consistent with $\mathcal{V}''(\pi)<0$).

\paragraph{Exponential+$\Lambda$}
The asymptotic $\lambda_\phi\to 0$ regime corresponds to $\mathcal{V}(\phi)\to 2\Lambda>0$. Hubble friction damps $\dot\phi$ and $\phi$ approaches the plateau. Therefore, the de Sitter point is a stable one provided the effective SF potential does not introduce a tachyonic direction.

\section{Poincar\'e compactification and critical points at infinity for $u<0$ branches}
\label{app:poincare_full}

The sign-safe variable:
\begin{equation}
u \equiv\frac{\mathcal{V}(\phi)}{3H^2}
\end{equation}
allows for negative values, but in such a case the state space is \emph{not} a priori compact. Divergences in $(x,u,\Omega_r)$ typically correspond to approaching a hypersurface where $H\to 0$ (turnaround/bounce), because all
Hubble-normalized densities scale as $H^{-2}$. Therefore, 'points at infinity' diagnose where the $N=\ln a$ time variable becomes ill-defined and where an $H$-regular extension (e.g. via a compact Hubble variable) would be required to continue the flow across $H=0$.

Throughout this appendix we work with the reduced 3D system on a \emph{constant $\lambda_{\phi*}$ branch} (exponential tail), i.e.:
\begin{equation}
    \lambda_\phi=\lambda_{\phi*}
\end{equation}
is treated as a constant parameter, and we also write:
\begin{equation}
    \Omega\equiv\Omega_r\,.
\end{equation}
for more compact form.

\subsection{Reduced polynomial system on a constant $\lambda_{\phi*}$ branch}
\label{app:poincare_system}

On a constant-$\lambda_{\phi*}$ branch, one could eliminate $\Omega_m$ via the constraint:
\begin{equation}
\Omega_m = 1-x^2-u-\Omega\,,
\end{equation}
and expand the reduced system \eqref{eq:DSfinal} for $(x,u,\Omega)$. One finds the following polynomial vector field:
\begin{subequations}
\label{eq:poly_reduced}
\begin{align}
x' &= \frac{3}{2}x^3-\frac{\sqrt6}{2}\beta\,x^2-\frac{3}{2}x\,u+\frac{1}{2}x\,\Omega-\frac{3}{2}x+\frac{\sqrt6}{2}(\lambda_{\phi*}-\beta)u-\frac{\sqrt6}{2}\beta\,\Omega+\frac{\sqrt6}{2}\beta\,,
\\
u' &= 3u\,x^2-\sqrt6\,\lambda_{\phi*}\,u\,x-3u^2+\Omega\,u+3u\,,
\\
\Omega' &= 3\Omega\,x^2-3\Omega\,u+\Omega^2-\Omega\,.
\end{align}
\end{subequations}
The maximal polynomial degree is $m=3$. It is useful to split the field into homogeneous segments:
\begin{equation}
\mathbf{F}(x,u,\Omega)=\mathbf{F}_3(x,u,\Omega)+\mathbf{F}_2(x,u,\Omega)+\mathbf{F}_1(x,u,\Omega)+\mathbf{F}_0\,,
\end{equation}
where the degree-3 and degree-2 parts are as follows:
\begin{align}
\label{eq:F3_full}
\mathbf{F}_3(x,u,\Omega)&=\left(\frac{3}{2}x^3,3u\,x^2,3\Omega\,x^2\right),
\\
\label{eq:F2_full}
\mathbf{F}_2(x,u,\Omega)&=\left(-\frac{\sqrt6}{2}\beta\,x^2-\frac{3}{2}x\,u+\frac{1}{2}x\,\Omega,-\sqrt6\,\lambda_{\phi*}\,u\,x-3u^2+\Omega u,\Omega^2-3\Omega u\right)\,.
\end{align}
The lower-degree parts only affect subleading behavior at infinity, \emph{except} where the leading homogeneous field vanishes (degenerate directions). This is precisely what occurs here on a large equatorial set.

\subsection{Poincar\'e sphere: embedding, desingularization, and induced flow}
\label{app:poincare_sphere}

Let us introduce the radial variable:
\begin{equation}
\label{eq:poincare_def}
r\equiv\sqrt{x^2+u^2+\Omega^2} \quad\land\quad (a,b,c)\equiv\left(\frac{x}{r},\frac{u}{r},\frac{\Omega}{r}\right) \quad\land\quad a^2+b^2+c^2=1\,.
\end{equation}
Infinity in $(x,u,\Omega)$ corresponds to $r\to\infty$, i.e. points on the unit sphere $S^2$ in $(a,b,c)$.

For a polynomial vector field of maximal degree $m=3$, Poincar\'e compactification uses a time rescaling:
\begin{equation}
\frac{d\tau}{dN}=r^{m-1}=r^2\,,
\end{equation}
so that the induced flow on the sphere depends only on the maximal-degree segment $\mathbf{F}_3$:
\begin{equation}
\label{eq:sphere_flow_general}
\dot{\mathbf{s}}=\mathbf{F}_3(\mathbf{s})-[\mathbf{s}\cdot\mathbf{F}_3(\mathbf{s})]\,\mathbf{s} \quad\land\quad \mathbf{s}=(a,b,c) \quad\land\quad \dot{}\equiv\frac{d}{d\tau}\,.
\end{equation}
Using \eqref{eq:F3_full}, one finds that:
\begin{equation}
    \mathbf{F}_3(\mathbf{s})=\left(\frac{3}{2}a^3,3b\,a^2,3c\,a^2\right)\,.
\end{equation}
Let us also define the scalar:
\begin{equation}
\label{eq:alpha_def}
\alpha(\mathbf{s})=\mathbf{s}\cdot\mathbf{F}_3(\mathbf{s})=\frac{3}{2}a^4+3a^2(b^2+c^2)=\frac{3}{2}a^2(2-a^2)\,.
\end{equation}
Then \eqref{eq:sphere_flow_general} yields the explicit sphere system:
\begin{subequations}
\label{eq:sphere_flow_explicit}
\begin{align}
\dot a &= \frac{3}{2}a^3(a^2-1)\,,
\\
\dot b &= \frac{3}{2}b\,a^4\,,
\\
\dot c &= \frac{3}{2}c\,a^4\,.
\end{align}
\end{subequations}

\subsection{Critical sets at infinity from the leading flow}
\label{app:poincare_critical_sets}

From \eqref{eq:sphere_flow_explicit}, the critical points on the sphere satisfy:
\begin{equation}
    \dot a=\dot b=\dot c=0\,.
\end{equation}
\paragraph{Poles (kinetic directions)}
There are two isolated points:
\begin{equation}
\mathbf{s}_{\pm}=(a,b,c)=\left(\pm1,0,0\right)\,,
\end{equation}
corresponding to:
\begin{equation}
    \frac{x}{r}\to\pm1\,,
\end{equation}
so that:
\begin{equation}
    |x|\gg |u| \quad\land\quad |x|\gg |\Omega|
\end{equation}
at infinity.

\paragraph{Equatorial critical circle (degenerate directions)}
Because:
\begin{equation}
    \dot b \propto a^4 \quad\land\quad \dot c \propto a^4\,,
\end{equation}
every point with $a=0$ is critical for the leading flow:
\begin{equation}
\mathcal{C}_\infty:\quad a=0 \quad\land\quad b^2+c^2=1\,,
\end{equation}
corresponding to directions where $|x|/r\to 0$ while $(u,\Omega)$ blow up proportionally.

This degeneracy means that the leading ($m=3$) compactification does not resolve the tangential dynamics along the equator. The degree-2 terms \eqref{eq:F2_full} provide the next non-vanishing contribution.

\subsection{Linearized tangential behavior on $S^2$: poles vs. equator}
\label{app:poincare_linear}

\paragraph{Pole tangential instability}
Linearizing \eqref{eq:sphere_flow_explicit} at:
\begin{equation}
    \mathbf{s}_+=(1,0,0)
\end{equation}
gives:
\begin{equation}
\dot b = \frac{3}{2}b \quad\land\quad \dot c=\frac{3}{2}c\,,
\end{equation}
and in similar way at:
\begin{equation}
    \mathbf{s}_{-}=(-1,0,0)\,.
\end{equation}
Therefore, both poles are \emph{repelling on the sphere} in the $(b,c)$ directions.

\paragraph{Equator normal attraction}
For $|a|\ll 1$ with $b^2+c^2\simeq 1$, one obtains:
\begin{equation}
    \dot a\simeq -\frac{3}{2}a^3\,,
\end{equation}
so that:
\begin{equation}
    a\to 0 \quad\land\quad \tau\to+\infty\,.
\end{equation}
Thus, the equator is \emph{normally attracting} for the desingularized flow.

However, to decide whether trajectories in the original variables actually approach infinity (radially), one must analyze the radial dynamics (growth of $r$) in the original $N$ time.

\subsection{Radial dynamics}
\label{app:poincare_radial}
A convenient diagnostic is the following evolution equation:
\begin{equation}
r'=\frac{x\,x'+u\,u'+\Omega\,\Omega'}{r}\,,
\end{equation}
where:
\begin{equation}
    r=\sqrt{x^2+u^2+\Omega^2}\,.
\end{equation}
At large $r$, the leading contribution depends on direction. Along the equator ($a=0$) the cubic terms vanish, so the \emph{quadratic} terms control the behavior of $r'$.

By using $x=0$ in \eqref{eq:poly_reduced} one finds that the leading (quadratic) subsystem is:
\begin{subequations}
\label{eq:equator_leading}
\begin{align}
u' &\simeq u\left(\Omega-3u\right),
\\
\Omega' &\simeq \Omega\left(\Omega-3u\right)\,,
\end{align}
\end{subequations}
which yields:
\begin{equation}
\label{eq:rprime_equator}
r' \simeq r(\Omega-3u)\,.
\end{equation}
Therefore, along the equatorial directions, the flow goes to infinity in $N$ time if and only if:
\begin{equation}
    \Omega-3u>0\,.
\end{equation}

\subsection{Equator refinement via degree-2 dynamics (full classification on $\mathcal{C}_\infty$)}
\label{app:poincare_equator_refine}

Because \eqref{eq:sphere_flow_explicit} leaves the entire equator critical, the correct classification of infinity requires a refined compactification on $\mathcal{C}_\infty$ using the degree-2 field \eqref{eq:F2_full}.

A direct way to see the equatorial direction field is to consider the following ratio by using \eqref{eq:equator_leading}:
\begin{equation}
    \frac{d\Omega}{du}=\frac{\Omega'}{u'} \quad\land\quad x=0\,.
\end{equation}
For $u\neq 0$ and $\Omega\neq 0$ one can obtain:
\begin{equation}
\frac{d\Omega}{du} = \frac{\Omega(\Omega-3u)}{u(\Omega-3u)}=\frac{\Omega}{u}\,,
\end{equation}
so that:
\begin{equation}
    \frac{\Omega}{u}=\mathrm{const}
\end{equation}
along the leading equatorial flow. This matches the geometric picture - equatorial directions are parameterized by the ratio:
\begin{equation}
    \frac{c}{b}=\frac{\Omega}{u}\,.
\end{equation}

Moreover, relation \eqref{eq:equator_leading} implies that the factor $(\Omega-3u)$ controls \emph{both} the direction field and the radial growth \eqref{eq:rprime_equator}. Therefore, $\Omega=3u$ is a distinguished line in the equatorial diagram. It corresponds to the point(s) where $c=3b$ intersect $b^2+c^2=1$:
\begin{equation}
(b,c)=\left(\pm \frac{1}{\sqrt{10}},\ \pm \frac{3}{\sqrt{10}}\right)\,.
\end{equation}
These are \emph{neutral} with respect to the leading radial growth at $x=0$ because $r'\simeq 0$ there. At such points one must proceed to next order (linear terms) to decide the radial stability.

\subsection{Specialization to the physical $u<0$ sector}
\label{app:poincare_u_negative}

The physically relevant negative SF potential is described by:
\begin{equation}
    u<0 \quad\land\quad \Omega=\Omega_r\ge 0\,,
\end{equation}
therefore:
\begin{equation}
    \Omega-3u>0\,.
\end{equation}
Therefore, from \eqref{eq:rprime_equator} we obtain that:
\begin{itemize}
\item every equatorial direction compatible with $u<0$ and $\Omega\ge 0$ is \emph{radially attracting to infinity} in 'e-fold time';
\item The equator is normally attracting on the Poincar\'e sphere (Appendix~\ref{app:poincare_linear}).
\end{itemize}

\paragraph{Main result (physical $u<0$ sector)}
For trajectories that enters the region $u<0$ with $\Omega\ge 0$, the reduced flow generically approaches the equator at infinity, which corresponds to $r\to\infty$ and thus $H\to 0$ (since $x,u,\Omega\propto H^{-2}$). In other words, in the negative SF potential sector, the reduced Hubble-normalized system is generically drawn to the $H\to 0$ boundary, and the $N$-time formulation becomes singular.

This is precisely why a bounce/cyclic completion requires an $H$-regular dynamical extension.

\subsection{Pole directions and their physical meaning}
\label{app:poincare_poles}

At the pole directions $\mathbf{s}_\pm=(\pm1,0,0)$ one obtains:
\begin{equation}
    \frac{x}{r}\to\pm1 \quad\land\quad |x|\to\infty\,.
\end{equation}

In terms of the original variables, this occurs when:
\begin{equation}
    |H|<\left|\dot\phi\right|
\end{equation}
(or when $\dot\phi$ grows while $H$ decays).

On the sphere, both poles are tangentially unstable (repelling in $(b,c)$), so that the generic orbits do not remain in pole directions. They are pushed toward equatorial directions, consistent with the generic approach to the $H\to 0$ boundary diagnosed above.

\subsection{Complete infinity classification}
\label{app:poincare_summary}
For the reduced constant $\lambda_{\phi*}$ system \eqref{eq:poly_reduced}:
\begin{enumerate}[label=(\arabic*)]
\item \textbf{Isolated pole points} $(a,b,c)=(\pm1,0,0)$: repelling on $S^2$ (tangentially unstable), not generic late-time infinity states;
\item \textbf{Equatorial critical circle} $a=0$, $b^2+c^2=1$: normally attracting on $S^2$. The refined ($m=2$) dynamics shows that in the physically relevant sector, the radial flow satisfies:
\begin{equation}
    r'\sim r(\Omega-3u)>0
\end{equation}
therefore, the equator is also \emph{radially attracting} to infinity in 'e-fold time'.
\end{enumerate}
Therefore, in the $u<0$ sector the reduced Hubble-normalized system generically runs into the $H\to 0$ boundary, and any continuation
requires an $H$-regular completion.

\subsection{Extension to the full system}
\label{app:poincare_4D}
On the constant $\lambda_{\phi*}$ branches, the above analysis is exact. Namely, in the 4D system:
\begin{equation}
\frac{d \lambda_{\phi}}{dN}=-\sqrt6\,\Bigl(\Gamma_\phi(\lambda_\phi)-1\Bigr)\,\lambda_\phi^2\,x\,,
\end{equation}
which yields that:
\begin{equation}
    \frac{d \lambda_{\phi}}{dN}\propto x
\end{equation}
Along the equator at infinity one obtains:
\begin{equation}
    a=\frac{x}{r}\to 0\,,
\end{equation}
and therefore $\lambda_\phi'$ is subleading in the infinity scaling and does not modify the leading global picture - the approach to equatorial infinity in the physical $u<0$ sector remains the dominant behavior. On the other hand, along large $|x|$ directions, $\lambda_\phi$ can be driven toward a constant branch (depending on the SF potential closure equation), which justifies using the constant $\lambda_{\phi*}$ reduced analysis as the universal leading description of the $u<0$ approach to $H\to 0$.

\section{Numerical evolution plots for the Planck 2018 initial conditions}
\label{app:num-plots}
\begin{figure}[htbp]
\centering
\includegraphics[width=1\linewidth]{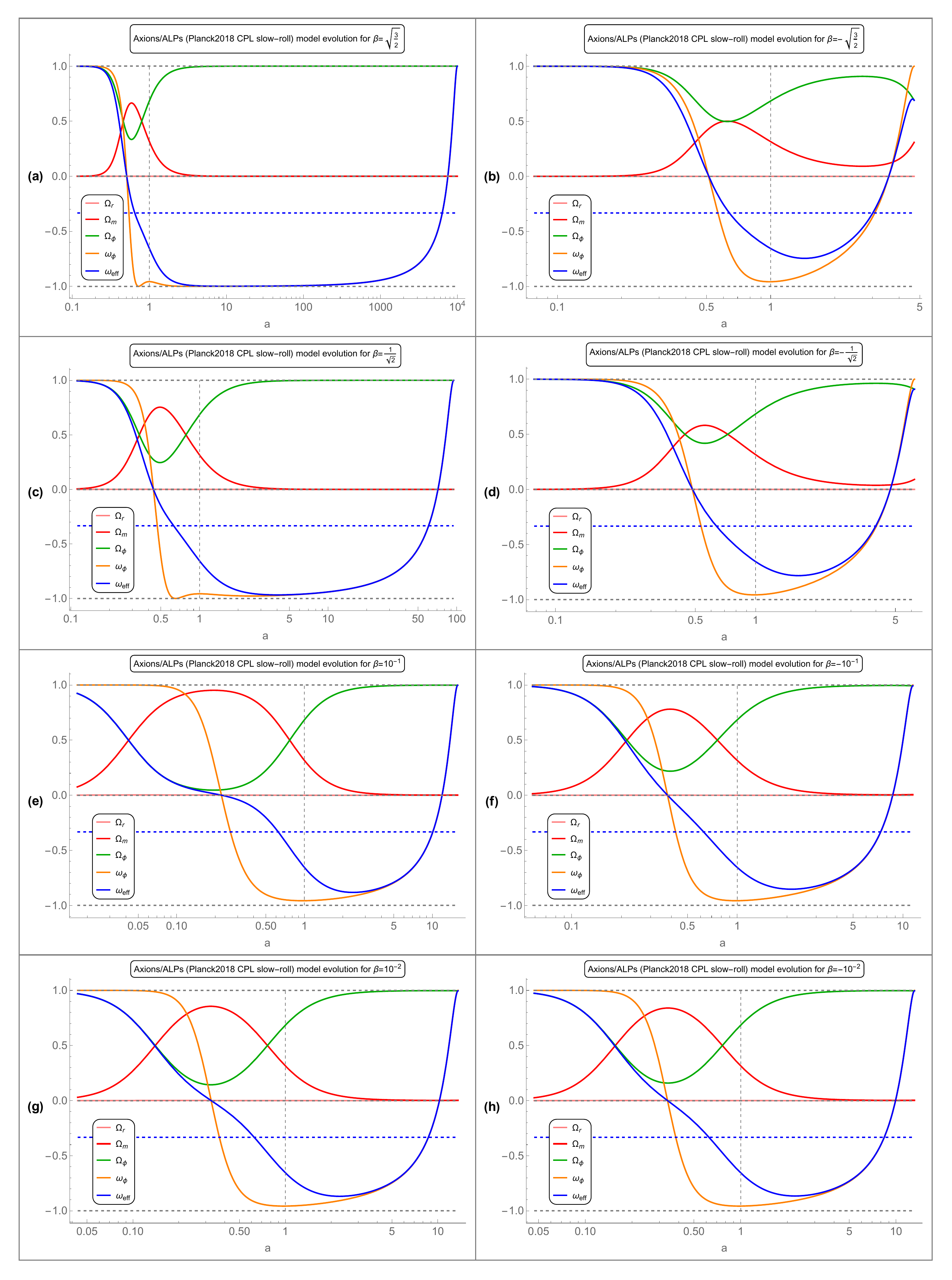}
\caption{Evolution of the NMC axions/ALPs model in the slow-roll approximation for Planck 2018 initial data.}
\label{fig:evol_planck2018_cplsl_axions}
\end{figure}

\begin{figure}[htbp]
\centering
\includegraphics[width=1\linewidth]{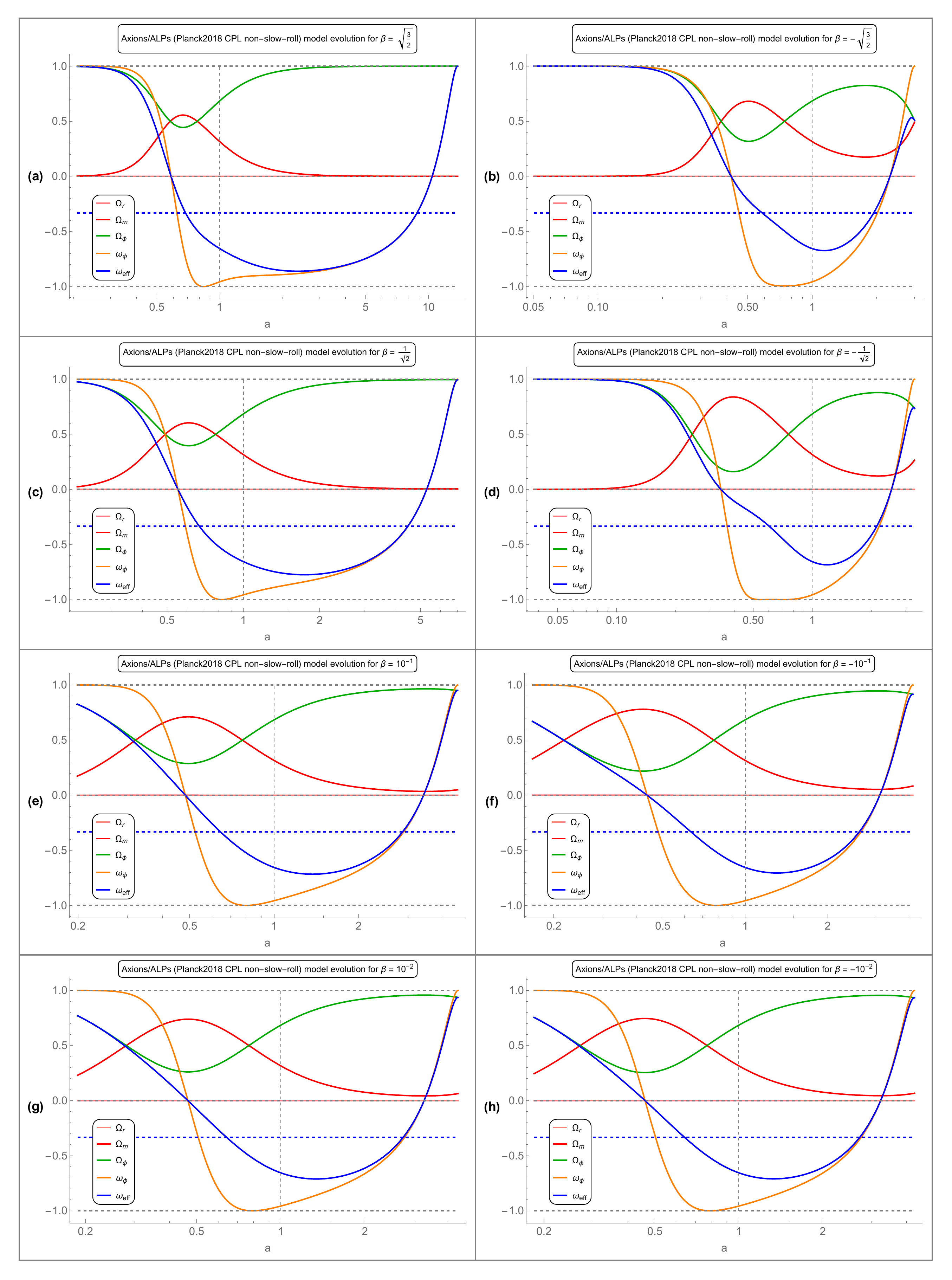}
\caption{Evolution of the NMC axions/ALPs model in the non-slow-roll CPL approximation for Planck 2018 initial data.}
\label{fig:evol_planck2018_cplnsl_axions}
\end{figure}

\begin{figure}[htbp]
\centering
\includegraphics[width=1\linewidth]{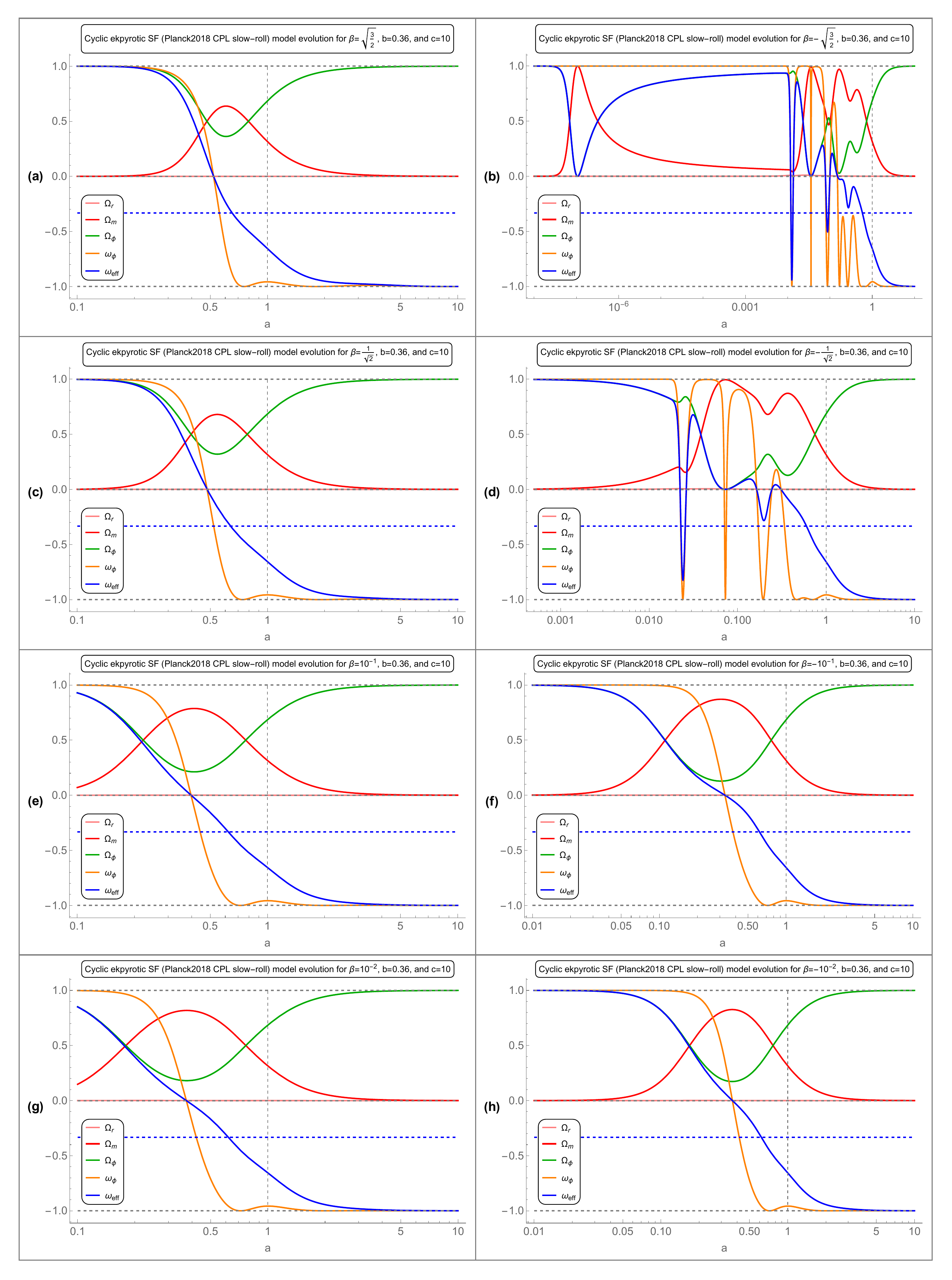}
\caption{Evolution of the NMC cyclic ekpyrotic model in the slow-roll approximation for Planck 2018 initial data.}
\label{fig:evol_planck2018_cplsl_cyclic}
\end{figure}

\begin{figure}[htbp]
\centering
\includegraphics[width=1\linewidth]{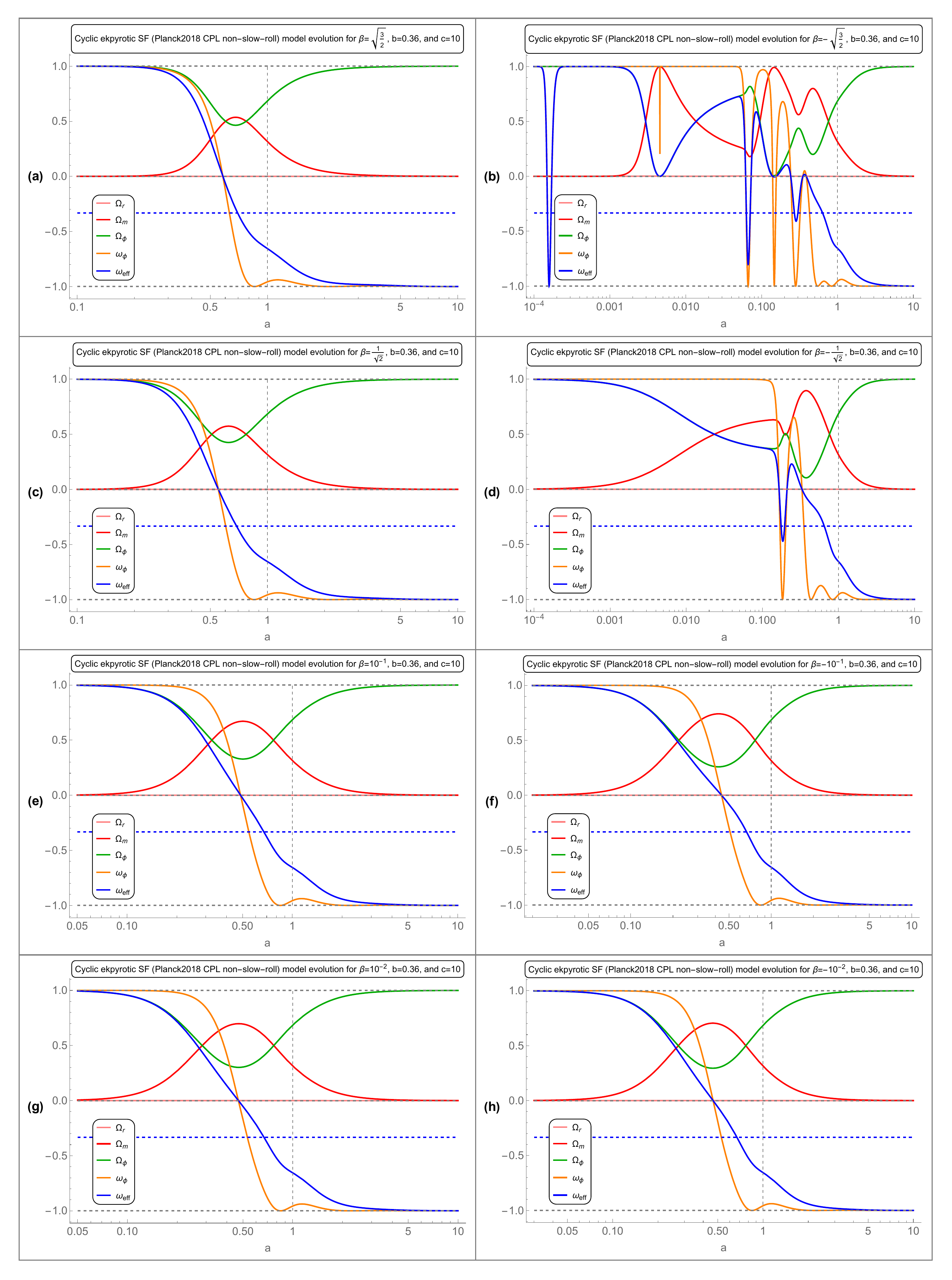}
\caption{Evolution of the NMC cyclic ekpyrotic model in the non-slow-roll CPL approximation for Planck 2018 initial data.}
\label{fig:evol_planck2018_cplnsl_cyclic}
\end{figure}

\begin{figure}[htbp]
\centering
\includegraphics[width=1\linewidth]{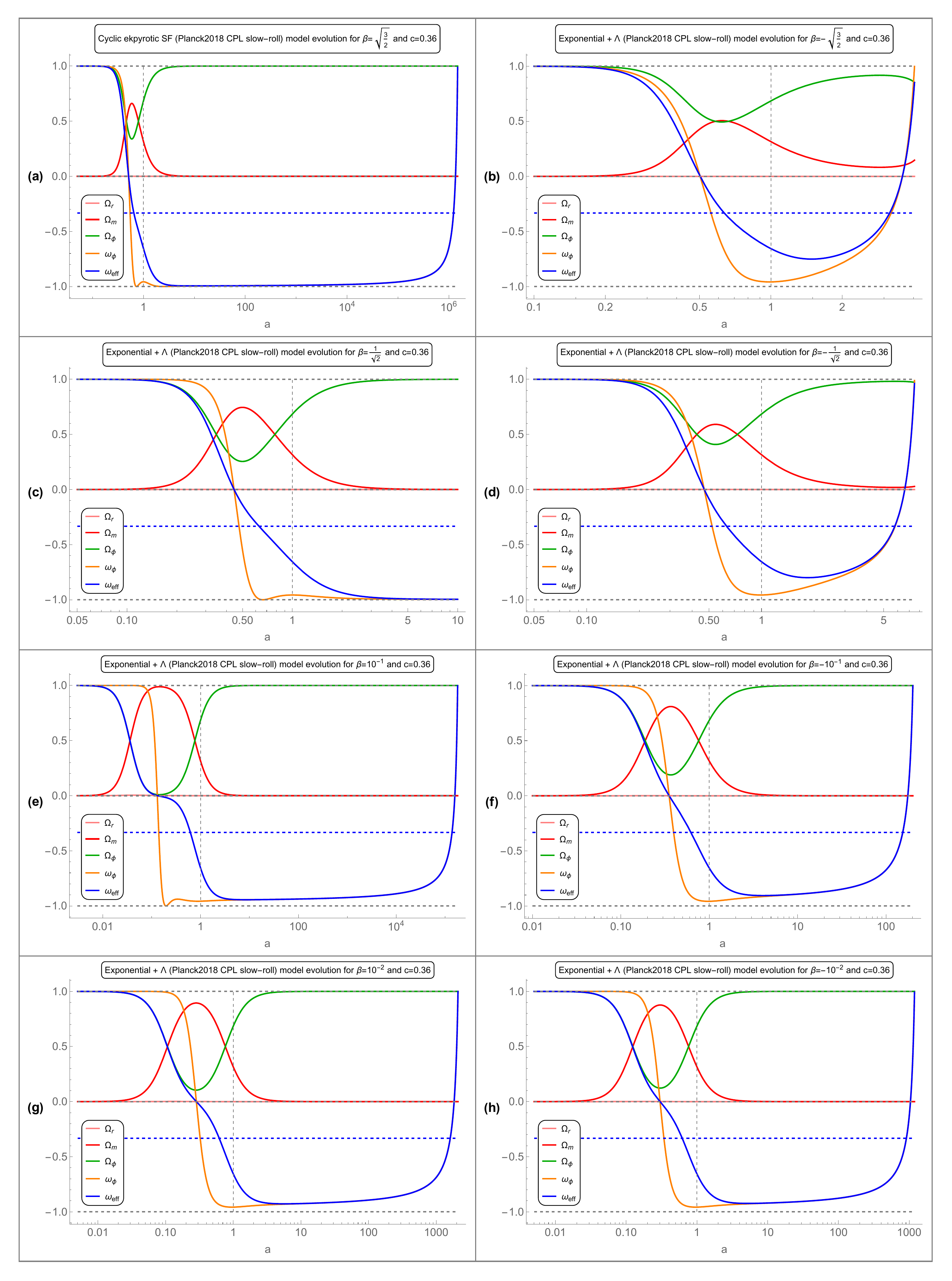}
\caption{Evolution of the NMC exponential with constant model in the slow-roll approximation for Planck 2018 initial data.}
\label{fig:evol_planck2018_cplsl_expl}
\end{figure}

\begin{figure}[htbp]
\centering
\includegraphics[width=1\linewidth]{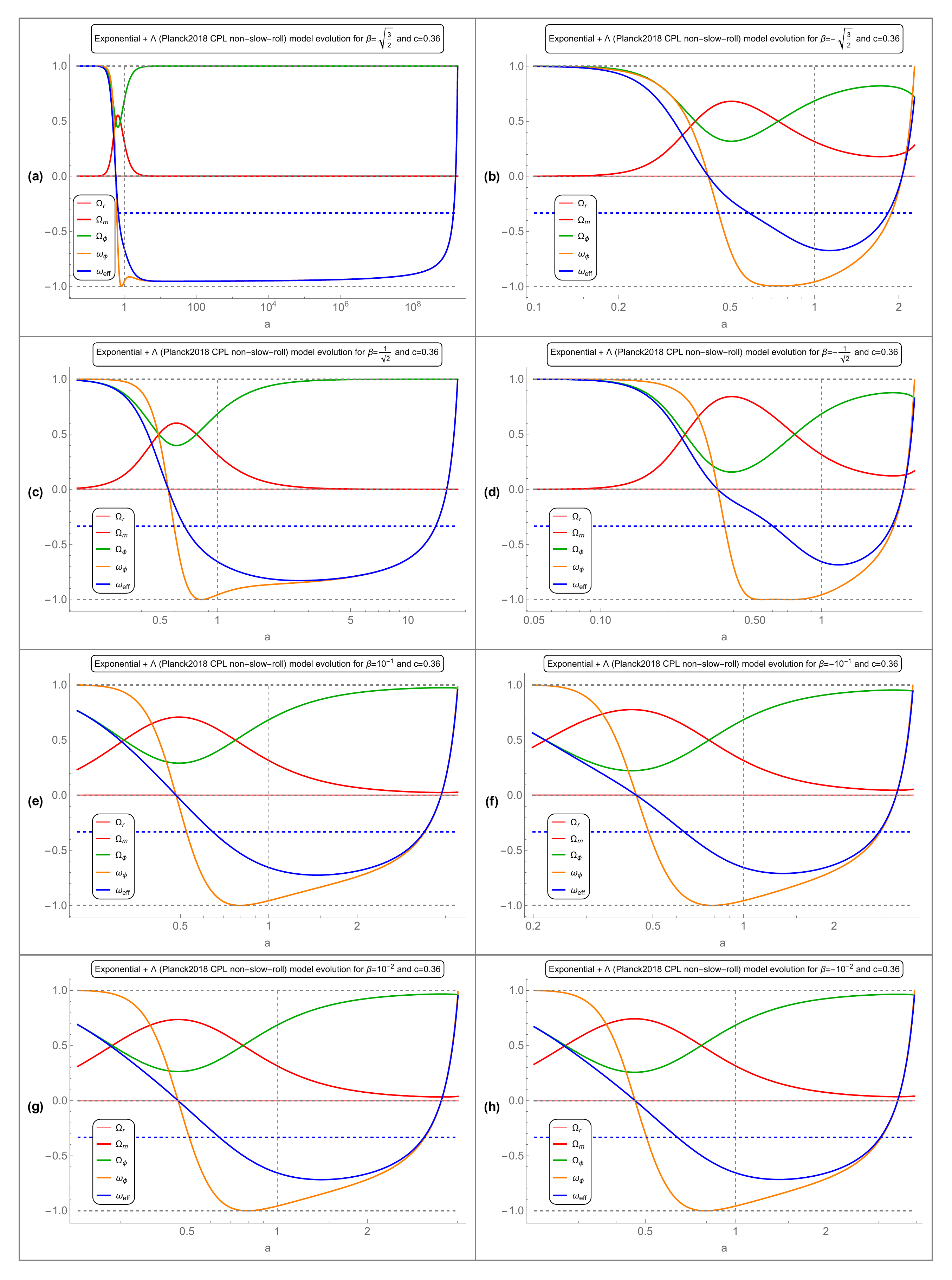}
\caption{Evolution of the NMC exponential with constant model in the non-slow-roll CPL approximation for Planck 2018 initial data.}
\label{fig:evol_planck2018_cplnsl_expl}
\end{figure}

\begin{figure}[htbp]
\centering
\includegraphics[width=1\linewidth]{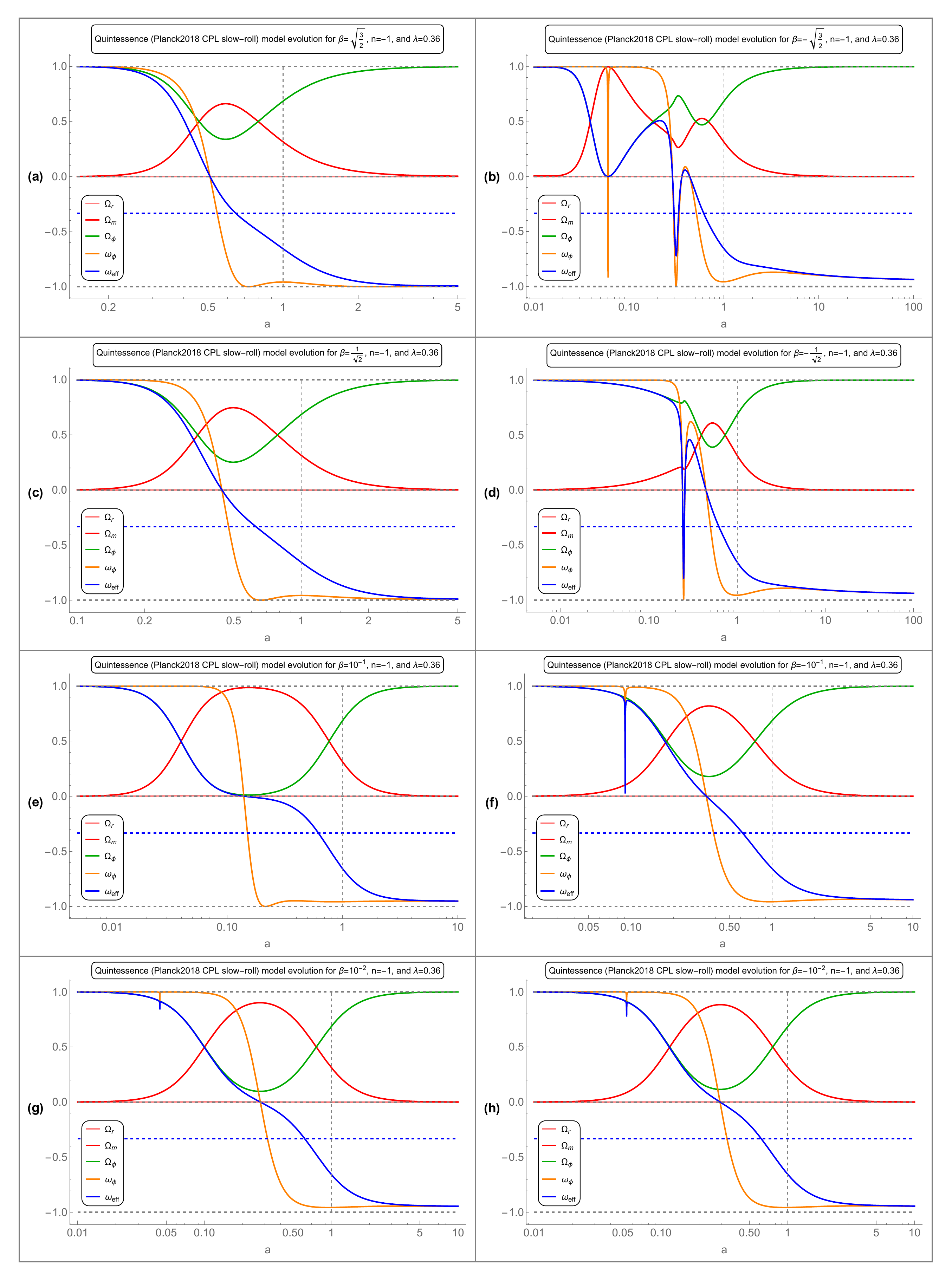}
\caption{Evolution of the NMC quintessence (n=-1) model in the slow-roll approximation for Planck 2018 initial data.}
\label{fig:evol_planck2018_cplsl_quint_n1}
\end{figure}

\begin{figure}[htbp]
\centering
\includegraphics[width=1\linewidth]{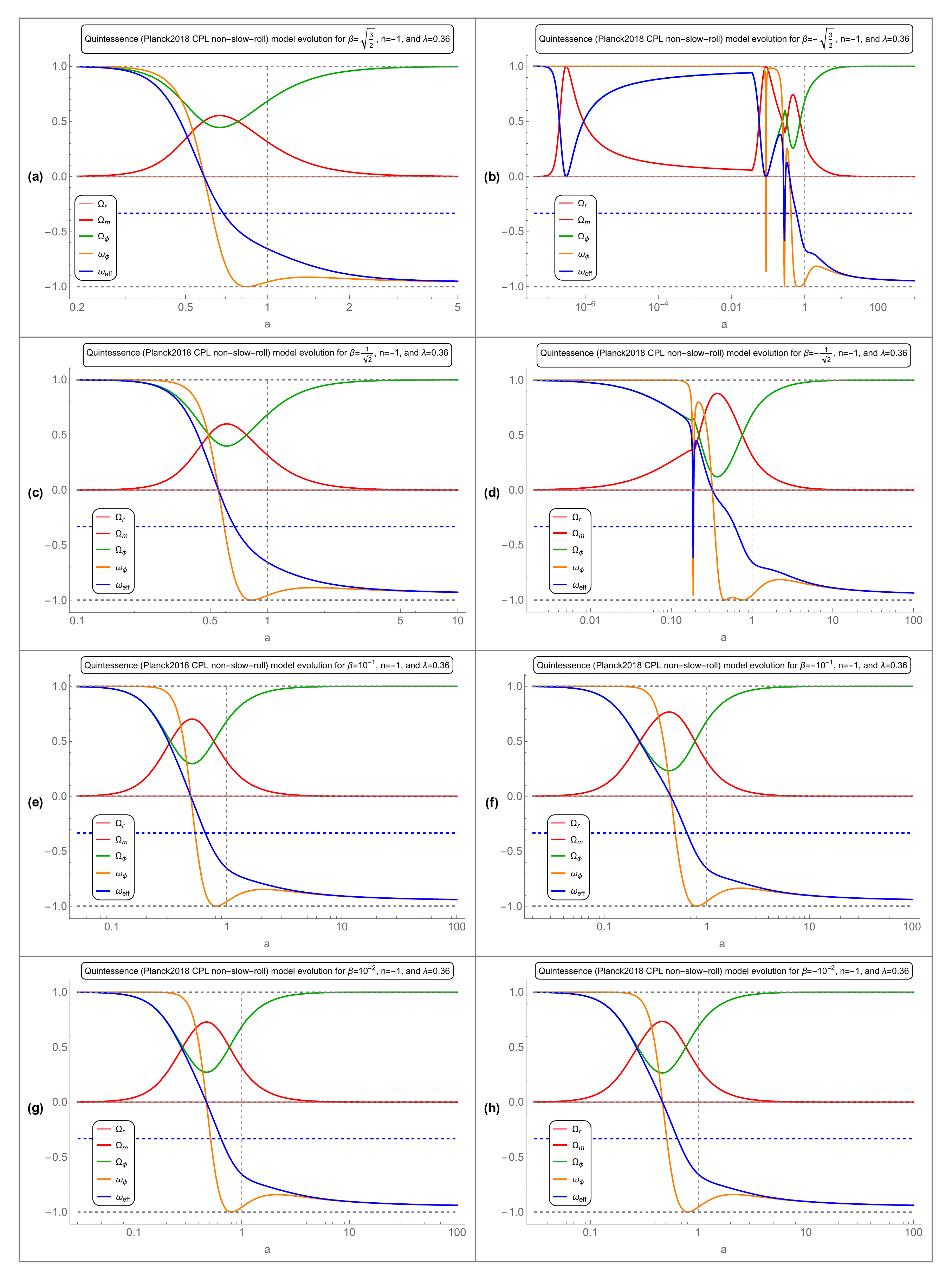}
\caption{Evolution of the NMC quintessence (n=-1) model in the non-slow-roll CPL approximation for Planck 2018 initial data.}
\label{fig:evol_planck2018_cplnsl_quint_n1}
\end{figure}

\begin{figure}[htbp]
\centering
\includegraphics[width=1\linewidth]{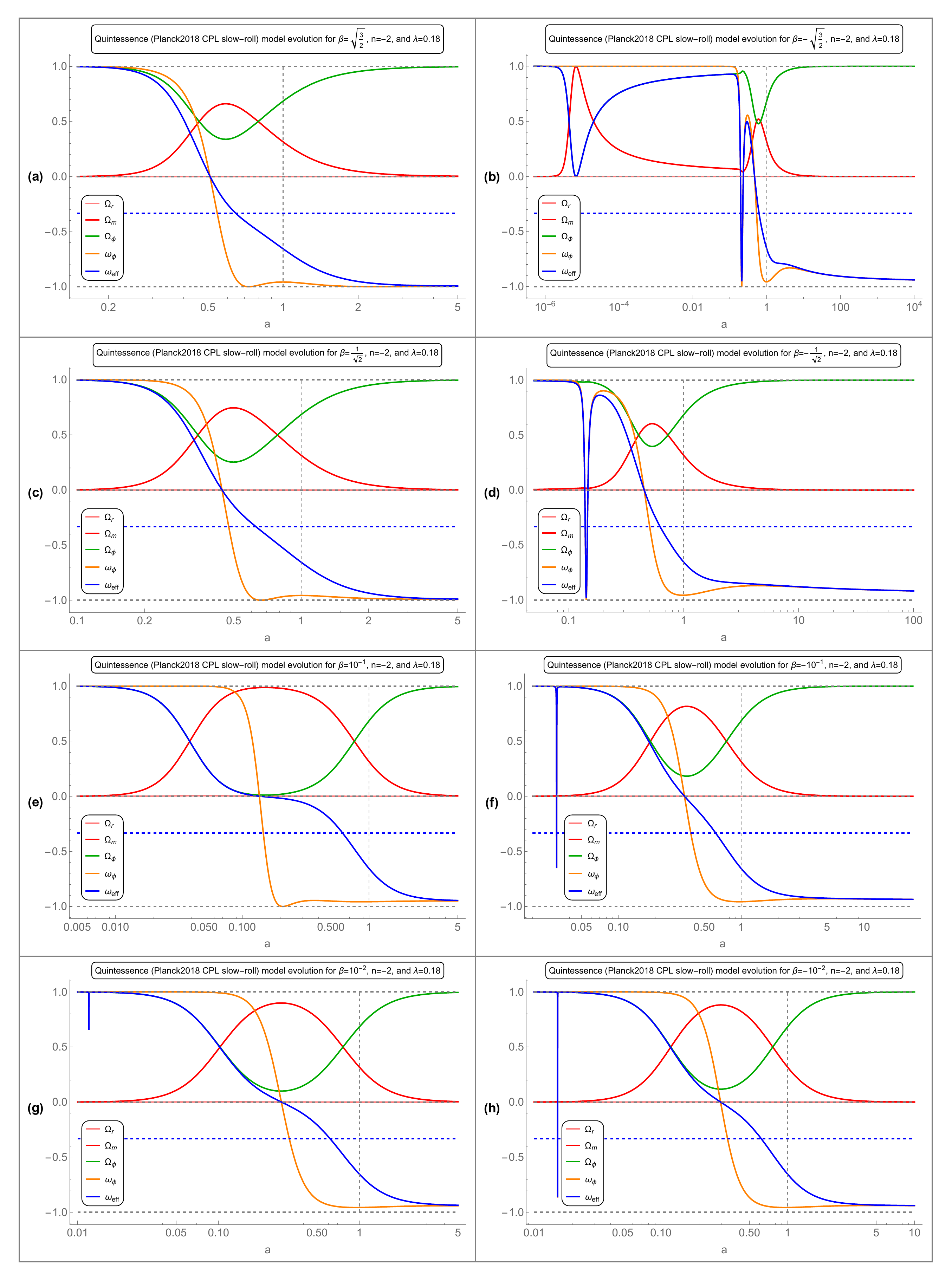}
\caption{Evolution of the NMC quintessence (n=-2) model in the slow-roll approximation for Planck 2018 initial data.}
\label{fig:evol_planck2018_cplsl_quint_n2}
\end{figure}

\begin{figure}[htbp]
\centering
\includegraphics[width=1\linewidth]{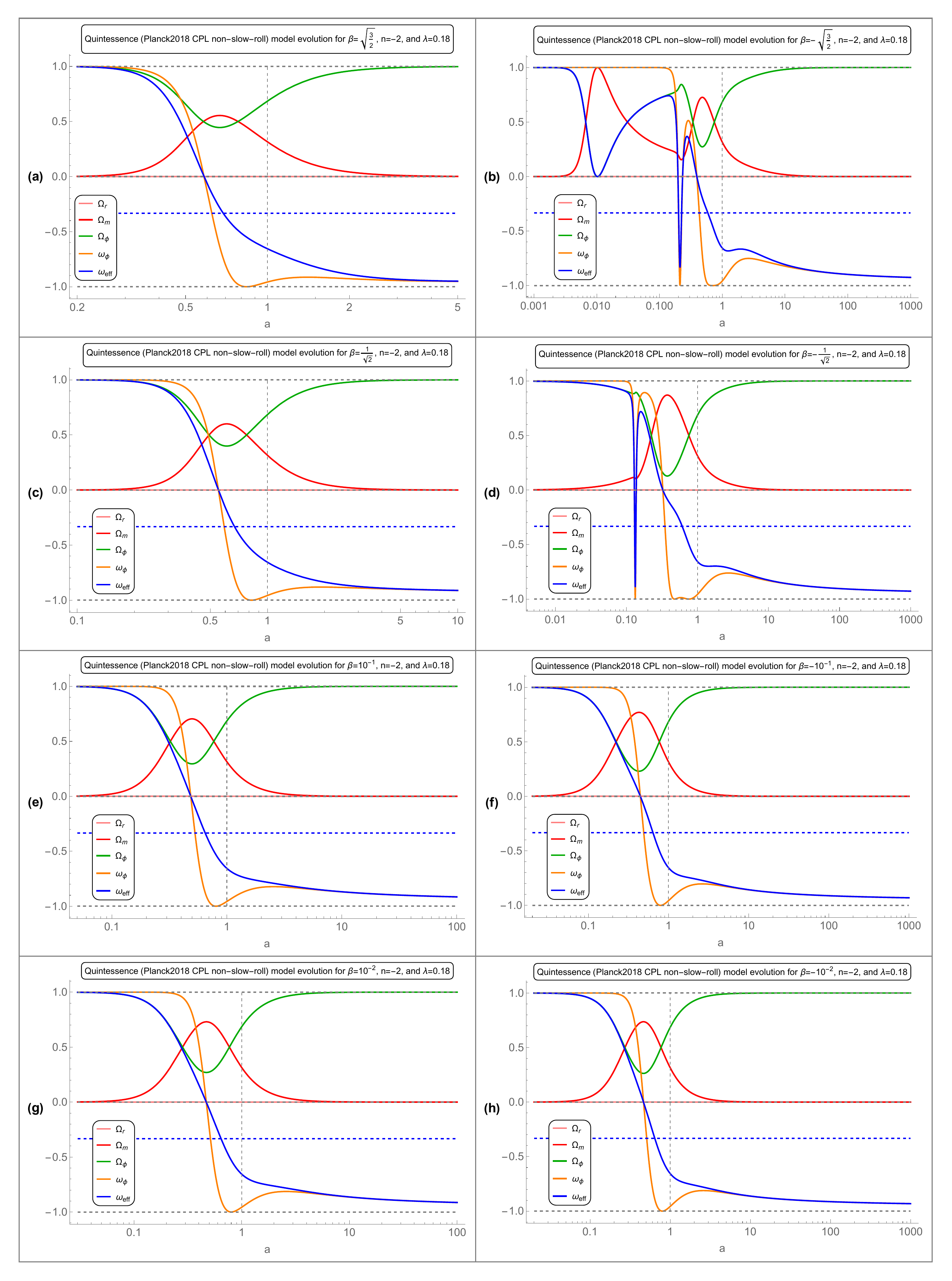}
\caption{Evolution of the NMC quintessence (n=-2) model in the non-slow-roll CPL approximation for Planck 2018 initial data.}
\label{fig:evol_planck2018_cplnsl_quint_n2}
\end{figure}

\begin{figure}[htbp]
\centering
\includegraphics[width=1\linewidth]{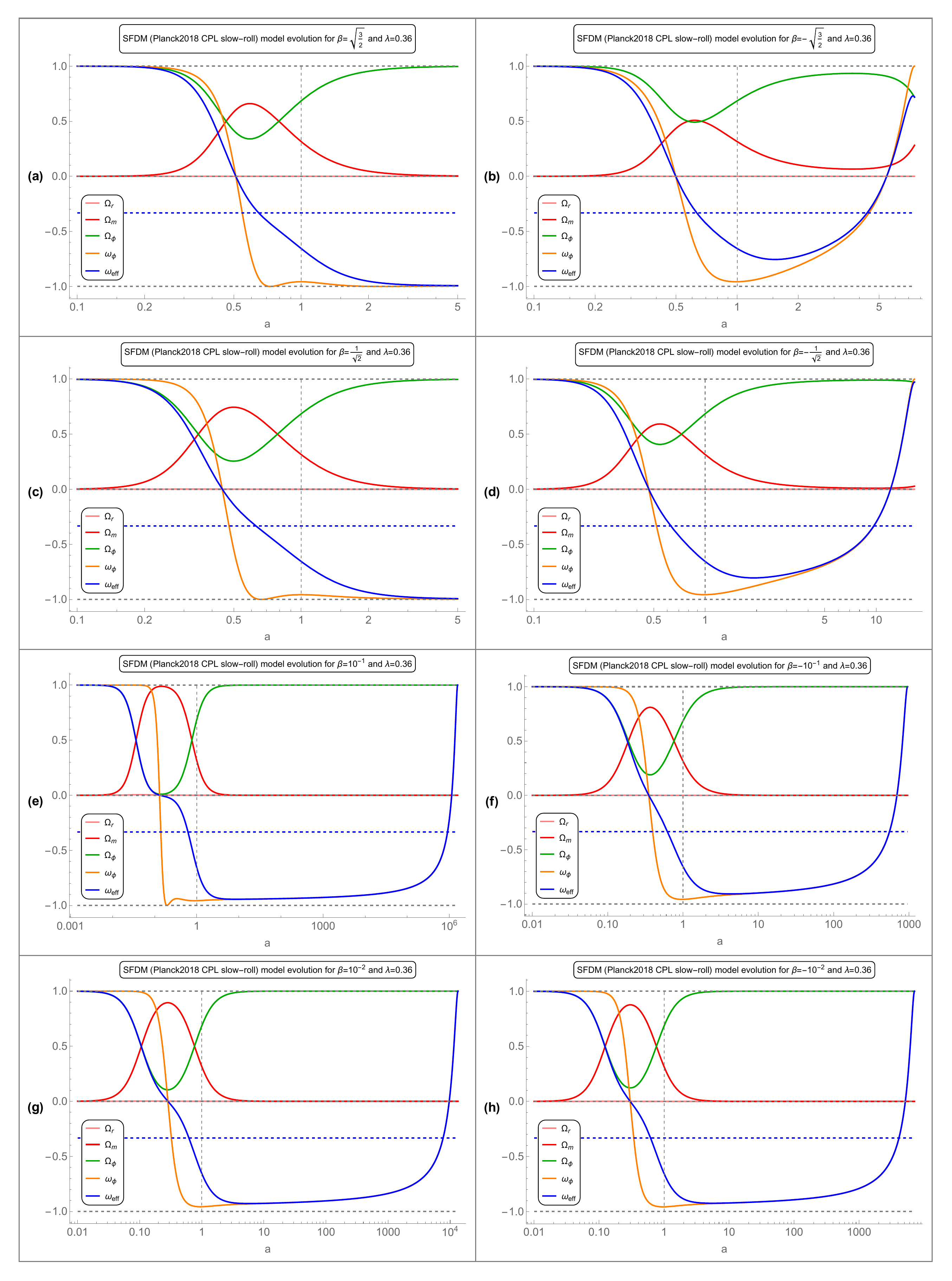}
\caption{Evolution of the NMC SFDM model in the slow-roll approximation for Planck 2018 initial data.}
\label{fig:evol_planck2018_cplsl_sfdm}
\end{figure}

\begin{figure}[htbp]
\centering
\includegraphics[width=1\linewidth]{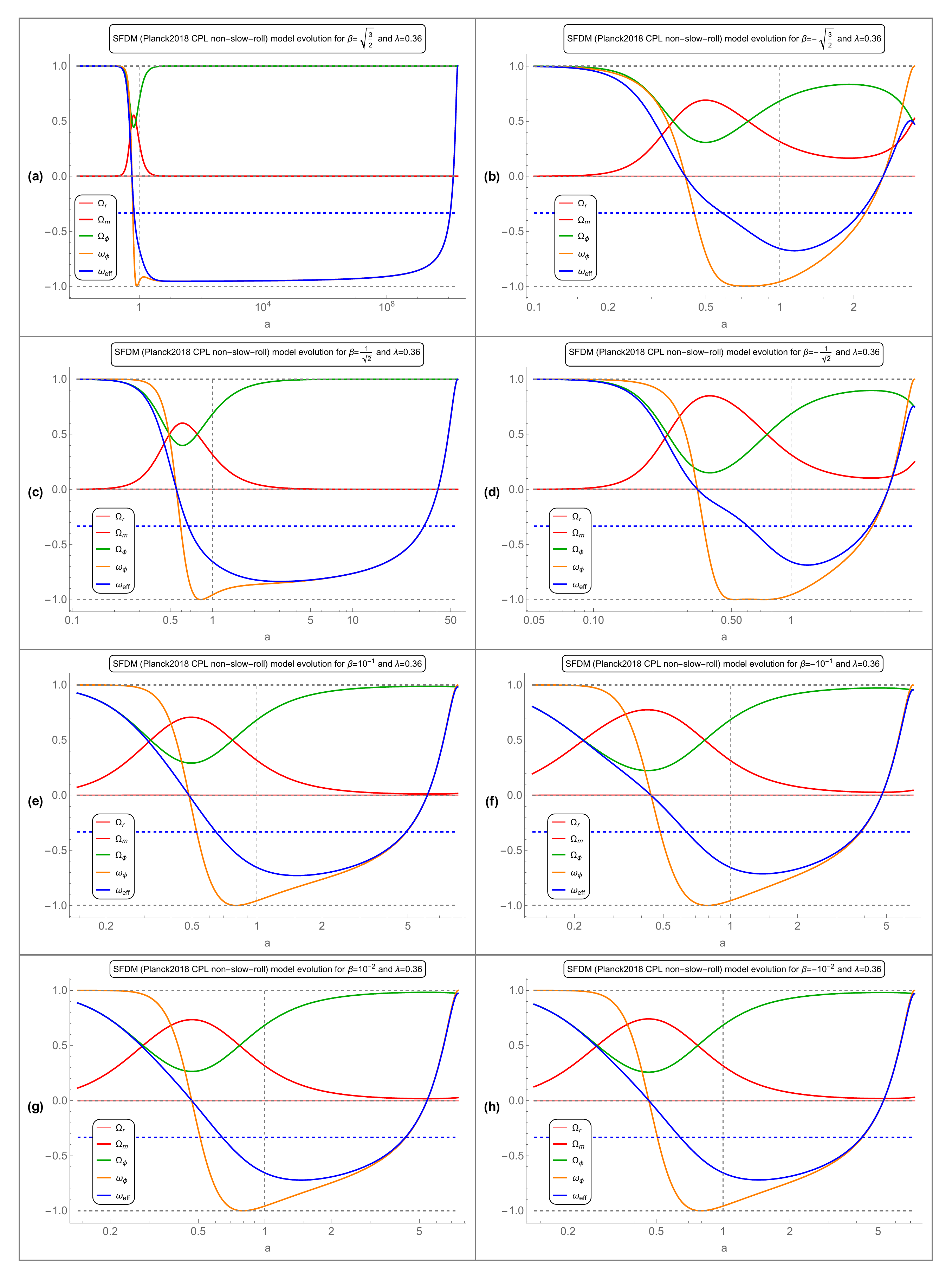}
\caption{Evolution of the NMC SFDM model in the non-slow-roll CPL approximation for Planck 2018 initial data.}
\label{fig:evol_planck2018_cplnsl_sfdm}
\end{figure}

\FloatBarrier

\bibliographystyle{iopart-num}
\bibliography{refs}

\end{document}